\documentclass[12pt]{article}
\usepackage{amsmath}

\usepackage{amssymb}
\usepackage{bm}
\usepackage{braket}
\usepackage[dvips]{graphicx}
\usepackage{comment}

\begin{document}

\baselineskip=17pt

\begin{titlepage}
\rightline{\tt arXiv:2305.11634}
\begin{center}
\vskip 1.5cm
\baselineskip=22pt
{\Large \bf {Correlation Functions Involving Dirac Fields}}\\
{\Large \bf {from Homotopy Algebras I: The Free Theory}}
\end{center}
\begin{center}
\vskip 1.0cm
{\large Keisuke Konosu and Yuji Okawa}
\vskip 1.0cm
{\it {Graduate School of Arts and Sciences, The University of Tokyo}}\\
{\it {3-8-1 Komaba, Meguro-ku, Tokyo 153-8902, Japan}}\\
konosu-keisuke@g.ecc.u-tokyo.ac.jp,
okawa@g.ecc.u-tokyo.ac.jp
\vskip 2.0cm

{\bf Abstract}
\end{center}

\noindent
We extend the formula for correlation functions of scalar field theories
in terms of quantum $A_\infty$ algebras, presented in arXiv:2203.05366,
to incorporate Dirac fields.
We use a description that is analogous to string field theory,
and the formula for correlation functions takes the same form
for both scalar fields and Dirac fields. 
We prove that correlation functions from our formula satisfy
the Schwinger-Dyson equations in the free theory.
The proof for interacting theories is presented
in the companion paper arXiv:2305.13103 by one of the authors.
We also explain the relation of our formula
to the definition of correlation functions
in the approach by Costello and Gwilliam
based on factorization algebras.

\end{titlepage}

\tableofcontents
\section{Introduction}

Homotopy algebras such as $A_\infty$ algebras~\cite{Stasheff:I, Stasheff:II, Getzler-Jones, Markl, Penkava:1994mu, Gaberdiel:1997ia}
and $L_\infty$ algebras~\cite{Zwiebach:1992ie, Markl:1997bj}
have been useful in constructing actions of string field theory.
Recently we recognized that homotopy algebras can also be useful
in solving the theory.
A key ingredient is a projection onto a subspace of the vector space $\mathcal{H}$
which we use in the homotopy algebra to describe the theory.
In, e.g.~the case of open string field theory,
the vector space $\mathcal{H}$ is the space of string fields,
and it is decomposed as
\begin{equation}
\mathcal{H} = \ldots \oplus \mathcal{H}_0 \oplus \mathcal{H}_1
\oplus \mathcal{H}_2 \oplus \mathcal{H}_3 \oplus \ldots \,,
\end{equation}
where $\mathcal{H}_n$ is the space for string fields of ghost number $n$.
The action is written in terms of string fields in~$\mathcal{H}_1$.
When we consider scattering amplitudes,
we use the projection onto on-shell states~\cite{Kajiura:2003ax}.
The projection onto massless fields is relevant
in the context of the low-energy effective action~\cite{Sen:2016qap, Erbin:2020eyc, Koyama:2020qfb, Arvanitakis:2020rrk, Arvanitakis:2021ecw}.
The projection onto physical states was used
in the discussion of the relation
between covariant and light-cone string field theories~\cite{Erler:2020beb}.
We use various projections depending on our purpose,
but homotopy algebras provide a universal formula for all cases.

To explain the universal formula for $A_\infty$ algebras,
it is convenient to consider linear operators
acting on the vector space $T \mathcal{H}$ defined by
\begin{equation}
  T\mathcal{H} = \mathcal{H}^{\otimes 0} \oplus \mathcal{H} \oplus \mathcal{H}^{\otimes 2} \oplus \mathcal{H}^{\otimes 3}\oplus\ldots\,,
\end{equation}
where
\begin{equation}
  \mathcal{H}^{\otimes n} = \underbrace{\mathcal{H}\otimes\mathcal{H}\otimes\ldots\otimes\mathcal{H}}_{n} \,.
\end{equation}
We denote the operator which describes the action of the theory by ${\bf Q}+{\bm m}$,
where ${\bf Q}$ is for the free part and ${\bm m}$ is for the interactions.
The projection onto a subspace of $\mathcal{H}$ can be described
by a projection operator on $T \mathcal{H}$, and we denote it by ${\bf P}$.
One more important operator acting on $T \mathcal{H}$ is ${\bm h}$,
which describes the propagator associated with the degrees of freedom that are projected out.
The universal formula is then given by
\begin{equation}
\pi_1 \, {\bf P} \, {\bf Q} \, {\bf P}
+\pi_1 \, {\bf P} \, {\bm m} \, \frac{1}{{\bf I}+{\bm h} \, {\bm m}} \, {\bf P} \,,
\label{universal-formula}
\end{equation}
where ${\bf I}$ is the identity operator on $T \mathcal{H}$
and $\pi_1$ is the projection operator onto $\mathcal{H}$.
In the context of, e.g.~the projection onto massless fields,
this operator describes the effective action for massless fields
where massive fields are classically integrated out.
For a projection onto physical states, the first term $\pi_1 \, {\bf P} \, {\bf Q} \, {\bf P}$
of Eq.~\eqref{universal-formula} vanishes, and the tree-level scattering amplitudes
are obtained from the second term of Eq.~\eqref{universal-formula}.\footnote{
See, e.g.~Refs.~\cite{Konopka:2015tta} and \cite{Kunitomo:2020xrl}
for discussions on the tree-level S-matrix of superstring field theory in this context.}

Furthermore, this universal formula can be extended to the quantum theory.
It is given by
\begin{equation}
\pi_1 \, {\bf P} \, {\bf Q} \, {\bf P}
+\pi_1 \, {\bf P} \, {\bm m} \,
\frac{1}{{\bf I}+{\bm h} \, {\bm m} +i \hbar \, {\bm h} \, {\bf U}} \, {\bf P} \,,
\label{quantum-universal-formula}
\end{equation}
where ${\bf U}$ can be thought of as being dual to the Laplacian
in the Batalin-Vilkovisky formalism~\cite{Batalin:1981jr, Batalin:1983ggl, Schwarz:1992nx}.
In the context of the projection onto massless fields,
this operator describes the effective action for massless fields
where massive fields are integrated out in the path integral.
For a projection onto physical states, the first term $\pi_1 \, {\bf P} \, {\bf Q} \, {\bf P}$
of Eq.~\eqref{quantum-universal-formula} again vanishes,
and the scattering amplitudes including contributions from loop diagrams
are obtained from the second term of Eq.~\eqref{quantum-universal-formula}.

Homotopy algebras can be used not only for string field theory
but also for quantum field theory~\cite{Hohm:2017pnh, Jurco:2018sby, Nutzi:2018vkl, Arvanitakis:2019ald, Macrelli:2019afx, Jurco:2019yfd, Saemann:2020oyz}.
The vector space $\mathcal{H}$ for quantum field theories
without gauge symmetries consists of two sectors.
In our convention which is analogous to open string field theory,
they are denoted by $\mathcal{H}_1$ and $\mathcal{H}_2$,
\begin{equation}
\mathcal{H} = \mathcal{H}_1 \oplus \mathcal{H}_2 \,,
\end{equation}
and the action is written in terms of elements of~$\mathcal{H}_1$.
The vector space~$\mathcal{H}_2$, on the other hand, is for antifields
in the Batalin-Vilkovisky formalism.
While we can describe the action of quantum field theory
in terms of the operator ${\bf Q}+{\bm m}$,
this description of the theory is rather trivial for theories
without gauge symmetries
and does not seem useful.
On the other hand, 
the formulas in Eqs.~\eqref{universal-formula} and~\eqref{quantum-universal-formula}
can be useful even for theories without gauge symmetries.
Furthermore, we expect that we can obtain insight into string field theory from the investigation
of quantum field theory, as the formulas are universal.

We are, e.g.~interested in correlation functions of gauge-invariant operators
in open superstring field theory
when we pursue the program of providing a framework
to prove the anti-de~Sitter/conformal field theory (AdS/CFT) correspondence~\cite{Okawa:2020llq}.
Calculations of such correlation functions would be complicated,
and we need efficient tools to reveal the structure of those calculations.
We consider that quantum $A_\infty$ algebras can do the job.
Since quantum $A_\infty$ algebras have not been fully explored,
it would be useful to develop our technology for quantum $A_\infty$ algebras
first in quantum field theory.
This was the motivation in Ref.~\cite{Okawa:2022sjf}
where it was found that
correlation functions can also be described using quantum $A_\infty$ algebras
and it was demonstrated explicitly in scalar field theories.\footnote{
The formula for correlation functions in Ref.~\cite{Okawa:2022sjf} was found from the fact
that Feynman diagrams can be generated algebraically
in the framework of homotopy algebras~\cite{Kajiura:2003ax, Doubek:2017naz, Masuda:2020tfa}.
See Refs.~\cite{Gwilliam:2012jg, Costello:2016vjw, Costello:2021jvx, Chiaffrino:2021pob} for discussions
on correlation functions in the framework of the Batalin-Vilkovisky formalism
and recent development in Ref.~\cite{DimitrijevicCiric:2023hua}.
We discuss the relation of our approach
to Refs.~\cite{Kajiura:2003ax, Doubek:2017naz, Masuda:2020tfa, Costello:2016vjw, Costello:2021jvx}
in Section~\ref{conclusion-section},
and we briefly explain the definition of correlation functions in the approach
by Costello and Gwilliam~\cite{Costello:2016vjw, Costello:2021jvx} based on factorization algebras
and its equivalence with ours
in Appendix~\ref{factorization-algebra-appendix}.
}
As open superstring field theory contains the Ramond sector
which describes spacetime fermions, we need to extend
the formula for correlation functions of scalar field theories in Ref.~\cite{Okawa:2022sjf}
to incorporate fermions.
In this paper, we initiate this generalization
and describe Dirac fields using quantum $A_\infty$ algebras.
Although we aim at applying the framework to open superstring field theory,
we expect that our approach would also be useful within the context of quantum field theory.

Technically we need to reproduce the antisymmetry of fermions in the correlation functions.
The vector space~$\mathcal{H}$ is graded with respect to $\mathbb{Z}_2$ degree,
and both scalar fields and Dirac fields are described by degree-even elements of~$\mathcal{H}_1$.
The naive extension of the formula for scalar fields in Ref.~\cite{Okawa:2022sjf}
to Dirac fields results in correlation functions that are symmetric
under the exchange of Dirac fields.
One possibility is to use super vector spaces.\footnote{
We would like to thank Kantaro Ohmori for helpful discussions on this point.
}
In addition to the grading from the homotopy algebra,
we introduce the $\mathbb{Z}_2$ grading from the super vector space
to distinguish bosons and fermions.
Mathematically, this approach would be more natural,
but superstring field theory tells us of a different approach
and we find that the latter is more convenient for our purposes.
In open superstring field theory, the Neveu-Schwarz sector for spacetime bosons
and the Ramond sector for spacetime fermions are both described
by degree-even string fields in~$\mathcal{H}_1$,
but the degree-even string field $\Phi$
is schematically expanded as
\begin{equation}
\Phi = \sum_{i} \int d^{10} k \, \varphi_i (k) \, | \, i \, ; k \, \rangle
+\sum_{\alpha} \int d^{10} k \, \psi_\alpha (k) \, | \, \alpha \, ; k \, \rangle \,,
\end{equation}
where $\varphi_i (k)$ are bosonic fields
and $\psi_\alpha (k)$ are fermionic fields
with $i$ and $\alpha$ collectively labeling various fields.
The important point is that
the states $| \, i \, ; k \, \rangle$ in the Neveu-Schwarz sector
are degree even
and the states $| \, \alpha \, ; k \, \rangle$ in the Ramond sector
are degree odd
so that the string field $\Phi$ is degree even
when these states are combined with degree-even fields $\varphi_i (k)$
and degree-odd fields $\psi_\alpha (k)$.\footnote{
As is written, e.g.~in section 10.6 of the textbook by Polchinski~\cite{Polchinski:1998rr},
the Grassmann parity of the spacetime field
is correlated with the Grassmann parity of the state on the world-sheet
in the superstring.
We define the degree $\mathrm{deg} (A)$ of the $A_\infty$ algebra
for a state $A$ on the world-sheet
by $\mathrm{deg} (A) = \epsilon (A) +1$ mod~$2$,
where $\epsilon (A)$ is the Grassmann parity of the state $A$.
The Grassmann parity of the state
corresponding to the unintegrated vertex operator in the open superstring
is odd in the Neveu-Schwarz sector
and even in the Ramond sector,
so the degree is even in the Neveu-Schwarz sector
and odd in the Ramond sector.
On the other hand, we do not shift the grading for the spacetime Grassmann parity.
Therefore, the degree of the spacetime field
is even in the Neveu-Schwarz sector and odd in the Ramond sector.
}
We can represent elements of $\mathcal{H}_1$ for quantum field theories
in the same way.\footnote{
See~subsection~3.3 of Ref.~\cite{Kajiura:2003ax}
and appendix~A of Ref.~\cite{Masuda:2020tfa} for related discussions.
}
For a scalar field $\varphi (x)$ in $d$ dimensions,
we expand $\Phi$ in $\mathcal{H}_1$ as
\begin{equation}
\Phi = \int d^d x \, \varphi (x) \, c(x) \,,
\label{scalar-expansion-introduction}
\end{equation}
where $c(x)$ is a degree-even basis vector of $\mathcal{H}_1$.
For a Dirac field $\Psi_\alpha (x)$ in $d$ dimensions,
we expand $\Phi$ in $\mathcal{H}_1$ as
\begin{equation}
\Phi = \int d^d x \, ( \, \overline{\theta}_\alpha (x) \, \Psi_\alpha (x)
+\overline{\Psi}_\alpha (x) \, \theta_\alpha (x) \, ) \,,
\label{Dirac-expansion-introduction}
\end{equation}
where $\theta (x)$ and $\overline{\theta} (x)$ are degree-odd basis vectors of $\mathcal{H}_1$.
We will present a formula for correlation functions of Dirac fields
based on this expansion and show that they are indeed antisymmetric
under the exchange of Dirac fields.
When we consider a theory of the scalar field $\varphi (x)$
and the Dirac field $\Psi_\alpha (x)$,
we combine Eqs.~\eqref{scalar-expansion-introduction} and~\eqref{Dirac-expansion-introduction}:
\begin{equation}
\Phi = \int d^d x \, \varphi (x) \, c(x)
+\int d^d x \, ( \, \overline{\theta}_\alpha (x) \, \Psi_\alpha (x)
+\overline{\Psi}_\alpha (x) \, \theta_\alpha (x) \, ) \,.
\end{equation}
Note that we use a single $\mathbb{Z}_2$ grading in this approach,
and it is not directly related to the $\mathbb{Z}$ grading of the homotopy algebra.
For $\mathcal{H}_1$,
$c(x)$ is degree even
but $\theta_\alpha (x)$ is degree odd,
while $\Phi$ is degree even for both cases.

In Ref.~\cite{Okawa:2022sjf} for scalar field theories,
elements of $\mathcal{H}_1$ are chosen to be functions of $x$ in $d$ dimensions.
The relation between the representation of $\mathcal{H}_1$
based on the expansion~\eqref{scalar-expansion-introduction}
and that in Ref.~\cite{Okawa:2022sjf}
is analogous to the relation between wave functions
and states in quantum mechanics.
The state $| \, \Psi \, \rangle$ in quantum mechanics
can be expanded in terms of position eigenstates $| \, x \, \rangle$ as
\begin{equation}
| \, \Psi \, \rangle = \int dx \, \psi (x) \, | \, x \, \rangle \,,
\end{equation}
and the wave function $\psi (x)$ appears as coefficients in this expansion.
The basis vector $c(x)$ in Eq.~\eqref{scalar-expansion-introduction}
plays the role of $| \, x \, \rangle$
in this analogy.

The rest of the paper is organized as follows.
After we introduce the necessary ingredients from $A_{\infty}$ algebras
in Section~\ref{A_infinity-section},
we explain the description of scalar field theories
based on the expansion~\eqref{scalar-expansion-introduction} in Section~\ref{scalar-section}.
We then present our formula for correlation functions involving Dirac fields
in Section~\ref{Dirac-section}.
Section~\ref{conclusion-section} is devoted to conclusions and discussion.
We only consider the free theory in this paper,
and the interacting theory involving Dirac fields is described in Ref.~\cite{Konosu:2023rkm}.

\section{$A_{\infty}$ algebras}
\label{A_infinity-section}

Let us begin with a brief explanation of $A_{\infty}$ algebras.
As we mentioned in the introduction,
the theory is described by elements of the vector space $\mathcal{H}$
which is decomposed as
\begin{equation}
  \mathcal{H} = \bigoplus_{n\in\mathbb{Z}} \mathcal{H}_{n} \,. \label{graded-vector-space}
\end{equation}
The action is written in terms of degree-even elements of $\mathcal{H}_1$,\footnote{
The action we consider here is the classical action.
The master action in the Batalin-Vilkovisky formalism
is written in terms of elements in the whole vector space $\mathcal{H}$.
}
and we consider an action of the form
\begin{equation}
S = {}-\frac{1}{2} \, \omega \, ( \, \Phi, Q \, \Phi \, )
-\sum_{n=0}^\infty \, \frac{1}{n+1} \,
\omega \, ( \, \Phi \,, m_n \, ( \, \Phi \otimes \ldots \otimes \Phi \, ) \, ) \,,
\end{equation}
where $\omega \, ( \, \Phi_1 \,, \Phi_2 \, )$
defined for $\Phi_1$ and $\Phi_2$ in $\mathcal{H}$
is the symplectic form satisfying the following property:
\begin{equation}
  \omega \, ( \, \Phi_1 \,, \Phi_2 \, )
  = {}-(-1)^{\mathrm{deg} (\Phi_1) \, \mathrm{deg} (\Phi_2)} \,
  \omega \, ( \,\Phi_2 \,, \Phi_1 \, )\,.
\label{omega-symplectic}
\end{equation}
Here and in what follows we denote the degree of $\Phi$ by $\mathrm{deg} \, (\Phi)$:
it is $0$ mod $2$ when $\Phi$ is degree even
and $1$ mod $2$ when $\Phi$ is degree odd.
The degree-odd operator $Q$ describes the kinetic terms of the free theory,
and interactions of $O(\Phi^{n+1})$ are described in terms of $m_{n}$
which is a degree-odd map from $\mathcal{H}^{\otimes n}$ to $\mathcal{H}$,\footnote{
Counterterms are included in the interactions described by the operators $m_n$.
}
where
\begin{equation}
  \mathcal{H}^{\otimes n}
  = \underbrace{\, \mathcal{H}\otimes\mathcal{H}\otimes\ldots\otimes\mathcal{H} \,}_{n}
\end{equation}
for $n > 0$.
The space $\mathcal{H}^{\otimes 0}$ is a 1D vector space
equipped with a single basis vector $\bf{1}$
which is degree even and satisfies
\begin{equation}
{\bf 1} \otimes \Phi = \Phi \,, \quad \Phi \otimes {\bf 1} = \Phi
\end{equation}
for any $\Phi$ in $\mathcal{H}$,
and elements of $\mathcal{H}^{\otimes 0}$ are given by multiplying $\bf{1}$ by complex numbers.

To describe the gauge invariance of the whole action we define $M_n$ by
\begin{equation}
M_n = m_n
\end{equation}
for $n \ne 1$ and
\begin{equation}
M_1 = Q +m_1 \,.
\end{equation}
The action is invariant under the gauge transformation
written in terms of $M_n$
when the operators $M_{n}$ satisfy a set of relations called $A_{\infty}$ \textit{relations}.
A few of them are
\begin{equation}
M_1 ( M_0 ( {\bf 1} ) ) = 0 \,, \label{A_1}
\end{equation}
\begin{equation}
M_1 ( M_1 ( \Phi_1 ) ) +M_2 ( M_0 ( {\bf 1} ) \otimes \Phi_1 )
+(-1)^{\mathrm{deg} ( \Phi_1 )} M_2 ( \Phi_1 \otimes M_0 ( {\bf 1} ) ) = 0 \,, \label{A_2}
\end{equation}
\begin{equation}
\begin{split}
& M_1 ( M_2 ( \Phi_1 \otimes \Phi_2 ) ) +M_2 ( M_1 ( \Phi_1 ) \otimes \Phi_2 )
+(-1)^{\mathrm{deg}(\Phi_1)} M_2 ( \Phi_1 \otimes M_1 ( \Phi_2 ) ) \\
& +M_3 ( M_0 ( {\bf 1} ) \otimes \Phi_1 \otimes \Phi_2 )
+(-1)^{\mathrm{deg}(\Phi_1)} M_3 ( \Phi_1 \otimes M_0 ( {\bf 1} ) \otimes \Phi_2 ) \\
& +(-1)^{\mathrm{deg}(\Phi_1)+\mathrm{deg}(\Phi_2)}
M_3 ( \Phi_1 \otimes \Phi_2 \otimes M_0 ( {\bf 1} ) ) = 0 \,, \label{A_3}
\end{split}
\end{equation}
\begin{equation}
\begin{split}
& M_1 ( M_3 ( \Phi_1 \otimes \Phi_2 \otimes \Phi_3 ) )
+M_2 ( M_2 ( \Phi_1 \otimes \Phi_2 ) \otimes \Phi_3 ) \\
& +(-1)^{\mathrm{deg}(\Phi_1)} M_2 ( \Phi_1 \otimes M_2 ( \Phi_2 \otimes \Phi_3 ) )
+M_3 ( M_1 ( \Phi_1 ) \otimes \Phi_2 \otimes \Phi_3 ) \\
&+(-1)^{\mathrm{deg}(\Phi_1)} M_3 ( \Phi_1 \otimes M_1 ( \Phi_2 ) \otimes \Phi_3 )
+(-1)^{\mathrm{deg}(\Phi_1)+\mathrm{deg}(\Phi_2)}
M_3 ( \Phi_1 \otimes \Phi_2 \otimes M_1 ( \Phi_3 ) ) \\
&+M_4 ( M_0 ( {\bf 1} ) \otimes \Phi_1 \otimes \Phi_2 \otimes \Phi_3 )
+(-1)^{\mathrm{deg}(\Phi_1)} M_4 ( \Phi_1 \otimes M_0 ( {\bf 1} ) \otimes \Phi_2 \otimes \Phi_3 ) \\
&+(-1)^{\mathrm{deg}(\Phi_1)+\mathrm{deg}(\Phi_2)}
M_4 ( \Phi_1 \otimes \Phi_2 \otimes M_0 ( {\bf 1} ) \otimes \Phi_3 ) \\
&+(-1)^{\mathrm{deg}(\Phi_1)+\mathrm{deg}(\Phi_2)+\mathrm{deg}(\Phi_3)}
M_4 ( \Phi_1 \otimes \Phi_2 \otimes \Phi_3 \otimes M_0 ( {\bf 1} ) ) = 0 \,. \label{A_4}
\end{split}
\end{equation}
The algebra is called a \textit{cyclic} $A_{\infty}$ \textit{algebra} 
when the operators $M_n$ have the following cyclic properties:
\begin{equation}
\omega \, ( \, \Phi_1 \,, M_n \, ( \, \Phi_2 \otimes \ldots \otimes \Phi_{n+1} \, ) \, )
= {}-(-1)^{\mathrm{deg}(\Phi_1)} \,
\omega \, ( \, M_n ( \, \Phi_1 \otimes \ldots \otimes \Phi_n \, ) \,, \Phi_{n+1} \, ) \,,
\label{cyclic}
\end{equation}
where $\Phi_1, \ldots, \Phi_n,$ and $\Phi_{n+1}$ are elements of $\mathcal{H}$.

To describe all the $A_{\infty}$ relations,
it is convenient to use the \textit{coalgebra representation}.\footnote{
The coalgebra representation of $A_\infty$ algebras
is explained in detail, e.g.~in appendix~A of Ref.~\cite{Erler:2015uba}
and in Ref.~\cite{Koyama:2020qfb}.
We mostly follow the conventions used in these papers.
}
In the coalgebra representation, we consider linear operators acting on $T \mathcal{H}$ defined by
\begin{equation}
  T\mathcal{H} = \mathcal{H}^{\otimes 0} \oplus \mathcal{H} \oplus \mathcal{H}^{\otimes 2} \oplus \mathcal{H}^{\otimes 3}\oplus\ldots\,.
\end{equation}
Associated with $c_n$ which is a map from $\mathcal{H}^{\otimes n}$ to $\mathcal{H}$,
we define an operator $\bm{c}_n$ acting on $T \mathcal{H}$ as follows:
\begin{align}
\bm{c}_n \pi_m & = 0 \quad \text{for} \quad m < n \,, \\
\bm{c}_n \pi_n & = c_n \, \pi_n \,, \\
\bm{c}_n \pi_{n+1} & = ( c_n \otimes \mathbb{I} + \mathbb{I} \otimes c_n ) \, \pi_{n+1} \,, \\
\bm{c}_n \pi_m
& = ( c_n \otimes \mathbb{I}^{\otimes(m-n)}
+\sum_{k=1}^{m-n-1} \mathbb{I}^{\otimes k} \otimes c_n \otimes \mathbb{I}^{\otimes (m-n-k)}
+\mathbb{I}^{\otimes(m-n)} \otimes c_n ) \, \pi_m \quad \text{for} \quad m > n+1 \,,
\end{align}
where $\pi_n$ is the projection operator onto $\mathcal{H}^{\otimes n}$,
$\mathbb{I}$ is the identity operator on $\mathcal{H}$,
and $\mathbb{I}^{\otimes n}$ is defined by
\begin{equation}
\mathbb{I}^{\otimes n} = \underbrace{\, \mathbb{I} \otimes \mathbb{I} \otimes \ldots \otimes \mathbb{I} \,}_n \,.
\end{equation}
Operators of this form are called \textit{coderivations}.
The degree of $\bm{c}_n$ is defined to be the same as that of $c_n$.
We define $\bf{M}$ by
\begin{equation}
  {\bf M} = \sum_{n=0}^{\infty} {\bf M}_n \,,
\end{equation}
where ${\bf M}_n$ is the coderivation associated with $M_n$,
then the $A_\infty$ relations are compactly described as
\begin{equation}
{\bf M}^2 = 0 \,.
\end{equation}
In fact, the relations in Eqs.~\eqref{A_1}, \eqref{A_2}, \eqref{A_3}, and~\eqref{A_4} are written as
\begin{align}
\pi_1 \, {\bf M}^2 ( {\bf 1} ) & = 0 \,, \\
\pi_1 \, {\bf M}^2 ( \Phi_1 ) & = 0 \,, \\
\pi_1 \, {\bf M}^2 ( \Phi_1 \otimes \Phi_2 ) & = 0 \,, \\
\pi_1 \, {\bf M}^2 (\Phi_1 \otimes \Phi_2 \otimes \Phi_3 ) & = 0 \,,
\end{align}
and we can show that $\pi_1 \, {\bf M}^2 = 0$ implies
${\bf M}^2 = 0$ when ${\bf M}$ is a degree-odd coderivation.

When we separate the kinetic term of the free theory described by $Q$
and describe interactions in terms of the operators $m_n$,
we introduce the coderivations ${\bf Q}$ and ${\bm m}_n$
associated with $Q$ and $m_n$, respectively,
and we define ${\bm m}$ by
\begin{equation}
  {\bm m} = \sum_{n=0}^{\infty} {\bm m}_n \,.
\end{equation}
The action is described by the coderivation ${\bf Q}+{\bm m}$,
and the gauge invariance of the action follows from the relation
\begin{equation}
( \, {\bf Q}+{\bm m} \, )^2 = 0 \,.
\end{equation}

As we mentioned in the introduction, we consider projections onto subspaces of $\mathcal{H}$.
We denote the projection operator by $P$, and it is degree even.
In the coalgebra representation, we use the projection operator ${\bf P}$ acting on $T \mathcal{H}$.
It is defined by
\begin{equation}
\begin{split}
{\bf P} \, \pi_0 & = \pi_0 \,, \\
{\bf P} \, \pi_n & = P^{\otimes n} \, \pi_n
\end{split}
\end{equation}
for $n > 0$, where
\begin{equation}
P^{\otimes n} = \underbrace{\, P \otimes P \otimes \ldots \otimes P \,}_n \,.
\end{equation}

Associated with the projection described by $P$,
we introduce the \textit{contracting homotopy} $h$
which is a degree-odd map from $\mathcal{H}$ to $\mathcal{H}$ and satisfies
\begin{equation}
Q \, h +h \, Q = \mathbb{I}-P \,, \quad
h \, P = 0 \,, \quad
P \, h = 0 \,, \quad
h^2 = 0 \,.
\label{chain-map}
\end{equation}
The contracting homotopy $h$ physically represents the propagator
for the kinetic term of the free theory described by $Q$.
We will present explicit forms of $h$
for the kinetic term of the scalar field in Section~\ref{scalar-section}
and for the kinetic term of the Dirac field in Section~\ref{Dirac-section}.
We then promote $h$ to the linear operator $\bm{h}$ acting on $T \mathcal{H}$.
The operator $\bm{h}$ is defined by
\begin{align}
\bm{h} \, \pi_0 & = 0 \,, \\
\bm{h} \, \pi_1 & = h \, \pi_1 \,, \\
\bm{h} \, \pi_2 & = ( \, h \otimes P +\mathbb{I} \otimes h \, ) \, \pi_2 \,, \\
\bm{h} \, \pi_n & = \Bigl( h \otimes P^{\otimes(n-1)}
+\sum_{m=1}^{n-2} \mathbb{I}^{\otimes m} \otimes h \otimes P^{\otimes(n-m-1)}
+\mathbb{I}^{\otimes(n-1)} \otimes h \Bigr) \, \pi_n
\end{align}
for $n > 2$.
We can show that the relations in Eq.~\eqref{chain-map} are promoted
to those of operators acting on $T \mathcal{H}$ given by
\begin{equation}
{\bf Q} \, {\bm h} +{\bm h} \, {\bf Q} = {\bf I}-{\bf P} \,, \quad
{\bm h} \, {\bf P} = 0 \,, \quad
{\bf P} \, {\bm h} = 0 \,, \quad
{\bm h}^2 = 0 \,,
\end{equation}
where ${\bf I}$ is the identity operator on $T \mathcal{H}$.

We have presented all the ingredients of the universal formula~\eqref{universal-formula}:
\begin{equation}
\pi_1 \, {\bf P} \, {\bf Q} \, {\bf P}
+\pi_1 \, {\bf P} \, {\bm m} \, \frac{1}{{\bf I}+{\bm h} \, {\bm m}} \, {\bf P} \,.
\end{equation}
For the formula for the quantum theory~\eqref{quantum-universal-formula},
\begin{equation}
\pi_1 \, {\bf P} \, {\bf Q} \, {\bf P}
+\pi_1 \, {\bf P} \, {\bm m} \,
\frac{1}{{\bf I}+{\bm h} \, {\bm m} +i \hbar \, {\bm h} \, {\bf U}} \, {\bf P} \,,
\end{equation}
we also need the operator ${\bf U}$.
We will present the definition of ${\bf U}$
for scalar fields in Section~\ref{scalar-section}
and for Dirac fields in Section~\ref{Dirac-section}.

\section{Scalar fields}
\label{scalar-section}

\subsection{$A_\infty$ algebras for scalar fields}

When we use $A_\infty$ algebras to describe theories without gauge symmetries,
we only need two sectors for the vector space $\mathcal{H}$,
and in our convention, which is analogous to open string field theory,
they are denoted by $\mathcal{H}_1$ and $\mathcal{H}_2$:
\begin{equation}
\mathcal{H} = \mathcal{H}_1 \oplus \mathcal{H}_2 \,.
\end{equation}
The formula for correlation functions of scalar field theories using quantum $A_\infty$ algebras
was presented in Ref.~\cite{Okawa:2022sjf},
where $\mathcal{H}_1$ is chosen to be the vector space of functions of $x$
and the function of $x$ is directly identified with the scalar field which appears in the action.
In this section we present a different description of the scalar field,
which is analogous to the description of spacetime fields in string field theory.

Let us denote the basis vector of $\mathcal{H}_1$ by $c(x)$,
where the label $x$ represents coordinates of Minkowski spacetime in $d$ dimensions,
and we define $c(x)$ to be degree even.
The important point is that we do \textit{not} identify $c(x)$ with the scalar field
which appears in the action.
The element $\Phi$ of $\mathcal{H}_1$ can be expanded as
\begin{equation}
\Phi = \int d^d x \, \varphi (x) \, c(x) \,,
\label{scalar-expansion}
\end{equation}
and we identify $\varphi (x)$ in this expansion with the scalar field
which appears in the action. We define $\varphi (x)$ to be degree even.

For the vector space $\mathcal{H}_2$,
we denote the basis vector by $d(x)$,
and we define $d(x)$ to be degree odd.
We then define the degree-odd operator $Q$ by
\begin{equation}
Q \, c(x) = ( {}-\partial^2 +m^2 \, ) \, d(x) \,, \qquad
Q \, d(x) = 0 \,,
\end{equation}
where $m$ is the mass of the scalar field.
The action of $Q$ on $\Phi$ in Eq.~\eqref{scalar-expansion} is given by
\begin{equation}
\begin{split}
Q \, \Phi & = \int d^d x \, \varphi (x) \, Q \, c(x)
= \int d^d x \, \varphi (x) \, ( {}-\partial^2 +m^2 \, ) \, d(x) \\
& = \int d^d x \, \bigl( {}-\partial^2 \varphi (x) +m^2 \varphi (x) \, \bigr) \, d(x) \,,
\end{split}
\end{equation}
where we used integration by parts in the last step.
Therefore, the equation of motion of the free scalar field,
\begin{equation}
{}-\partial^2 \varphi (x) +m^2 \varphi (x) = 0 \,,
\end{equation}
can be written in terms of $Q \, \Phi$ in~$\mathcal{H}_2$ as
\begin{equation}
Q \, \Phi = 0
\end{equation}
with $\Phi$ as given in Eq.~\eqref{scalar-expansion}.
Note that the operator $Q$ is nilpotent:
\begin{equation}
Q^2 = 0 \,.
\end{equation}

The symplectic form $\omega$ is defined by
\begin{equation}
\biggl(
\begin{array}{cc}
\omega \, ( \, c(x_1) \,, c(x_2) \, ) & \omega \, ( \, c(x_1) \,, d(x_2) \, ) \\
\omega \, ( \, d(x_1) \,, c(x_2) \, ) & \omega \, ( \, d(x_1) \,, d(x_2) \, )
\end{array}
\biggr)
= \biggl(
\begin{array}{cc}
0 & \delta^d ( x_1-x_2 ) \\
{}-\delta^d ( x_1-x_2 ) & 0
\end{array}
\biggr) \,.
\end{equation}
This definition is consistent with the property given in Eq.~\eqref{omega-symplectic}.
The symplectic form involving $Q$ is nonvanishing for
\begin{equation}
\omega \, ( \, Q \, c(x) \,, c(y) \, )
= {}-( {}-\partial_x^2 +m^2 \, ) \, \delta^d ( x-y )
\end{equation}
and
\begin{equation}
\omega \, ( \, c(x) \,, Q \, c(y) \, )
= ( {}-\partial_y^2 +m^2 \, ) \, \delta^d ( x-y ) \,.
\end{equation}
Since
\begin{equation}
\partial_x^2 \, \delta^d ( x-y ) = \partial_y^2 \, \delta^d ( x-y ) \,,
\end{equation}
the definitions of $Q$ and $\omega$ are consistent with the property
\begin{equation}
\omega \, ( \, Q \, \Phi_1 \,, \Phi_2 \, )
= {}-(-1)^{\mathrm{deg} (\Phi_1)} \, \omega \, ( \, \Phi_1 \,, Q \, \Phi_2 \, )
\label{Q-cyclicity}
\end{equation}
for $\Phi_1$ and $\Phi_2$ in~$\mathcal{H}$,
which is required for cyclic $A_\infty$ algebras.
Using this symplectic form, the action of the free theory
\begin{equation}
  S = {}-\frac{1}{2} \int d^d x \, \bigl[ \, \partial_\mu \varphi (x) \, \partial^\mu \varphi (x)
  +m^2 \, \varphi (x)^2 \, \bigr]
\end{equation}
can be written as
\begin{equation}
S = {}-\frac{1}{2} \, \omega \, ( \, \Phi, Q \, \Phi \, )
\end{equation}
with $\Phi$ as given in Eq.~\eqref{scalar-expansion}.

As an example of interacting theories, let us consider $\varphi^3$ theory.
The cubic term in the action is given by
\begin{equation}
\frac{g}{6} \int d^{d}x \, \varphi(x)^3 \,,
\label{cubic-interaction}
\end{equation}
where $g$ is the coupling constant.
This interaction term can be described by $m_2$ in the following form:
\begin{equation}
\begin{split}
& m_2 \, ( \, c \, (x_1) \otimes c \, (x_2) \, )
= {}-\frac{g}{2} \int d^d x \, \delta^d (x-x_1) \, \delta^d (x-x_2) \, d \, (x) \,, \\
& m_2 \, ( \, c \, (x_1) \otimes d \, (x_2) \, ) = 0 \,, \quad
m_2 \, ( \, d \, (x_1) \otimes c \, (x_2) \, ) = 0 \,, \quad
m_2 \, ( \, d \, (x_1) \otimes d \, (x_2) \, ) = 0 \,.
\end{split}
\end{equation}
We can carry out the integral over $x$ for $m_2 \, ( \, c \, (x_1) \otimes c \, (x_2) \, )$
to obtain less symmetric expressions:
\begin{equation}
m_2 \, ( \, c \, (x_1) \otimes c \, (x_2) \, )
= {}-\frac{g}{2} \, \delta^d (x_1-x_2) \, d \, (x_1)
\end{equation}
or
\begin{equation}
m_2 \, ( \, c \, (x_1) \otimes c \, (x_2) \, )
= {}-\frac{g}{2} \, \delta^d (x_1-x_2) \, d \, (x_2) \,.
\end{equation}
Since
\begin{equation}
\begin{split}
\omega \, ( \, c \, (x_1) \,,\, m_2 \, ( \, c \, (x_2) \otimes c \, (x_3) \, ) \, )
& = {}-\frac{g}{2} \int d^d x \, \delta^d (x-x_1) \, \delta^d (x-x_2) \, \delta^d (x-x_3) \,, \\
\omega \, ( \, m_2 \, ( \, c \, (x_1) \otimes c \, (x_2) \, ) \,,\, c \, (x_3) \, )
& = \frac{g}{2} \int d^d x \, \delta^d (x-x_1) \, \delta^d (x-x_2) \, \delta^d (x-x_3) \,,
\end{split}
\end{equation}
we find
\begin{equation}
\omega \, ( \, m_2 \, ( \, \Phi_1 \otimes \Phi_2 \, ) \,, \Phi_3 \, )
= {}-(-1)^{\mathrm{deg} (\Phi_1)} \, \omega \, ( \, \Phi_1 \,, m_2 \, ( \, \Phi_2 \otimes \Phi_3 \, ) \, )
\end{equation}
for $\Phi_1$, $\Phi_2$, and $\Phi_3$ in~$\mathcal{H}$,
which is required for cyclic $A_\infty$ algebras.
The cubic interaction~\eqref{cubic-interaction} can be written using $m_2$ as
\begin{equation}
\frac{g}{6} \int d^{d}x \, \varphi(x)^3
= {}-\frac{1}{3} \, \omega \, ( \, \Phi \,, m_2 \, ( \, \Phi \otimes \Phi \, ) \, )
\end{equation}
with $\Phi$ as given in Eq.~\eqref{scalar-expansion}.

Let us now consider correlation functions.
It was emphasized in Ref.~\cite{Okawa:2022sjf} that the projection operator we should use
when we calculate correlation functions is
\begin{equation}
P = 0 \,.
\end{equation}
Then the conditions we impose on the contracting homotopy $h$ are
\begin{equation}
Q \, h +h \, Q = \mathbb{I} \,, \qquad
h^2 = 0 \,.
\end{equation}
We can use the Feynman propagator
\begin{equation}
\Delta (x-y)
= \int \frac{d^d k}{(2 \pi)^d} \,
\frac{e^{ik \, (x-y)}}{k^2+m^2-i \epsilon}
\end{equation}
to construct $h$ satisfying these conditions as follows:
\begin{equation}
h \, c(x) = 0 \,, \qquad
h \, d(x) = \int d^d y \, \Delta (x-y) \, c (y) \,.
\label{scalar-h}
\end{equation}
Since $P=0$, the associated operator $\bm{h}$ is given by
\begin{equation}
\bm{h} = h \, \pi_1
+\sum_{n=2}^\infty ( \, \mathbb{I}^{\otimes (n-1)} \otimes h \, ) \, \pi_n \,.
\end{equation}

The last ingredient for correlation functions is the operator ${\bf U}$.
For the scalar field theory that we are considering we define ${\bf U}$ by 
\begin{equation}
{\bf U} = \int d^d x \, {\bm c} (x) \, {\bm d} (x) \,,
\end{equation}
where ${\bm c} (x)$ is a degree-even coderivation with $\pi_1 \, {\bm c} (x)$ given by
\begin{equation}
\pi_1 \, {\bm c} (x) \, {\bf 1} = c(x) \,, \quad
\pi_1 \, {\bm c} (x) \, \pi_n = 0
\end{equation}
for $n > 0$
and ${\bm d} (x)$ is a degree-odd coderivation with $\pi_1 \, {\bm d} (x)$ given by
\begin{equation}
\pi_1 \, {\bm d} (x) \, {\bf 1} = d(x) \,, \quad
\pi_1 \, {\bm d} (x) \, \pi_n = 0
\end{equation}
for $n > 0 \,$.
We can show that ${\bm c} (x)$ and ${\bm d} (x)$ commute so that their order in ${\bf U}$
does not matter.
The operator ${\bf U}$ is normalized such that
\begin{equation}
( \, \omega \otimes \mathbb{I} \, ) \, ( \, \mathbb{I} \otimes U \, ) = \mathbb{I}
\end{equation}
is satisfied,\footnote{
See, e.g.~Ref.~\cite{Erler:2019loq} for the discussion on the normalization of ${\bf U}$.
}
where $U$ is a map from $\mathcal{H}^{\otimes 0}$ to $\mathcal{H}^{\otimes 2}$ given by
\begin{equation}
U = \pi_2 \, {\bf U} \, \pi_0
\end{equation}
and $\omega$ is a map from $\mathcal{H}^{\otimes 2}$ to $\mathcal{H}^{\otimes 0}$ with
\begin{equation}
\omega \, ( \, \Phi_1 \otimes \Phi_2 \, ) = \omega \, ( \, \Phi_1 \,,\, \Phi_2 \, ) \, {\bf 1}
\end{equation}
for $\Phi_1$ and $\Phi_2$ in $\mathcal{H}$.

\subsection{Correlation functions}

We claim that correlation functions are given by
\begin{equation}
\langle \, \Phi^{\otimes n} \, \rangle = \pi_n \, {\bm f} \, {\bf 1}
\label{scalar-correlation-functions}
\end{equation}
with
\begin{equation}
\Phi^{\otimes n} = \underbrace{\, \Phi \otimes \Phi \otimes \ldots \otimes \Phi \,}_n
\end{equation}
and
\begin{equation}
{\bm f} = \frac{1}{{\bf I} +{\bm h} \, {\bm m} +i \hbar \, {\bm h} \, {\bf U}} \,,
\end{equation}
where the inverse of ${\bf I} +{\bm h} \, {\bm m} +i \hbar \, {\bm h} \, {\bf U}$ is defined by
\begin{equation}
\frac{1}{{\bf I} +{\bm h} \, {\bm m} +i \hbar \, {\bm h} \, {\bf U}}
= {\bf I} +\sum_{n=1}^\infty \, (-1)^n \,
( \, {\bm h} \, {\bm m} +i \hbar \, {\bm h} \, {\bf U} \, )^n \,.
\end{equation}
Since the left-hand side of Eq.~\eqref{scalar-correlation-functions} can be expanded as
\begin{equation}
\begin{split}
\langle \, \Phi^{\otimes n} \, \rangle
& = \langle \, \underbrace{\, \Phi \otimes \Phi \otimes \ldots \otimes \Phi \,}_n \, \rangle \\
& = \int d^d x_1 d^d x_2 \ldots d^d x_n \,
\langle \, \varphi (x_1) \, \varphi (x_2) \, \ldots \, \varphi (x_n) \, \rangle \,
c (x_1) \otimes c(x_2) \otimes \ldots \otimes c(x_n) \,,
\end{split}
\end{equation}
the formula~\eqref{scalar-correlation-functions} states
that the correlation functions appear
as coefficients when we expand the right-hand side of Eq.~\eqref{scalar-correlation-functions}:
\begin{equation}
\pi_n \, {\bm f} \, {\bf 1}
= \int d^d x_1 d^d x_2 \ldots d^d x_n \,
\langle \, \varphi (x_1) \, \varphi (x_2) \, \ldots \, \varphi (x_n) \, \rangle \,
c (x_1) \otimes c(x_2) \otimes \ldots \otimes c(x_n) \,.
\end{equation}
Since
\begin{equation}
\omega \, ( \, c(x') \,, d(x) \, ) = \delta^d (x'-x) \,,
\end{equation}
the correlation functions
$\langle \, \varphi (x_1) \, \varphi (x_2) \, \ldots \, \varphi (x_n) \, \rangle$
can be extracted from $\langle \, \Phi^{\otimes n} \, \rangle$ as
\begin{equation}
\langle \, \varphi (x_1) \, \varphi (x_2) \, \ldots \, \varphi (x_n) \, \rangle
= \omega_n \, ( \, \langle \, \Phi^{\otimes n} \, \rangle \,,
d (x_1) \otimes d (x_2) \otimes \ldots \otimes d (x_n) \, ) \,,
\end{equation}
where
\begin{equation}
\omega_n \, ( \, c (x_1)  \otimes c (x_2) \otimes \ldots \otimes c (x_n) \,,
d (x'_1) \otimes d (x'_2) \otimes \ldots \otimes d (x'_n) \, )
= \prod_{i=1}^n \, \omega \, ( \, c (x_i) \,, d (x'_i) \, ) \,.
\label{omega_n-c-d}
\end{equation}
Then the formula~\eqref{scalar-correlation-functions} can be expressed as
\begin{equation}
\langle \, \varphi (x_1) \, \varphi (x_2) \, \ldots \, \varphi (x_n) \, \rangle
= \omega_n \, ( \, \pi_n \, {\bm f} \, {\bf 1} \,,
d (x_1) \otimes d (x_2) \otimes \ldots \otimes d (x_n) \, ) \,.
\label{scalar-component-correlation-functions}
\end{equation}

Let us calculate correlation functions of the free theory,
which corresponds to the case where ${\bm m} = 0$.
The two-point function can be calculated from $\, \pi_{2} \, {\bm f} \, {\bf 1} \,$.
The operator ${\bm h} \, {\bf U}$ can be decomposed as
\begin{equation}
{\bm h} \, {\bf U} = \sum_{n=0}^\infty \, \pi_{n+2} \, {\bm h} \, {\bf U} \, \pi_n \,,
\end{equation}
so $\, \pi_{2} \, {\bm f} \, {\bf 1} \,$ for the free theory is given by
\begin{equation}
\pi_{2} \, {\bm f} \, {\bf 1} \,
= {}-i \hbar \, \pi_{2} \, {\bm h} \, {\bf U} \, {\bf 1} \,.
\end{equation}
The operator ${\bf U}$ acting on ${\bf 1}$ generates
the following element of $\mathcal{H} \otimes \mathcal{H}$:
\begin{equation}
{\bf U} \, {\bf 1} = \int d^d x \, ( \, c(x) \otimes d(x) +d(x) \otimes c(x) \, ) \,.
\label{U1}
\end{equation}
The action of ${\bm h}$ on $\mathcal{H} \otimes \mathcal{H}$ is given by
\begin{equation}
{\bm h} \, \pi_2 = ( \, \mathbb{I} \otimes h \, ) \, \pi_2 \,.
\end{equation}
Since $h$ annihilates $c(x)$, one of the two terms in Eq.~\eqref{U1} survives:
\begin{equation}
{\bm h} \, {\bf U} \, {\bf 1} = \int d^d x \, c(x) \otimes h \, d(x) \,.
\label{hU1}
\end{equation}
We thus find
\begin{equation}
\begin{split}
\pi_{2} \, {\bm f} \, {\bf 1}
& = {}-i \hbar \, \pi_{2} \, {\bm h} \, {\bf U} \, {\bf 1} \\
& = {}-i \hbar \int d^d x \int d^d y \, [ \, c(x) \otimes \Delta(x-y) \, c(y) \, ] \,,
\end{split}
\end{equation}
and $\omega_2 \, ( \, \pi_2 \, {\bm f} \, {\bf 1} \,,d (x_1) \otimes d (x_2) \, )$ is given by
\begin{equation}
      \omega_2 \, ( \, \pi_2 \, {\bm f} \, {\bf 1} \,,d (x_1) \otimes d (x_2) \, )
      = {}-i \hbar \,\Delta\,(x_1\,-\,x_2\,) \,.
\end{equation}
This correctly reproduces the two-point function $\langle \, \varphi (x_1) \, \varphi (x_2) \,  \rangle$:
\begin{equation}
\langle \, \varphi (x_1) \, \varphi (x_2) \,  \rangle
= \frac{\hbar}{i}\,\Delta\,(x_1\,-\,x_2\,) \,.
\end{equation}

The four-point function can be calculated from $\, \pi_{4} \, {\bm f} \, {\bf 1} \,$.
In the free theory, we find
\begin{equation}
\pi_{4} \, {\bm f} \, {\bf 1} = {}-\hbar^2 \, \pi_{4} \, {\bm h}  \, {\bf U} \, {\bm h} \, {\bf U} \,{\bf 1} \,.
\end{equation}
The operator ${\bf U}$ acting on ${\bm h} \, {\bf U} \, {\bf 1}$ in~\eqref{hU1} generates
many terms in $\mathcal{H}^{\otimes 4}$.
The action of ${\bm h}$ on $\mathcal{H}^{\otimes 4}$ is given by
\begin{equation}
{\bm h} \, \pi_4 = ( \, \mathbb{I} \otimes \mathbb{I} \otimes \mathbb{I} \otimes h \, ) \, \pi_4 \,,
\end{equation}
and $h$ annihilates $c(x)$ and $h \, d(x)$ so that only three terms survive:
\begin{equation}
\begin{split}
{\bm h} \, {\bf U} \, {\bm h} \, {\bf U} \, {\bf 1}
& = \int d^d x \int d^d x' \, ( \, 
c(x') \otimes c(x) \otimes h \, d(x) \otimes h \, d(x') \\
& \qquad \qquad \qquad \quad
+c(x) \otimes c(x') \otimes h \, d(x) \otimes h \, d(x') \\
& \qquad \qquad \qquad \quad
+c(x) \otimes h \, d(x) \otimes c(x') \otimes h \, d(x') \, ) \,.
\end{split}
\label{hUhU1}
\end{equation}
Using the definition of $h$ in~\eqref{scalar-h},
we thus find that
\begin{equation}
\begin{split}
& \pi_{4} \, {\bm f} \, {\bf 1}
= {}-\hbar^2 \, \pi_{4} \, {\bm h} \, {\bf U} \, {\bm h} \, {\bf U} \, {\bf 1} \\
& = {}-\hbar^2 \int d^d x \int d^d x' \int d^d y \int d^d y' \,
[ \, c(x') \otimes c(x)\otimes \Delta (x-y) \, c(y) \otimes \Delta (x'-y') \, c(y') \\
& \qquad \qquad \qquad \qquad \qquad \qquad \qquad \quad~\,
+c(x) \otimes c(x') \otimes \Delta (x-y) \, c(y) \otimes \Delta (x'-y') \, c(y') \\
& \qquad \qquad \qquad \qquad \qquad \qquad \qquad \quad~\,
+c(x) \otimes \Delta (x-y) \, c(y) \otimes c(x') \otimes \Delta (x'-y') \, c(y') \, ] \,,
\end{split}
\end{equation}
and 
$\omega_4 \, ( \, \pi_4 \, {\bm f} \, {\bf 1} \,,
d (x_1) \otimes d (x_2) \otimes d(x_3) \otimes d(x_4) \, )$
is given by
\begin{equation}
\begin{split}
& \omega_4 \, ( \, \pi_4 \, {\bm f} \, {\bf 1} \,,
d (x_1) \otimes d (x_2) \otimes d(x_3) \otimes d(x_4) \, ) \\
& = {}-\hbar^2 \, [ \, \Delta (x_2-x_3) \, \Delta (x_1-x_4)
+\Delta (x_1-x_3) \, \Delta (x_2-x_4)
+\Delta (x_1-x_2) \, \Delta (x_3-x_4) \, ] \,.
\end{split}
\end{equation}
This takes the form of
\begin{equation}
\begin{split}
& \omega_4 \, ( \, \pi_4 \, {\bm f} \, {\bf 1} \,,
d (x_1) \otimes d (x_2) \otimes d(x_3) \otimes d(x_4) \, ) \\ 
& = \langle \, \varphi (x_2) \, \varphi (x_3) \, \rangle \,
\langle \, \varphi (x_1) \, \varphi (x_4) \, \rangle
+\langle \, \varphi (x_1) \, \varphi (x_3) \, \rangle \,
\langle \, \varphi (x_2) \, \varphi (x_4) \, \rangle \\
& \quad~ +\langle \, \varphi (x_1) \, \varphi (x_2) \, \rangle \,
\langle \, \varphi (x_3) \, \varphi (x_4) \, \rangle \,,
\end{split}
\end{equation}
and Wick's theorem for the four-point function,
\begin{equation}
\begin{split}
& \langle \, \varphi (x_1) \, \varphi (x_2) \, \varphi (x_3) \, \varphi (x_4) \,\rangle \\
& = \langle \, \varphi (x_2) \, \varphi (x_3) \, \rangle \,
\langle \, \varphi (x_1) \, \varphi (x_4) \, \rangle
+\langle \, \varphi (x_1) \, \varphi (x_3) \, \rangle \,
\langle \, \varphi (x_2) \, \varphi (x_4) \, \rangle \\
& \quad~ +\langle \, \varphi (x_1) \, \varphi (x_2) \, \rangle \,
\langle \, \varphi (x_3) \, \varphi (x_4) \, \rangle \,,
\end{split}
\end{equation}
is correctly reproduced.

\subsection{The Schwinger-Dyson equations}

It was shown in Ref.~\cite{Okawa:2022sjf}
that correlation functions from the formula based on the quantum $A_\infty$ algebra
satisfy the Schwinger-Dyson equations
as an immediate consequence of the relation
\begin{equation}
( \, {\bf I} +{\bm h} \, {\bm m} +i \hbar \, {\bm h} \, {\bf U} \, ) \, 
\frac{1}{{\bf I} +{\bm h} \, {\bm m} +i \hbar \, {\bm h} \, {\bf U}} \, {\bf 1} = {\bf 1} \,.
\label{f=inverse}
\end{equation}
In this subsection we present the proof for the free theory
using our description~\eqref{scalar-expansion} of the scalar field,
and we generalize it to the free Dirac field in the next section.
The proof that correlation functions from our formula satisfy
the Schwinger-Dyson equations in the case of interacting theories
involving scalar fields and Dirac fields is presented in Ref.~\cite{Konosu:2023rkm}.

In the path integral formalism, correlation functions are defined by
\begin{equation}
\langle \, \varphi (x_1) \, \varphi (x_2) \, \ldots \, \varphi (x_n) \, \rangle
= \frac{1}{Z} \int \mathcal{D} \varphi \,
\varphi (x_1) \, \varphi (x_2) \, \ldots \, \varphi (x_n) \, e^{\frac{i}{\hbar}S},
\end{equation}
where
\begin{equation}
Z = \int \mathcal{D}\varphi \, e^{\frac{i}{\hbar}S}.
\end{equation}
Since
\begin{equation}
\frac{1}{Z} \int \mathcal{D} \varphi \,
\frac{\delta}{\delta \varphi(x_n)} \Bigl[ \,
\varphi (x_1) \, \varphi (x_2) \, \ldots \, \varphi (x_{n-1}) \, e^{\frac{i}{\hbar}S} \, \Bigr] = 0 \,,
\end{equation}
the Schwinger-Dyson equations are given by
\begin{equation}
\sum_{i=1}^{n-1} \delta^d(x_i-x_n) \,
\langle \, \varphi (x_1) \ldots \varphi (x_{i-1}) \,
\varphi (x_{i+1}) \ldots \varphi (x_{n-1}) \, \rangle
+\frac{i}{\hbar} \, \langle \, \varphi (x_1) \ldots \varphi (x_{n-1}) \,
\frac{\delta S}{\delta\varphi(x_n)} \, \rangle = 0 \,.
\end{equation}
Let us show
that correlation functions from the formula~\eqref{scalar-correlation-functions}
satisfy the Schwinger-Dyson equations for the free theory.
The free theory corresponds to the case where ${\bm m}=0$,
and the relation~\eqref{f=inverse} is given by
\begin{equation}
( \, {\bf I} +i \hbar \, {\bm h} \, {\bf U} \, ) \,
\frac{1}{{\bf I} +i \hbar \, {\bm h} \, { \bf U}} \, {\bf 1} = {\bf 1} \,.
\end{equation}
Since
\begin{equation}
\pi_n \, {\bf 1} = 0
\end{equation}
for $n \geq 1$,
we have
\begin{equation}
\pi_n \, { \bm f} \, {\bf 1}
+i \hbar \, \pi_n \, {\bm h} \, {\bf U} \, {\bm f} \, {\bf 1} = 0
\end{equation}
and consider the relation
\begin{equation}
\omega_n \, ( \, \pi_n \, {\bm f} \, {\bf 1} \,,\,
d (x_1)  \otimes \ldots \otimes d (x_n) \, )
+i \hbar \, \omega_n \, ( \, \pi_n \, {\bm h} \, {\bf U} \, {\bm f} \, {\bf 1} \,,\,
d (x_1)  \otimes \ldots \otimes d (x_n) \, ) = 0 \,,
\label{sd_scalar}
\end{equation}
which holds for $n \geq 1$.
The first term on the left-hand side corresponds to the $n$-point function given by
\begin{equation}
\omega_n \, ( \, \pi_n \, {\bm f} \, {\bf 1} \,,\,
d (x_1)  \otimes \ldots \otimes d (x_n) \, )
= \langle \, \varphi (x_1) \, \ldots \, \varphi (x_n) \, \rangle \,.
\end{equation}
For the second term on the left-hand side,
\begin{equation}
i \hbar \, \omega_n \, ( \, \pi_n \, {\bm h} \, {\bf U} \, {\bm f} \, {\bf 1} \,,\,
d (x_1)  \otimes \ldots \otimes d (x_n) \, ) \,,
\label{sd2}
\end{equation}
$\pi_n \, {\bm h} \, {\bf U}$ is given by
\begin{equation}
\begin{split}
\pi_1 \, {\bm h} \, {\bf U} & = 0 \,, \\
\pi_n \, {\bm h} \, {\bf U}
& = \int d^d x \sum_{i=1}^{n-1} \, ( \,
\mathbb{I}^{\otimes (i-1)} \otimes c(x) \otimes \mathbb{I}^{\otimes (n-i-1)} \otimes h \, d(x) \, ) \,
\pi_{n-2}
\end{split}
\end{equation}
for $n > 1$.
Since
\begin{equation}
\begin{split}
& \omega_n \, ( \, 
( \, \mathbb{I}^{\otimes (i-1)} \otimes c(x) \otimes \mathbb{I}^{\otimes (n-i-1)} \otimes h \, d(x) \, ) \,
\pi_{n-2} \, {\bm f} \, {\bf 1} \,,
d (x_1)  \otimes \ldots \otimes d (x_n) \, ) \\
& = \omega_2 \, ( \, c(x) \otimes h \, d(x) \,,\, d (x_i)  \otimes d(x_n) \, ) \\
& \quad~ \times
\omega_{n-2} \, ( \, \pi_{n-2} \, {\bm f} \, {\bf 1} \,,
d (x_1)  \otimes \ldots \otimes d (x_{i-1}) \otimes
d (x_{i+1}) \otimes \ldots \otimes d (x_{n-1}) \, )
\end{split}
\end{equation}
and
\begin{equation}
\int d^d x \, \omega_2 \, ( \, c(x) \otimes h \, d(x) \,,\, d(x_i)  \otimes d(x_n) \, )
= \Delta (x_i-x_n) \,,
\end{equation}
we obtain
\begin{equation}
\begin{split}
& i \hbar \, \omega_n \, ( \, \pi_n \, {\bm h} \, {\bf U} \, {\bm f} \, {\bf 1} \,,
d (x_1) \otimes \ldots \otimes d (x_n) \, ) \\
& = {}-\frac{\hbar}{i} \, \sum_{i=1}^{n-1}
\Delta (x_i-x_n) \, \langle \, \varphi (x_1) \, \ldots \, \varphi (x_{i-1}) \,
\varphi (x_{i+1}) \, \ldots \, \varphi (x_{n-1}) \, \rangle \,.
\end{split}
\label{sd_scalar2}
\end{equation}
Therefore, the relation (\ref{sd_scalar}) is translated into
\begin{equation}
\langle \, \varphi(x_1) \, \rangle = 0
\label{scalar-one-point}
\end{equation}
and
\begin{equation}
\langle \, \varphi (x_1) \, \ldots \, \varphi (x_n) \, \rangle
-\frac{\hbar}{i} \, \sum_{i=1}^{n-1} \Delta(x_i-x_n) \,
\langle \, \varphi (x_1) \, \ldots \, \varphi (x_{i-1}) \,
\varphi (x_{i+1}) \, \ldots \, \varphi (x_{n-1}) \, \rangle = 0
\label{scalar-recursion}
\end{equation}
for $n > 0$.
We then act the operator ${}-\partial_{x_n}^2 +m^2$ to find
\begin{equation}
\begin{split}
& \langle \, \varphi (x_1) \, \ldots \, \varphi (x_{n-1}) \,
( {}-\partial_{x_n}^2 +m^2 \, ) \, \varphi(x_n) \, \rangle \\
& {}-\frac{\hbar}{i} \, \sum_{i=1}^{n-1} \delta^d (x_i-x_n) \,
\langle \, \varphi (x_1) \, \ldots \, \varphi (x_{i-1}) \,
\varphi (x_{i+1}) \, \ldots \, \varphi (x_{n-1}) \, \rangle = 0 \,.
\end{split}
\end{equation}
Since
\begin{equation}
\frac{\delta S}{\delta \varphi(x_n)} = {}-( {}-\partial_{x_n}^2 +m^2 \, ) \,\varphi(x_n) \,,
\end{equation}
we find
\begin{equation}
\begin{split}
& {}-\frac{\hbar}{i} \, \sum_{i=1}^{n-1} \delta^d (x_i-x_n) \,
\langle \, \varphi (x_1) \, \ldots \, \varphi (x_{i-1}) \,
\varphi (x_{i+1}) \, \ldots \, \varphi (x_{n-1}) \, \rangle \\
& {}-\langle \, \varphi (x_1) \, \ldots \, \varphi (x_{n-1}) \,
\frac{\delta S}{\delta \varphi (x_n)} \, \rangle = 0 \,.
\end{split}
\end{equation}
We have thus shown that the Schwinger-Dyson equations are satisfied.
It is clear that all the correlation functions of the free theory
are uniquely determined from Eqs.~\eqref{scalar-one-point} and~\eqref{scalar-recursion}.

\section{Dirac fields}
\label{Dirac-section}

\subsection{$A_\infty$ algebras for Dirac fields}

Let us move on to the description of Dirac fields using $A_\infty$ algebras.\footnote{
As in Ref.~\cite{Okawa:2022sjf}, we mostly follow the conventions of the textbook
by Srednicki~\cite{Srednicki:2007qs}.
}
We consider theories without gauge symmetries
so that the vector space $\mathcal{H}$ is given by
\begin{equation}
\mathcal{H} = \mathcal{H}_1 \oplus \mathcal{H}_2 \,.
\end{equation}
Let us denote the basis vector of $\mathcal{H}_1$ by $\theta_\alpha (x)$,
where the label $x$ represents coordinates of Minkowski spacetime in $d$ dimensions
and $\alpha$ is the spinor index.
We define $\theta_\alpha (x)$ to be degree odd.
We also use the Dirac adjoint $\overline{\theta}_\alpha (x)$ of $\theta_\alpha (x)$
to represent quantities satisfying the reality condition,
and it is also degree odd.
The element $\Phi$ of $\mathcal{H}_1$ can be expanded in this basis as
\begin{equation}
\Phi = \int d^d x \, ( \, \overline{\theta}_\alpha (x) \, \Psi_\alpha (x)
+\overline{\Psi}_\alpha (x) \, \theta_\alpha (x) \, ) \,.
\label{Dirac-expansion}
\end{equation}
We identify $\Psi_\alpha (x)$ with the Dirac field which appears in the action,
and we define $\Psi_\alpha (x)$ to be degree odd.
In this expansion, $\overline{\Psi}_\alpha (x)$ has to be the Dirac adjoint of $\Psi_\alpha (x)$
so that $\Phi$ satisfies the reality condition,
and $\overline{\Psi}_\alpha (x)$ is also degree odd.
Note that $\Phi$ is degree even.

For the vector space $\mathcal{H}_2$,
we denote the basis vector by $\lambda_\alpha (x)$,
and we define $\lambda_\alpha (x)$ to be degree even.
We also use the Dirac adjoint $\overline{\lambda}_\alpha (x)$ of $\lambda_\alpha (x)$,
which is degree even.
We then define the degree-odd operator $Q$ by
\begin{equation}
\begin{split}
& Q \, \theta_\alpha (x)
= {}( {}-i \, \partial\!\!\!/ +m \, )_{\alpha \beta} \, \lambda_\beta (x) \,, \quad
Q \, \lambda_\alpha (x) = 0 \,, \\
& Q \, \overline{\theta}_\alpha (x)
= {}-\overline{\lambda}_\beta (x) \, ( \, i \overleftarrow{\partial\!\!\!/} +m \, )_{\beta \alpha} \,, \quad
Q \, \overline{\lambda}_\alpha (x) = 0 \,,
\end{split}
\end{equation}
where $m$ is the mass of the Dirac field.
The action of $Q$ on $\Phi$ in Eq.~\eqref{Dirac-expansion} is given by
\begin{equation}
\begin{split}
Q \, \Phi & = {}-\int d^d x \, ( \, \overline{\lambda}_\beta (x) \,
( \, i \overleftarrow{\partial\!\!\!/} +m \, )_{\beta \alpha}
\Psi_\alpha (x)
+\overline{\Psi}_\alpha (x) \, ( {}-i \, \partial\!\!\!/ +m \, )_{\alpha \beta} \,
\lambda_\beta (x) \, ) \\
& = \int d^d x \, ( \, \overline{\lambda}_\beta (x) \,
( \, i \, \partial\!\!\!/ -m \, )_{\beta \alpha} \Psi_\alpha (x)
+\overline{\Psi}_\alpha (x) \, ( {}-i \overleftarrow{\partial\!\!\!/} -m \, )_{\alpha \beta} \, \lambda_\beta (x) \, ) \,,
\end{split}
\end{equation}
where we used integration by parts in the last step.
Note that $Q \, \Phi$ satisfies the reality condition
when $\Phi$ satisfies the reality condition.
The Dirac equation
\begin{equation}
( \, {}-i \, \partial\!\!\!/ +m \, ) \, \Psi (x) = 0
\end{equation}
and its Dirac adjoint
\begin{equation}
\overline{\Psi} (x) \, ( \, i \overleftarrow{\partial\!\!\!/} +m \, ) = 0
\end{equation}
can therefore be written in terms of $Q \, \Phi$ in~$\mathcal{H}_2$ as
\begin{equation}
Q \, \Phi = 0
\end{equation}
with $\Phi$ as given in Eq.~\eqref{Dirac-expansion}.
Note that the operator $Q$ is nilpotent:
\begin{equation}
Q^2 = 0 \,.
\end{equation}

Let us next consider the symplectic form $\omega$.
We define the symplectic form $\omega$ by
\begin{align}
\omega \, ( \, \theta_{\alpha_1} (x_1) \,, \overline{\lambda}_{\alpha_2} (x_2) \, )
& = \delta_{\alpha_1 \alpha_2} \, \delta^d ( x_1-x_2 ) \,, \\
\omega \, ( \, \overline{\theta}_{\alpha_1} (x_1) \,, \lambda_{\alpha_2} (x_2) \, )
& = \delta_{\alpha_1 \alpha_2} \, \delta^d ( x_1-x_2 ) \,, \\
\omega \, ( \,  \overline{\lambda}_{\alpha_1} (x_1)\,, \theta_{\alpha_2}(x_2)  \, )
& = {}-\delta_{\alpha_1 \alpha_2} \, \delta^d ( x_1-x_2 ) \,, \\
\omega \, ( \, \lambda_{\alpha_1} (x_1)  \,, \overline{\theta}_{\alpha_2} (x_2)\, )
& = {}-\delta_{\alpha_1 \alpha_2} \, \delta^d ( x_1-x_2 ) \,,
\end{align}
and the symplectic form vanishes for all other cases.
The symplectic form $\omega \, ( \, e_1 \,,\, e_2 \, )$
for $e_1$ and $e_2$ in the basis of $\mathcal{H}$ gives a number,
which is obviously degree even,
and it can be nonvanishing
only when the degree of $e_1$ and the degree of $e_2$ are different.
Therefore, the symplectic form $\omega$ is regarded as being degree odd,
and we should consider that a degree-odd object is inserted
somewhere in the symplectic form.
When we consider $\omega \, ( \, \Phi_1 \,,\, \Phi_2 \, )$
with $\Phi_1$ and $\Phi_2$ given by
\begin{equation}
\Phi_1 = f_1 \, e_1 \,, \qquad \Phi_2 = f_2 \, e_2 \,,
\label{Phi-form}
\end{equation}
the degree-odd object may give signs
in the process of taking out $f_1$ and $f_2$ from
$\omega \, ( \, f_1 \, e_1 \,,\, f_2 \, e_2 \, )$
to obtain, e.g.~$f_1 \, f_2 \, \omega \, ( \, e_1 \,,\, e_2 \, )$.
Our description of scalar fields corresponds to the case
where $f_1$ and $f_2$ are degree even
so that the location of the degree-odd object does not matter.
Our description of Dirac fields in Eq.~\eqref{Dirac-expansion}, on the other hand,
corresponds to the case
where $f_1$ and $f_2$ are degree odd,
and we need to specify the location of the degree-odd object in the symplectic form.
Consistency of the definition of the symplectic form
with the property
\begin{equation}
\omega \, ( \, \Phi_1 \,,\, \Phi_2 \, )
= {}-(-1)^{\mathrm{deg} (\Phi_1) \, \mathrm{deg} (\Phi_2)} \,
\omega \, ( \, \Phi_2 \,,\, \Phi_1 \, )
\end{equation}
for $\Phi_1$ and $\Phi_2$ in the form of Eq.~\eqref{Phi-form} is studied
in Appendix~\ref{omega-appendix},
and in this paper we choose the convention
\begin{equation}
\omega \, ( \, f_1 \, e_1 \,,\, f_2 \, e_2 \, )
= (-1)^{\mathrm{deg} (f_1)+\mathrm{deg} (f_2)+\mathrm{deg} (e_1) \, \mathrm{deg} (f_2)} \,
f_1 \, f_2 \, \omega \, ( \, e_1 \,,\, e_2 \, ) \,.
\label{omega-convention}
\end{equation}
This corresponds to the case where the degree-odd object is inserted
to the left of $\Phi_1$ in $\omega \, ( \, \Phi_1 \,,\, \Phi_2 \, )$.
Our mnemonic is that the symbol $\omega$ in $\omega \, ( \, \Phi_1 \,,\, \Phi_2 \, )$
is considered to be degree odd.
In Eq.~\eqref{omega-convention},
the factor $(-1)^{\mathrm{deg} (f_1)}$
comes from $f_1$ passing through $\omega$,
the factor $(-1)^{\mathrm{deg} (e_1) \, \mathrm{deg} (f_2)}$
comes from $f_2$ passing through $e_1$,
and the factor $(-1)^{\mathrm{deg} (f_2)}$
comes from $f_2$ passing through $\omega$.

When we consider degree-odd coefficients in front of basis vectors,
we also need to refine the definition of $\omega_n$
because the naive identification of
$\omega_n \, ( \Phi_1 \otimes \Phi_2 \otimes \ldots \otimes \Phi_n \,,
\widetilde{\Phi}_1 \otimes \widetilde{\Phi}_2 \otimes \ldots \otimes \widetilde{\Phi}_n \, )$
with
$\omega \, ( \, \Phi_1 \,,\, \widetilde{\Phi}_1 \, ) \,
\omega \, ( \, \Phi_2 \,,\, \widetilde{\Phi}_2 \, ) \ldots
\omega \, ( \, \Phi_n \,,\, \widetilde{\Phi}_n \, )$
suffers from inconsistency.
For example, consider $\omega_2 \, ( \, f e_1 \otimes e_2 \,,\, \tilde{e}_1 \otimes \tilde{e}_2 \, )$
with both $f$ and $e_1$ being degree odd and $\tilde{e}_1$ being degree even. 
In this case, we have
\begin{align}
\omega_2 \, ( \, f e_1 \otimes e_2 \,,\, \tilde{e}_1 \otimes \tilde{e}_2 \, )
& = {}-\omega_2 \, ( \, e_1 \otimes f e_2 \,,\, \tilde{e}_1 \otimes \tilde{e}_2 \, ) \,, \\
\omega \, ( \, f e_1 \,,\, \tilde{e}_1 \, ) \, \omega \, ( \, e_2 \,,\, \tilde{e}_2 \, )
& = \omega \, ( \, e_1 \,,\, \tilde{e}_1 \, ) \, \omega \, ( \, f e_2 \,,\, \tilde{e}_2 \, )
\end{align}
so that we cannot identify
$\omega_2 \, ( \Phi_1 \otimes \Phi_2 \,, \widetilde{\Phi}_1 \otimes \widetilde{\Phi}_2 \, )$
with $\omega \, ( \, \Phi_1 \,,\, \widetilde{\Phi}_1 \, ) \,
\omega \, ( \, \Phi_2 \,,\, \widetilde{\Phi}_2 \, )$.
To avoid this inconsistency,
we define $\omega_n$ by
\begin{equation}
\begin{split}
&\omega_n \, ( \, \Phi_1 \otimes \Phi_2 \otimes \ldots \otimes \Phi_n \,,
\widetilde{\Phi}_1 \otimes \widetilde{\Phi}_2 \otimes \ldots \otimes \widetilde{\Phi}_n \, ) \\
& = (-1)^\sigma \, \omega \, ( \, \Phi_1 \,, \widetilde{\Phi}_1 \, ) \,
\omega \, ( \, \Phi_2 \,, \widetilde{\Phi}_2 \, ) \ldots \,
\omega \, ( \, \Phi_n \,, \widetilde{\Phi}_n \, ) \,,
\end{split}
\label{omega_n}
\end{equation}
where
\begin{equation}
\sigma
= \sum_{i=1}^{n-1} \sum_{j=i+1}^n \mathrm{deg} (\widetilde{\Phi}_i) \, \mathrm{deg} (\Phi_j)
+\sum_{k=1}^{n-1} \, (n-k) \,
( \, \mathrm{deg} (\Phi_k) +\mathrm{deg} (\widetilde{\Phi}_k) -1 \, ) \mod 2 \,.
\end{equation}
Let us explain the sign factor $(-1)^\sigma$ by dividing it into three pieces.
The first one is the sign from the following reordering:
\begin{equation}
\Phi_1 \, \Phi_2 \ldots \Phi_n \,
\widetilde{\Phi}_1 \, \widetilde{\Phi}_2 \ldots \widetilde{\Phi}_n
\to \Phi_1 \, \widetilde{\Phi}_1 \, \Phi_2 \, \widetilde{\Phi}_2 \ldots
\Phi_n \, \widetilde{\Phi}_n \,. 
\end{equation}
This generates the contribution to $\sigma$ given by
\begin{equation}
\begin{split}
& \mathrm{deg} (\widetilde{\Phi}_1) \,
( \, \mathrm{deg} (\Phi_2) + \ldots +\mathrm{deg} (\Phi_n) \, ) \\
& +\mathrm{deg} (\widetilde{\Phi}_2) \, ( \, \mathrm{deg} (\Phi_3) + \ldots +\mathrm{deg} (\Phi_n) \, )
+ \ldots
+\mathrm{deg} (\widetilde{\Phi}_{n-1}) \, \mathrm{deg} (\Phi_n) \\
& = \sum_{i=1}^{n-1} \sum_{j=i+1}^n \mathrm{deg} (\widetilde{\Phi}_i) \, \mathrm{deg} (\Phi_j) \,.
\end{split}
\end{equation}
Second, we consider the degree of the symbol $\omega_n$ to be $n$ mod $2$
and distribute $n$ degree-odd objects as follows:
\begin{equation}
\omega \, \omega \ldots \omega \,
\Phi_1 \, \widetilde{\Phi}_1 \, \Phi_2 \, \widetilde{\Phi}_2 \ldots
\Phi_n \, \widetilde{\Phi}_n
\to \omega \, \Phi_1 \, \widetilde{\Phi}_1 \, \omega \, \Phi_2 \, \widetilde{\Phi}_2 \ldots
\omega \, \Phi_n \, \widetilde{\Phi}_n
\end{equation}
with $\omega$ being degree odd.
This generates the contribution to $\sigma$ given by
\begin{equation}
\begin{split}
& (n-1) \, ( \, \mathrm{deg} (\Phi_1) +\mathrm{deg} (\widetilde{\Phi}_1) \, ) \\
& +(n-2) \, ( \, \mathrm{deg} (\Phi_2) +\mathrm{deg} (\widetilde{\Phi}_2) \, )
+ \ldots
+( \, \mathrm{deg} (\Phi_{n-1}) +\mathrm{deg} (\widetilde{\Phi}_{n-1}) \, ) \\
& = \sum_{k=1}^{n-1} \, (n-k) \,
( \, \mathrm{deg} (\Phi_k) +\mathrm{deg} (\widetilde{\Phi}_k) \, ) \,.
\end{split}
\label{sigma_2}
\end{equation}
Finally, there is a contribution to $(-1)^\sigma$ which is independent of
degrees of $\Phi_i$'s and $\widetilde{\Phi}_i$'s. Since
\begin{equation}
{}-\sum_{k=1}^{n-1} \, (n-k) = {}-\frac{n (n-1)}{2} \,,
\label{sigma_3}
\end{equation}
the explicit contribution to $(-1)^\sigma$ is $(-1)^{-\frac{n(n-1)}{2}}$.
While this is not necessary for consistency,
the contribution~\eqref{sigma_3} cancels the second contribution~\eqref{sigma_2}
when $\mathrm{deg} (\Phi_k) +\mathrm{deg} (\widetilde{\Phi}_k) = 1$ mod~$2$
for any $k$.
This definition of $\omega_n$ coincides with that in Eq.~\eqref{omega_n-c-d}.
For Dirac fields, we have
\begin{equation}
\begin{split}
& \omega_n \, ( \, \theta_{\alpha_1} (x_1)  \otimes
\theta_{\alpha_2} (x_2) \otimes \ldots \otimes
\theta_{\alpha_n} (x_n) \,,
\overline{\lambda}_{\alpha'_1} (x'_1) \otimes
\overline{\lambda}_{\alpha'_2} (x'_2) \otimes \ldots \otimes
\overline{\lambda}_{\alpha'_n} (x'_n) \, ) \\
& = \omega \, ( \, \theta_{\alpha_1} (x_1) \,, \overline{\lambda}_{\alpha'_1} (x'_1) \, ) \,
\omega \, ( \, \theta_{\alpha_2} (x_2) \,, \overline{\lambda}_{\alpha'_2} (x'_2) \, ) \ldots
\omega \, ( \, \theta_{\alpha_n} (x_n) \,, \overline{\lambda}_{\alpha'_n} (x'_n) \, )
\end{split}
\end{equation}
without any sign on the right-hand side.
Incidentally, we have not so far used $\omega_2$ of the kind
$\omega_2 \, ( \, c (x_1)  \otimes d (x_2) \,,
d (x'_1) \otimes c (x'_2)  \, )$, but it is given by
\begin{equation}
\omega_2 \, ( \, c (x_1)  \otimes d (x_2) \,,
d (x'_1) \otimes c (x'_2)  \, )
= {}-\omega \, ( \, c (x_1) \,, d (x'_1) \, ) \, \omega \, ( \, d (x_2) \,, c (x'_2) \, )
\end{equation}
according to the refined definition.
We will use this and its generalizations to $\omega_n$ in Appendix~\ref{factorization-algebra-appendix}.

Let us come back to $\omega$ and confirm that the operator $Q$ has the correct cyclic property
with respect to $\omega$.
Since
\begin{equation}
\begin{split}
& \omega \, ( \, Q \, \theta_\alpha (x) \,,\, \overline{\theta}_\beta (y) \, )
= {}-( {}-i \, \partial\!\!\!/_x +m \, )_{\alpha \beta} \, \delta^d (x-y) \,, \\
& \omega \, ( \, \theta_\alpha (x) \,,\, Q \, \overline{\theta}_\beta (y) \, )
= {}-\delta^d (x-y) \, ( \, i \overleftarrow{\partial\!\!\!/}_y +m \, )_{\alpha \beta}
= {}-( \, i \, \partial\!\!\!/_y +m \, )_{\alpha \beta} \, \delta^d (x-y) \,, \\
& \omega \, ( \, Q \, \overline{\theta}_\alpha (x) \,,\, \theta_\beta (y) \, )
= \delta^d (x-y) \, ( \, i \overleftarrow{\partial\!\!\!/}_x +m \, )_{\beta \alpha}
= ( \, i \, \partial\!\!\!/_x +m \, )_{\beta \alpha} \, \delta^d (x-y) \,, \\
& \omega \, ( \, \overline{\theta}_\alpha (x) \,,\, Q \, \theta_\beta (y) \, )
= {}( {}-i \, \partial\!\!\!/_y +m \, )_{\beta \alpha} \, \delta^d (x-y) \,, \\
\end{split}
\end{equation}
the definitions of $Q$ and $\omega$ are consistent with the property
\begin{equation}
\omega \, ( \, Q \, \Phi_1 \,, \Phi_2 \, )
= {}-(-1)^{\mathrm{deg} (\Phi_1)} \, \omega \, ( \, \Phi_1 \,, Q \, \Phi_2 \, )
\end{equation}
for $\Phi_1$ and $\Phi_2$ in~$\mathcal{H}$,
which is required for cyclic $A_\infty$ algebras.

Having defined the symplectic form, let us express the action
of the Dirac field
\begin{equation}
S = \int d^d x \, \bigl[ \, i \, \overline{\Psi} (x) \, \partial\!\!\!/ \, \Psi (x)
-m \, \overline{\Psi} (x) \, \Psi (x) \, \bigr]
\label{Dirac-action}
\end{equation}
using $\Phi$ as given in Eq.~\eqref{Dirac-expansion}.
Since
\begin{equation}
\begin{split}
\Phi & = \int d^d x \, ( \, \overline{\theta}_\alpha (x) \, \Psi_\alpha (x)
+\overline{\Psi}_\alpha (x) \, \theta_\alpha (x) \, ) \,, \\
Q \, \Phi & = \int d^d x \, ( \,
\overline{\lambda}_\beta (x) \,
( \, i \, \partial\!\!\!/ -m \, )_{\beta \alpha} \Psi_\alpha (x)
+\overline{\Psi}_\alpha (x) \, ( {}-i \overleftarrow{\partial\!\!\!/} -m \, )_{\alpha \beta} \,
\lambda_\beta (x) \, ) \,,
\end{split}
\end{equation}
we find
\begin{equation}
\begin{split}
{}-\frac{1}{2} \, \omega \, ( \, \Phi \,,\, Q \, \Phi \, )
& = \frac{1}{2} \, \int d^d x \, ( \,
\overline{\Psi} (x) \, ( {}-i \overleftarrow{\partial\!\!\!/} -m \, ) \, \Psi (x)
+\overline{\Psi} (x) \, ( \, i \, \partial\!\!\!/ -m \, ) \Psi (x) \, ) \\
& = \int d^d x \, \overline{\Psi} (x) \, ( \, i \, \partial\!\!\!/ -m \, ) \, \Psi (x) \,.
\end{split}
\end{equation}
We have thus seen that the action~\eqref{Dirac-action}
can be written as
\begin{equation}
S = {}-\frac{1}{2} \, \omega \, ( \, \Phi \,,\, Q \, \Phi \, )
\end{equation}
with $\Phi$ as given in Eq.~\eqref{Dirac-expansion}.

When we calculate correlation functions,
we should consider the projection with
\begin{equation}
P = 0
\end{equation}
as in the case of scalar field theories.
The conditions we impose on the contracting homotopy $h$ are again
\begin{equation}
Q \, h +h \, Q = \mathbb{I} \,, \qquad
h^2 = 0 \,,
\end{equation}
and we can use the propagator
\begin{equation}
S(x-y)_{\alpha \beta}
= \int \frac{d^d p}{(2 \pi)^d} \, e^{ip \, (x-y)} \,
\frac{( {}-p\!\!\!/ +m \, )_{\alpha \beta}}{p^2 +m^2 -i \epsilon}
\end{equation}
to construct $h$ satisfying these conditions as follows:
\begin{equation}
\begin{split}
& h \, \theta_\alpha (x) = 0 \,, \qquad
h \, \lambda_\alpha (x) = {}\int d^d y \, S(x-y)_{\alpha \beta} \, \theta_\beta (y) \,, \\
& h \, \overline{\theta}_\alpha (x) = 0 \,, \qquad
h \, \overline{\lambda}_\alpha (x)
= -\int d^d y \, \overline{\theta}_\beta (y) \, S(y-x)_{\beta \alpha} \,.
\end{split}
\label{Dirac-h}
\end{equation}
The associated operator $\bm{h}$ is given by
\begin{equation}
\bm{h} = h \, \pi_1
+\sum_{n=2}^\infty ( \, \mathbb{I}^{\otimes (n-1)} \otimes h \, ) \, \pi_n \,.
\end{equation}

The operator ${\bf U}$ for the theory we are considering is defined by
\begin{equation}
{\bf U} = -\int d^d x \, ( \, \overline{\bm \theta}_\alpha (x) \, {\bm \lambda}_\alpha (x)
+{\bm \theta}_\alpha (x) \, \overline{\bm \lambda}_\alpha (x) \, ) \,,
\label{U-Dirac}
\end{equation}
where ${\bm \theta}_\alpha (x)$ and $\overline{\bm \theta}_\alpha (x)$ are
degree-odd coderivations with $\pi_1 \, {\bm \theta}_\alpha (x)$
and $\pi_1 \, \overline{\bm \theta}_\alpha (x)$ specified by
\begin{equation}
\pi_1 \, {\bm \theta}_\alpha (x) \, {\bf 1} = \theta_\alpha (x) \,, \quad
\pi_1 \, {\bm \theta}_\alpha (x) \, \pi_n = 0 \,, \quad
\pi_1 \, \overline{\bm \theta}_\alpha (x) \, {\bf 1} = \overline{\theta}_\alpha (x) \,, \quad
\pi_1 \, \overline{\bm \theta}_\alpha (x) \, \pi_n = 0
\end{equation}
for $n > 0$ and both of ${\bm \lambda}_\alpha (x)$ and $\overline{\bm \lambda}_\alpha (x)$ are
degree-even coderivations with $\pi_1 \, {\bm \lambda}_\alpha (x)$
and $\pi_1 \, \overline{\bm \lambda}_\alpha (x)$ specified by
\begin{equation}
\pi_1 \, {\bm \lambda}_\alpha (x) \, {\bf 1} = \lambda_\alpha (x) \,, \quad
\pi_1 \, {\bm \lambda}_\alpha (x) \, \pi_n = 0 \,, \quad
\pi_1 \, \overline{\bm \lambda}_\alpha (x) \, {\bf 1} = \overline{\lambda}_\alpha (x) \,, \quad
\pi_1 \, \overline{\bm \lambda}_\alpha (x) \, \pi_n = 0
\end{equation}
for $n > 0 \,$.
We can show that ${\bm \lambda}_\alpha (x)$ and $\overline{\bm \theta}_\alpha (x)$ commute
so that their order in ${\bf U}$ does not matter.
We can also show that ${\bm \theta}_\alpha (x)$ and $\overline{\bm \lambda}_\alpha (x)$ commute
so that their order in ${\bf U}$ does not matter, either.
As in the case of the scalar field,
the operator ${\bf U}$ is normalized such that
\begin{equation}
( \, \omega \otimes \mathbb{I} \, ) \, ( \, \mathbb{I} \otimes U \, ) = \mathbb{I}
\end{equation}
is satisfied.
Note that the minus sign in Eq.~\eqref{U-Dirac} is necessary for this relation to hold.

\subsection{Correlation functions}

We claim that the formula for correlation functions takes the same form
as Eq.~\eqref{scalar-correlation-functions} when it is expressed in terms of $\Phi$:
\begin{equation}
\langle \, \Phi^{\otimes n} \, \rangle = \pi_n \, {\bm f} \, {\bf 1} \,,
\label{Dirac-correlation-functions}
\end{equation}
where
\begin{equation}
{\bm f} = \frac{1}{{\bf I} +{\bm h} \, {\bm m} +i \hbar \, {\bm h} \, {\bf U}} \,.
\end{equation}
In the case of the Dirac field, minus signs can appear
when we expand the left-hand side of Eq.~\eqref{Dirac-correlation-functions}.
When, e.g.~$n=1$ and $n=2$, we find
\begin{equation}
\langle \, \Phi \, \rangle
= \int d^d x_1 \, \Big[ \, 
{}-\langle \, \Psi_{\alpha_1} (x_1) \, \rangle \, \overline{\theta}_{\alpha_1} (x_1) \, 
+\langle \, \overline{\Psi}_{\alpha_1} (x_1) \, \rangle \, \theta_{\alpha_1} (x_1) \, \Big] \,,
\end{equation}
and
\begin{equation}
\begin{split}
\langle \, \Phi \otimes \Phi \, \rangle
= \int d^d x_1 \, d^d x_2 \, \Big[ \,
& {}-\langle \, \Psi_{\alpha_1} (x_1) \, \Psi_{\alpha_2} (x_2) \, \rangle \,
\overline{\theta}_{\alpha_1} (x_1) \otimes \overline{\theta}_{\alpha_2} (x_2) \\
& +\langle \, \Psi_{\alpha_1} (x_1) \, \overline{\Psi}_{\alpha_2} (x_2) \, \rangle \,
\overline{\theta}_{\alpha_1} (x_1) \otimes \theta_{\alpha_2} (x_2) \\
& +\langle \, \overline{\Psi}_{\alpha_1} (x_1) \, \Psi_{\alpha_2} (x_2) \, \rangle \,
\theta_{\alpha_1} (x_1) \otimes \overline{\theta}_{\alpha_2} (x_2) \\
& {}-\langle \, \overline{\Psi}_{\alpha_1} (x_1) \, \overline{\Psi}_{\alpha_2} (x_2) \, \rangle \,
\theta_{\alpha_1} (x_1) \otimes \theta_{\alpha_2} (x_2) \, \Big] \,.
\end{split}
\label{Phi^2-expansion}
\end{equation}
The formula~\eqref{Dirac-correlation-functions} therefore states that
\begin{equation}
\pi_1 \, {\bm f} \, {\bf 1}
= \int d^d x_1 \, \Big[ \, 
{}-\langle \, \Psi_{\alpha_1} (x_1) \, \rangle \, \overline{\theta}_{\alpha_1} (x_1) \, 
+\langle \, \overline{\Psi}_{\alpha_1} (x_1) \, \rangle \, \theta_{\alpha_1} (x_1) \, \Big] \,,
\end{equation}
and
\begin{equation}
\begin{split}
\pi_2 \, {\bm f} \, {\bf 1}
= \int d^d x_1 \, d^d x_2 \, \Big[ \,
& {}-\langle \, \Psi_{\alpha_1} (x_1) \, \Psi_{\alpha_2} (x_2) \, \rangle \,
\overline{\theta}_{\alpha_1} (x_1) \otimes \overline{\theta}_{\alpha_2} (x_2) \\
& +\langle \, \Psi_{\alpha_1} (x_1) \, \overline{\Psi}_{\alpha_2} (x_2) \, \rangle \,
\overline{\theta}_{\alpha_1} (x_1) \otimes \theta_{\alpha_2} (x_2) \\
& +\langle \, \overline{\Psi}_{\alpha_1} (x_1) \, \Psi_{\alpha_2} (x_2) \, \rangle \,
\theta_{\alpha_1} (x_1) \otimes \overline{\theta}_{\alpha_2} (x_2) \\
& {}-\langle \, \overline{\Psi}_{\alpha_1} (x_1) \, \overline{\Psi}_{\alpha_2} (x_2) \, \rangle \,
\theta_{\alpha_1} (x_1) \otimes \theta_{\alpha_2} (x_2) \, \Big] \,.
\end{split}
\end{equation}
Since
\begin{equation}
\omega \, ( \, \theta_{\alpha'} (x') \,, \overline{\lambda}_{\alpha} (x) \, )
= \delta_{\alpha' \alpha} \, \delta^d ( x'-x ) \,, \quad
\omega \, ( \, \overline{\theta}_{\alpha'} (x') \,, \lambda_{\alpha} (x) \, ) 
= \delta_{\alpha' \alpha} \, \delta^d ( x'-x ) \,,
\end{equation}
the one-point functions are given by
\begin{equation}
\begin{split}
\langle \, \Psi_{\alpha_1} (x_1) \, \rangle
& = \omega \, ( \, \pi_1 \, {\bm f} \, {\bf 1} \,,
\lambda_{\alpha_1} (x_1) \, ) \,, \\
\langle \, \overline{\Psi}_{\alpha_1} (x_1) \, \rangle
& = {}-\omega \, ( \, \pi_1 \, {\bm f} \, {\bf 1} \,,
\overline{\lambda}_{\alpha_1} (x_1) \, ) \,, \\
\end{split}
\end{equation}
and the two-point functions are
\begin{equation}
\begin{split}
\langle \, \Psi_{\alpha_1} (x_1) \, \Psi_{\alpha_2} (x_2) \, \rangle
& = {}-\omega_2 \, ( \, \pi_2 \, {\bm f} \, {\bf 1} \,,
\lambda_{\alpha_1} (x_1) \otimes \lambda_{\alpha_2} (x_2) \, ) \,, \\
\langle \, \Psi_{\alpha_1} (x_1) \, \overline{\Psi}_{\alpha_2} (x_2) \, \rangle
& = \omega_2 \, ( \, \pi_2 \, {\bm f} \, {\bf 1} \,,
\lambda_{\alpha_1} (x_1) \otimes \overline{\lambda}_{\alpha_2} (x_2) \, ) \,, \\
\langle \, \overline{\Psi}_{\alpha_1} (x_1) \, \Psi_{\alpha_2} (x_2) \, \rangle
& = \omega_2 \, ( \, \pi_2 \, {\bm f} \, {\bf 1} \,,
\overline{\lambda}_{\alpha_1} (x_1) \otimes \lambda_{\alpha_2} (x_2) \, ) \,, \\
\langle \, \overline{\Psi}_{\alpha_1} (x_1) \, \overline{\Psi}_{\alpha_2} (x_2) \, \rangle
& = {}-\omega_2 \, ( \, \pi_2 \, {\bm f} \, {\bf 1} \,,
\overline{\lambda}_{\alpha_1} (x_1) \otimes \overline{\lambda}_{\alpha_2} (x_2) \, ) \,.
\end{split}
\label{two-point-functions-formula}
\end{equation}
Note that such minus signs do not appear
when the number of $\Psi$'s and the number of $\overline{\Psi}$'s are the same,
and this is when correlation functions can be nonvanishing.
Thus, e.g.~the correlation function
$\langle \, \Psi_{\alpha_1} (x_1) \, \ldots \Psi_{\alpha_n} (x_n) \,
\overline{\Psi}_{\beta_1} (y_1) \, \ldots \overline{\Psi}_{\beta_n} (y_n) \, \rangle$
can be extracted as
\begin{equation}
\begin{split}
& \langle \, \Psi_{\alpha_1} (x_1) \, \ldots \Psi_{\alpha_n} (x_n) \,
\overline{\Psi}_{\beta_1} (y_1) \, \ldots \overline{\Psi}_{\beta_n} (y_n) \, \rangle \\
& = \omega_{2n} \, ( \, \pi_{2n} \, {\bm f} \, {\bf 1} \,,
\lambda_{\alpha_1} (x_1) \otimes \ldots \otimes \lambda_{\alpha_n} (x_n)
\otimes \overline{\lambda}_{\beta_1} (y_1) \otimes \ldots \otimes \overline{\lambda}_{\beta_n} (y_n) \, )
\label{main-formula}
\end{split}
\end{equation}
with no minus sign on the right-hand side.
We only consider the free theory involving Dirac fields in this paper,
which corresponds to the case where ${\bm m} = 0$.
Interacting theories involving Dirac fields are discussed in Ref.~\cite{Konosu:2023rkm}.

To tame various combinations of $\Psi$'s and $\overline{\Psi}$'s,
it is convenient to introduce $n$ degree-odd spinors $\epsilon^i_\alpha$
with $i=1,2, \ldots , n$
and their Dirac-adjoint spinors $\overline{\epsilon}^{\, i}_\alpha$
for $n$-point functions,
and we define $\widehat{\Psi}^i (x)$ by
\begin{equation}
\widehat{\Psi}^i (x) = \overline{\epsilon}^{\, i}_\alpha \Psi_\alpha (x)
+\overline{\Psi}_\alpha (x) \, \epsilon^i_\alpha \,.
\label{Psi-hat}
\end{equation}
By expanding $\langle \, \widehat{\Psi}^1 (x_1) \, \widehat{\Psi}^2 (x_2) \, \rangle$ as
\begin{equation}
\begin{split}
\langle \, \widehat{\Psi}^1 (x_1) \, \widehat{\Psi}^2 (x_2) \, \rangle
& = \langle \, ( \, \overline{\epsilon}^{\, 1}_{\alpha_1} \Psi_{\alpha_1} (x_1)
+\overline{\Psi}_{\alpha_1} (x_1) \, \epsilon^1_{\alpha_1} \, ) \,
( \, \overline{\epsilon}^{\, 2}_{\alpha_2} \Psi_{\alpha_2} (x_2)
+\overline{\Psi}_{\alpha_2} (x_2) \, \epsilon^2_{\alpha_2} \, ) \, \rangle \\
& = {}-\langle \, \Psi_{\alpha_1} (x_1) \, \Psi_{\alpha_2} (x_2) \, \rangle \,
\overline{\epsilon}^{\, 1}_{\alpha_1} \, \overline{\epsilon}^{\, 2}_{\alpha_2}
+\langle \, \Psi_{\alpha_1} (x_1) \, \overline{\Psi}_{\alpha_2} (x_2) \, \rangle \,
\overline{\epsilon}^{\, 1}_{\alpha_1} \, \epsilon^2_{\alpha_2} \\
& \quad~ {}+\langle \, \overline{\Psi}_{\alpha_1} (x_1) \, \Psi_{\alpha_2} (x_2) \, \rangle \, 
\epsilon^1_{\alpha_1} \, \overline{\epsilon}^{\, 2}_{\alpha_2}
-\langle \, \overline{\Psi}_{\alpha_1} (x_1) \, \overline{\Psi}_{\alpha_2} (x_2) \, \rangle \,
\epsilon^1_{\alpha_1} \, \epsilon^2_{\alpha_2} \,,
\end{split}
\label{Psi-hat-expansion}
\end{equation}
we see that the two-point function
$\langle \, \widehat{\Psi}^1 (x_1) \, \widehat{\Psi}^2 (x_2) \, \rangle$ encodes
the information on all the two-point functions
$\langle \, \Psi_{\alpha_1} (x_1) \, \Psi_{\alpha_2} (x_2) \, \rangle$,
$\langle \, \Psi_{\alpha_1} (x_1) \, \overline{\Psi}_{\alpha_2} (x_2) \, \rangle$,
$\langle \, \overline{\Psi}_{\alpha_1} (x_1) \, \Psi_{\alpha_2} (x_2) \, \rangle$,
and $\langle \, \overline{\Psi}_{\alpha_1} (x_1) \, \overline{\Psi}_{\alpha_2} (x_2) \, \rangle \,$.
Note that minus signs appear in the same way as in Eq.~\eqref{Phi^2-expansion}.
Since
\begin{equation}
\omega \, ( \, \Phi \,, \lambda_\alpha (x) \, ) = \Psi_\alpha (x) \,, \qquad
\omega \, ( \, \Phi \,, \overline{\lambda}_\alpha (x) \, ) = {}-\overline{\Psi}_\alpha (x) \,,
\end{equation}
we have
\begin{equation}
\widehat{\Psi}^i (x) = \omega \, ( \, \Phi \,,
{}-\overline{\epsilon}^{\, i}_\alpha \, \lambda_\alpha (x)
-\overline{\lambda}_\alpha (x) \, \epsilon^i_\alpha \, ) \,.
\end{equation}
Motivated by this, we define $\widehat{\lambda}^{\, i} (x)$ by
\begin{equation}
\widehat{\lambda}^{\, i} (x) = {}-\overline{\epsilon}^{\, i}_\alpha \, \lambda_\alpha (x)
-\overline{\lambda}_\alpha (x) \, \epsilon^i_\alpha
\end{equation}
so that we can extract $\widehat{\Psi}^i (x)$ from $\Phi$ as
\begin{equation}
\widehat{\Psi}^i (x) = \omega \, ( \, \Phi \,, \widehat{\lambda}^{\, i} (x) \, ) \,.
\end{equation}
Then the $n$-point function
$\langle \, \widehat{\Psi}^1 (x_1) \, \widehat{\Psi}^2 (x_2) \, \ldots \,
\widehat{\Psi}^n (x_n) \, \rangle$
can be extracted from $\langle \, \Phi^{\otimes n} \, \rangle$ as
\begin{equation}
\langle \, \widehat{\Psi}^1 (x_1) \, \widehat{\Psi}^2 (x_2) \, \ldots \,
\widehat{\Psi}^n (x_n) \, \rangle
= \omega_n \, ( \, \langle \, \Phi^{\otimes n} \, \rangle \,,\,
\widehat{\lambda}^{\, 1} (x_1) \otimes \widehat{\lambda}^{\, 2} (x_2) \otimes \ldots
\otimes \widehat{\lambda}^{\, n} (x_n) \, ) \,.
\label{extraction}
\end{equation}
Let us demonstrate this for the two-point function. We combine
\begin{equation}
\begin{split}
\omega_2 \, ( \, \overline{\theta}_{\alpha'_1} (x'_1) \otimes \overline{\theta}_{\alpha'_2} (x'_2) \,,
\widehat{\lambda}^{\, 1} (x_1) \otimes \widehat{\lambda}^{\, 2} (x_2) \, ) 
& = \overline{\epsilon}^{\, 1}_{\alpha'_1} \, \delta^d ( x'_1-x_1 ) \,
\overline{\epsilon}^{\, 2}_{\alpha'_2} \, \delta^d ( x'_2-x_2 ) \,, \\ 
\omega_2 \, ( \, \overline{\theta}_{\alpha'_1} (x'_1) \otimes \theta_{\alpha'_2} (x'_2) \,,
\widehat{\lambda}^{\, 1} (x_1) \otimes \widehat{\lambda}^{\, 2} (x_2) \, ) 
& = \overline{\epsilon}^{\, 1}_{\alpha'_1} \, \delta^d ( x'_1-x_1 ) \,
\epsilon^{\, 2}_{\alpha'_2} \, \delta^d ( x'_2-x_2 ) \,, \\ 
\omega_2 \, ( \, \theta_{\alpha'_1} (x'_1) \otimes \overline{\theta}_{\alpha'_2} (x'_2) \,,
\widehat{\lambda}^{\, 1} (x_1) \otimes \widehat{\lambda}^{\, 2} (x_2) \, ) 
& = \epsilon^{\, 1}_{\alpha'_1} \, \delta^d ( x'_1-x_1 ) \,
\overline{\epsilon}^{\, 2}_{\alpha'_2} \, \delta^d ( x'_2-x_2 ) \,, \\ 
\omega_2 \, ( \, \theta_{\alpha'_1} (x'_1) \otimes \theta_{\alpha'_2} (x'_2) \,,
\widehat{\lambda}^{\, 1} (x_1) \otimes \widehat{\lambda}^{\, 2} (x_2) \, )
&= \epsilon^1_{\alpha'_1} \, \delta^d ( x'_1-x_1 ) \,
\epsilon^2_{\alpha'_2} \, \delta^d ( x'_2-x_2 )
\end{split}
\end{equation}
with the expansion~\eqref{Phi^2-expansion} to find
\begin{equation}
\begin{split}
& \omega_2 \, ( \, \langle \, \Phi \otimes \Phi \, \rangle \,,
\widehat{\lambda}^{\, 1} (x_1) \otimes \widehat{\lambda}^{\, 2} (x_2) \, ) \\
& = {}-\langle \, \Psi_{\alpha_1} (x_1) \, \Psi_{\alpha_2} (x_2) \, \rangle \,
\overline{\epsilon}^{\, 1}_{\alpha_1} \, \overline{\epsilon}^{\, 2}_{\alpha_2}
+\langle \, \Psi_{\alpha_1} (x_1) \, \overline{\Psi}_{\alpha_2} (x_2) \, \rangle \,
\overline{\epsilon}^{\, 1}_{\alpha_1} \, \epsilon^2_{\alpha_2} \\
& \quad~ {}+\langle \, \overline{\Psi}_{\alpha_1} (x_1) \, \Psi_{\alpha_2} (x_2) \, \rangle \, 
\epsilon^1_{\alpha_1} \, \overline{\epsilon}^{\, 2}_{\alpha_2}
-\langle \, \overline{\Psi}_{\alpha_1} (x_1) \, \overline{\Psi}_{\alpha_2} (x_2) \, \rangle \,
\epsilon^1_{\alpha_1} \, \epsilon^2_{\alpha_2} \,.
\end{split}
\end{equation}
By comparing this with Eq.~\eqref{Psi-hat-expansion},
we confirm that the two-point function
$\langle \, \widehat{\Psi}^1 (x_1) \, \widehat{\Psi}^2 (x_2) \, \rangle$
is indeed extracted from $\langle \, \Phi \otimes \Phi \, \rangle$ as
\begin{equation}
\langle \, \widehat{\Psi}^1 (x_1) \, \widehat{\Psi}^2 (x_2) \, \rangle
= \omega_2 \, ( \, \langle \, \Phi \otimes \Phi \, \rangle \,,
\widehat{\lambda}^{\, 1} (x_1) \otimes \widehat{\lambda}^{\, 2} (x_2) \, ) \,.
\end{equation}
Using Eq.~\eqref{extraction},
the formula~\eqref{Dirac-correlation-functions} can be expressed as
\begin{equation}
\langle \, \widehat{\Psi}^1 (x_1) \, \widehat{\Psi}^2 (x_2) \, \ldots \,
\widehat{\Psi}^n (x_n) \, \rangle
= \omega_n \, ( \, \pi_n \, {\bm f} \, {\bf 1} \,,\,
\widehat{\lambda}^{\, 1} (x_1) \otimes \widehat{\lambda}^{\, 2} (x_2) \otimes \ldots
\otimes \widehat{\lambda}^{\, n} (x_n) \, ) \,.
\label{extended-formula}
\end{equation}

Let us calculate two-point functions and four-point functions.
The two-point functions can be calculated from $\, \pi_{2} \, {\bm f} \, {\bf 1} \,$.
For the free theory, it is given by
\begin{equation}
\, \pi_{2} \, {\bm f} \, {\bf 1} \,
= {}-i \hbar \, \pi_{2} \, {\bm h} \, {\bf U} \, {\bf 1}
\end{equation}
as in the case of the scalar field.
The operator ${\bf U}$ acting on ${\bf 1}$ generates
the following element of $\mathcal{H} \otimes \mathcal{H}$:
\begin{equation}
{\bf U} \, {\bf 1} = {}-\int d^d x \, ( \,
\overline{\theta}_\alpha (x) \otimes \lambda_\alpha (x)
+\lambda_\alpha (x) \otimes \overline{\theta}_\alpha (x)
+\theta_\alpha (x) \otimes \overline{\lambda}_\alpha (x)
+\overline{\lambda}_\alpha (x) \otimes \theta_\alpha (x) \, ) \,.
\end{equation}
The action of ${\bm h}$ on $\mathcal{H} \otimes \mathcal{H}$ is given by
\begin{equation}
{\bm h} \, \pi_2 = ( \, \mathbb{I} \otimes h \, ) \, \pi_2 \,.
\end{equation}
Since $h$ annihilates $\theta_\alpha (x)$ and $\overline{\theta}_\alpha (x)$,
two of the four terms survive:
\begin{equation}
{\bm h} \, {\bf U} \, {\bf 1} = \int d^d x \, ( \,
\overline{\theta}_\alpha (x) \otimes h \, \lambda_\alpha (x)
+\theta_\alpha (x) \otimes h \, \overline{\lambda}_\alpha (x) \, ) \,.
\label{Dirac-hU1}
\end{equation}
We thus find
\begin{equation}
\begin{split}
\pi_{2} \, {\bm f} \, {\bf 1}
& = {}-i \hbar \, \pi_{2} \, {\bm h} \, {\bf U} \, {\bf 1} \\
& = {}-i \hbar \int d^d x \int d^d y \, [ \,
\overline{\theta}_{\alpha} (x) \otimes S(x-y)_{\alpha \beta} \, \theta_{\beta} (y)
-\theta_{\alpha}(x) \otimes \overline{\theta}_{\beta} (y) \, S(y-x)_{\beta\alpha } \, ] \,,
\end{split}
\label{Dirac-pi_2-f1}
\end{equation}
and 
$\omega_2 \, ( \, \pi_2 \, {\bm f} \, {\bf 1} \,,
\lambda_\alpha (x) \otimes \overline{\lambda}_\beta (y) \, )$
is given by
\begin{equation}
\omega_2 \, ( \, \pi_2 \, {\bm f} \, {\bf 1} \,,
\lambda_\alpha (x) \otimes \overline{\lambda}_\beta (y) \, )
= {}-i \hbar \, S(x-y)_{\alpha \beta} \,.
\end{equation}
This correctly reproduces the two-point function
$\langle \, \Psi_\alpha (x) \, \overline{\Psi}_\beta (y) \, \rangle$:
\begin{equation}
\langle \, \Psi_\alpha (x) \, \overline{\Psi}_\beta (y) \, \rangle
= \frac{\hbar}{i} \, S(x-y)_{\alpha \beta} \,.
\end{equation}
We also see from Eq.~\eqref{Dirac-pi_2-f1} that
$\omega_2 \, ( \, \pi_2 \, {\bm f} \, {\bf 1} \,,
\overline{\lambda}_\beta (y) \otimes \lambda_\alpha (x) \, )$
is nonvanishing, and it is given by
\begin{equation}
\omega_2 \, ( \, \pi_2 \, {\bm f} \, {\bf 1} \,,
\overline{\lambda}_\beta (y) \otimes \lambda_\alpha (x) \, )
= i \hbar \, S(x-y)_{\alpha \beta} \,.
\end{equation}
This correctly reproduces the two-point function
$\langle \, \overline{\Psi}_\beta (y) \, \Psi_\alpha (x) \, \rangle$:
\begin{equation}
\langle \, \overline{\Psi}_\beta (y) \, \Psi_\alpha (x) \, \rangle
= {}-\frac{\hbar}{i} \, S(x-y)_{\alpha \beta} \,.
\end{equation}
Note that the antisymmetry under the exchange of fermions
is realized:
\begin{equation}
\langle \, \overline{\Psi}_\beta (y) \, \Psi_\alpha (x) \, \rangle
= {}-\langle \, \Psi_\alpha (x) \, \overline{\Psi}_\beta (y) \, \rangle \,.
\end{equation}

The results of the two-point functions can be incorporated into
$\langle \, \widehat{\Psi}^1 (x_1) \, \widehat{\Psi}^2 (x_2) \, \rangle$ as
\begin{equation}
\begin{split}
\langle \, \widehat{\Psi}^1 (x_1) \, \widehat{\Psi}^2 (x_2) \, \rangle
& = \frac{\hbar}{i} \, S (x_1-x_2)_{\alpha_1 \alpha_2} \,
\overline{\epsilon}^{\, 1}_{\alpha_1} \, \epsilon^2_{\alpha_2}
-\frac{\hbar}{i} \, S (x_2-x_1)_{\alpha_2 \alpha_1} \,
\epsilon^1_{\alpha_1} \, \overline{\epsilon}^{\, 2}_{\alpha_2} \\
& = \frac{\hbar}{i} \, S (x_1-x_2)_{\alpha_1 \alpha_2} \,
\overline{\epsilon}^{\, 1}_{\alpha_1} \, \epsilon^2_{\alpha_2}
+\frac{\hbar}{i} \, S (x_2-x_1)_{\alpha_2 \alpha_1} \,
\overline{\epsilon}^{\, 2}_{\alpha_2} \, \epsilon^1_{\alpha_1} \,.
\end{split}
\end{equation}
We define
\begin{equation}
\widehat{S}^{\, ij} (x_i-x_j)
= \overline{\epsilon}^{\, i}_\alpha \, S (x_i-x_j)_{\alpha \beta} \, \epsilon^j_\beta
+\overline{\epsilon}^{\, j}_\alpha \, S (x_j-x_i)_{\alpha \beta} \, \epsilon^i_\beta
\end{equation}
to express $\langle \, \widehat{\Psi}^1 (x_1) \, \widehat{\Psi}^2 (x_2) \, \rangle$ as
\begin{equation}
\langle \, \widehat{\Psi}^1 (x_1) \, \widehat{\Psi}^2 (x_2) \, \rangle
= \frac{\hbar}{i} \, \widehat{S}^{\, 12} (x_1-x_2) \,.
\end{equation}
Note that $\widehat{S}^{\, ij} (x_i-x_j)$ is symmetric in the following sense:
\begin{equation}
\widehat{S}^{\, ij} (x_i-x_j) = \widehat{S}^{\, ji} (x_j-x_i) \,.
\end{equation}

The four-point functions can be calculated from $\, \pi_{4} \, {\bm f} \, {\bf 1} \,$.
For the free theory, it is given by
\begin{equation}
\pi_{4} \, {\bm f} \, {\bf 1}
= {}-\hbar^2 \, \pi_{4} \, {\bm h} \, {\bf U} \, {\bm h} \, {\bf U} \,{\bf 1}
\end{equation}
as in the case of the scalar field.
The operator ${\bf U}$ acting on ${\bm h} \, {\bf U} \, {\bf 1}$ in Eq.~\eqref{Dirac-hU1} generates
many terms in $\mathcal{H}^{\otimes 4}$.
Let us decompose ${\bf U}$ as
\begin{equation}
{\bf U} = {\bf V}+\overline{\bf V} \,,
\end{equation}
where
\begin{equation}
{\bf V} = -\int d^d x \, \overline{\bm \theta}_\alpha (x) \, {\bm \lambda}_\alpha (x) \,, \qquad
\overline{\bf V} = -\int d^d x \, {\bm \theta}_\alpha (x) \, \overline{\bm \lambda}_\alpha (x) \,,
\end{equation}
and we first consider ${\bm h} {\bf V} \, {\bm h} {\bf V} \, {\bf 1}$.
The action of ${\bm h}$ on $\mathcal{H}^{\otimes 4}$ is given by
\begin{equation}
{\bm h} \, \pi_4 = ( \, \mathbb{I} \otimes \mathbb{I} \otimes \mathbb{I} \otimes h \, ) \, \pi_4 \,,
\end{equation}
and $h$ annihilates $\overline{\theta}_\alpha (x)$
and $h \, \lambda_\alpha (x)$
so that the following three terms survive:
\begin{equation}
\begin{split}
{\bm h} \, {\bf V} \, {\bm h} \, {\bf V} \, {\bf 1}
& = \int d^d x \int d^d x' \, ( \, 
\overline{\theta}_{\alpha'} (x') \otimes \overline{\theta}_\alpha (x)
\otimes h \, \lambda_\alpha (x) \otimes h \, \lambda_{\alpha'} (x') \\
& \qquad \qquad \qquad \quad
{}-\overline{\theta}_\alpha (x) \otimes \overline{\theta}_{\alpha'} (x')
\otimes h \, \lambda_\alpha (x) \otimes h \, \lambda_{\alpha'} (x') \\
& \qquad \qquad \qquad \quad
+\overline{\theta}_\alpha (x) \otimes h \, \lambda_\alpha (x)
\otimes \overline{\theta}_{\alpha'} (x') \otimes h \, \lambda_{\alpha'} (x') \, ) \,.
\end{split}
\label{hVhV1}
\end{equation}
This should be contrasted with ${\bm h} \, {\bf U} \, {\bm h} \, {\bf U} \, {\bf 1}$
for the scalar field in Eq.~\eqref{hUhU1}.
Whereas $c(x)$ and $h \, d(x)$ for the scalar field are degree even,
$\overline{\theta}_\alpha (x)$ and $h \, \lambda_\alpha (x)$
for the Dirac field are degree odd
so that we have the minus sign in the second line of Eq.~\eqref{hVhV1}.
This minus sign is necessary for the antisymmetry of fermions
in the correlation functions.
We can similarly calculate
${\bm h} \, {\bf V} \, {\bm h} \, \overline{\bf V} \, {\bf 1}$,
${\bm h} \, \overline{\bf V} \, {\bm h} \, {\bf V} \, {\bf 1}$,
and ${\bm h} \, \overline{\bf V} \, {\bm h} \, \overline{\bf V} \, {\bf 1}$:
\begin{equation}
\begin{split}
{\bm h} \, {\bf V} \, {\bm h} \, \overline{\bf V} \, {\bf 1}
& = \int d^d x \int d^d x' \, ( \, 
\overline{\theta}_{\alpha'} (x') \otimes \theta_\alpha (x)
\otimes h \, \overline{\lambda}_\alpha (x) \otimes h \, \lambda_{\alpha'} (x') \\
& \qquad \qquad \qquad \quad
{}-\theta_\alpha (x) \otimes \overline{\theta}_{\alpha'} (x')
\otimes h \, \overline{\lambda}_\alpha (x) \otimes h \, \lambda_{\alpha'} (x') \\
& \qquad \qquad \qquad \quad
+\theta_\alpha (x) \otimes h \, \overline{\lambda}_\alpha (x)
\otimes \overline{\theta}_{\alpha'} (x') \otimes h \, \lambda_{\alpha'} (x') \, ) \,, \\
{\bm h} \, \overline{\bf V} \, {\bm h} \, {\bf V} \, {\bf 1}
& = \int d^d x \int d^d x' \, ( \, 
\theta_{\alpha'} (x') \otimes \overline{\theta}_\alpha (x)
\otimes h \, \lambda_\alpha (x) \otimes h \, \overline{\lambda}_{\alpha'} (x') \\
& \qquad \qquad \qquad \quad
{}-\overline{\theta}_\alpha (x) \otimes \theta_{\alpha'} (x')
\otimes h \, \lambda_\alpha (x) \otimes h \, \overline{\lambda}_{\alpha'} (x') \\
& \qquad \qquad \qquad \quad
+\overline{\theta}_\alpha (x) \otimes h \, \lambda_\alpha (x)
\otimes \theta_{\alpha'} (x') \otimes h \, \overline{\lambda}_{\alpha'} (x') \, ) \,, \\
{\bm h} \, \overline{\bf V} \, {\bm h} \, \overline{\bf V} \, {\bf 1}
& = \int d^d x \int d^d x' \, ( \, 
\theta_{\alpha'} (x') \otimes \theta_\alpha (x)
\otimes h \, \overline{\lambda}_\alpha (x) \otimes h \, \overline{\lambda}_{\alpha'} (x') \\
& \qquad \qquad \qquad \quad
{}-\theta_\alpha (x) \otimes \theta_{\alpha'} (x')
\otimes h \, \overline{\lambda}_\alpha (x) \otimes h \, \overline{\lambda}_{\alpha'} (x') \\
& \qquad \qquad \qquad \quad
+\theta_\alpha (x) \otimes h \, \overline{\lambda}_\alpha (x)
\otimes \theta_{\alpha'} (x') \otimes h \, \overline{\lambda}_{\alpha'} (x') \, ) \,.
\end{split}
\end{equation}
Using the definition of $h$ in Eq.~\eqref{Dirac-h},
$\pi_{4} \, {\bm f} \, {\bf 1}$ is given by
\begin{equation}
\pi_{4} \, {\bm f} \, {\bf 1}
= {}-\hbar^2 \, \pi_{4} \, {\bm h} \, {\bf U} \, {\bm h} \, {\bf U} \,{\bf 1}
= {}-\hbar^2 \int d^d x \int d^d x' \int d^d y \int d^d y' \, \mathcal{F} \, (x,y,x',y') \,,
\end{equation}
where
\begin{equation}
\begin{split}
\mathcal{F} \, (x,y,x',y')
& =
\overline{\theta}_{\alpha'} (x') \otimes \overline{\theta}_\alpha (x)
\otimes S(x-y)_{\alpha \beta} \, \theta_\beta (y) \otimes S(x'-y')_{\alpha' \beta'} \, \theta_{\beta'} (y') \\
& \quad~
{}-\overline{\theta}_\alpha (x) \otimes \overline{\theta}_{\alpha'} (x')
\otimes S(x-y)_{\alpha \beta} \, \theta_\beta (y) \otimes S(x'-y')_{\alpha' \beta'} \, \theta_{\beta'} (y') \\
& \quad~
+\overline{\theta}_\alpha (x) \otimes S(x-y)_{\alpha \beta} \, \theta_\beta (y)
\otimes \overline{\theta}_{\alpha'} (x') \otimes S(x'-y')_{\alpha' \beta'} \, \theta_{\beta'} (y') \\
& \quad~
{}-\overline{\theta}_{\alpha'} (x') \otimes \theta_\alpha (x)
\otimes \overline{\theta}_\beta (y) \, S(y-x)_{\beta \alpha} \otimes S(x'-y')_{\alpha' \beta'} \, \theta_{\beta'} (y') \\
& \quad~
+\theta_\alpha (x) \otimes \overline{\theta}_{\alpha'} (x')
\otimes \overline{\theta}_\beta (y) \, S(y-x)_{\beta \alpha} \otimes S(x'-y')_{\alpha' \beta'} \, \theta_{\beta'} (y') \\
& \quad~
{}-\theta_\alpha (x) \otimes \overline{\theta}_\beta (y) \, S(y-x)_{\beta \alpha}
\otimes \overline{\theta}_{\alpha'} (x') \otimes S(x'-y')_{\alpha' \beta'} \, \theta_{\beta'} (y') \\
& \quad~
{}-\theta_{\alpha'} (x') \otimes \overline{\theta}_\alpha (x)
\otimes S(x-y)_{\alpha \beta} \, \theta_\beta (y) \otimes \overline{\theta}_{\beta'} (y') \, S(y'-x')_{\beta' \alpha'} \\
& \quad~
+\overline{\theta}_\alpha (x) \otimes \theta_{\alpha'} (x')
\otimes S(x-y)_{\alpha \beta} \, \theta_\beta (y) \otimes \overline{\theta}_{\beta'} (y') \, S(y'-x')_{\beta' \alpha'} \\
& \quad~
{}-\overline{\theta}_\alpha (x) \otimes S(x-y)_{\alpha \beta} \, \theta_\beta (y)
\otimes \theta_{\alpha'} (x') \otimes \overline{\theta}_{\beta'} (y') \, S(y'-x')_{\beta' \alpha'} \\
& \quad~
+\theta_{\alpha'} (x') \otimes \theta_\alpha (x)
\otimes \overline{\theta}_\beta (y) \, S(y-x)_{\beta \alpha} \otimes \overline{\theta}_{\beta'} (y') \, S(y'-x')_{\beta' \alpha'} \\
& \quad~
{}-\theta_\alpha (x) \otimes \theta_{\alpha'} (x')
\otimes \overline{\theta}_\beta (y) \, S(y-x)_{\beta \alpha} \otimes \overline{\theta}_{\beta'} (y') \, S(y'-x')_{\beta' \alpha'} \\
& \quad~
+\theta_\alpha (x) \otimes \overline{\theta}_\beta (y) \, S(y-x)_{\beta \alpha}
\otimes \theta_{\alpha'} (x') \otimes \overline{\theta}_{\beta'} (y') \, S(y'-x')_{\beta' \alpha'} \,.
\end{split}
\end{equation}
This contains all the information about the four-point functions.
For example, the four-point function
$\langle \, \Psi_{\alpha_1} (x_1) \, \Psi_{\alpha_2} (x_2) \,
\overline{\Psi}_{\beta_1} (y_1) \, \overline{\Psi}_{\beta_2} (y_2) \, \rangle$ is given by
\begin{equation}
\begin{split}
& \omega_4 \, ( \, \pi_4 \, {\bm f} \, {\bf 1} \,,
\lambda_{\alpha_1} (x_1) \otimes \lambda_{\alpha_2} (x_2)
\otimes \overline{\lambda}_{\beta_1} (y_1) \otimes \overline{\lambda}_{\beta_2} (y_2) \, ) \\
&= {}-\hbar^2 \, [ \,
S (x_2-y_1)_{\alpha_2 \beta_1} \, S (x_1-y_2)_{\alpha_1 \beta_2}
-S (x_1-y_1)_{\alpha_1 \beta_1} \, S (x_2-y_2)_{\alpha_2 \beta_2} \, ] \,,
\end{split}
\end{equation}
and Wick's theorem is correctly reproduced as follows:
\begin{equation}
\begin{split}
& \langle \, \Psi_{\alpha_1} (x_1) \, \Psi_{\alpha_2} (x_2) \,
\overline{\Psi}_{\beta_1} (y_1) \, \overline{\Psi}_{\beta_2} (y_2) \, \rangle \\
& = \langle \, \Psi_{\alpha_2} (x_2) \, \overline{\Psi}_{\beta_1} (y_1) \, \rangle \,
\langle \, \Psi_{\alpha_1} (x_1) \, \overline{\Psi}_{\beta_2} (y_2) \,  \rangle
-\langle \, \Psi_{\alpha_1} (x_1) \, \overline{\Psi}_{\beta_1} (y_1) \, \rangle \,
\langle \, \Psi_{\alpha_2} (x_2) \, \overline{\Psi}_{\beta_2} (y_2) \, \rangle \,.
\end{split}
\end{equation}
On the other hand, the four-point function
$\langle \, \Psi_{\alpha_1} (x_1) \, 
\overline{\Psi}_{\beta_1} (y_1) \, 
\Psi_{\alpha_2} (x_2) \,
\overline{\Psi}_{\beta_2} (y_2) \, \rangle$ is given by
\begin{equation}
\begin{split}
& \omega_4 \, ( \, \pi_4 \, {\bm f} \, {\bf 1} \,,
\lambda_{\alpha_1} (x_1) \otimes \overline{\lambda}_{\beta_1} (y_1)
\otimes \lambda_{\alpha_2} (x_2) \otimes \overline{\lambda}_{\beta_2} (y_2) \, ) \\
&= {}-\hbar^2 \, [ \,
S (x_1-y_1)_{\alpha_1 \beta_1} \, S (x_2-y_2)_{\alpha_2 \beta_2}
-S (x_2-y_1)_{\alpha_2 \beta_1} \, S (x_1-y_2)_{\alpha_1 \beta_2} \, ] \,,
\end{split}
\end{equation}
and we see that the antisymmetry under the exchange of fermions is realized:
\begin{equation}
\langle \, \Psi_{\alpha_1} (x_1) \, \overline{\Psi}_{\beta_1} (y_1) \, 
\Psi_{\alpha_2} (x_2) \, \overline{\Psi}_{\beta_2} (y_2) \, \rangle
= {}-\langle \, \Psi_{\alpha_1} (x_1) \, \Psi_{\alpha_2} (x_2) \,
\overline{\Psi}_{\beta_1} (y_1) \, \overline{\Psi}_{\beta_2} (y_2) \, \rangle \,.
\end{equation}

All the information on the four-point functions can be compactly expressed
in terms of $\widehat{\lambda}^{\, i} (x)$ and $\widehat{S}^{\, ij} (x_i-x_j)$.
We find
\begin{equation}
\begin{split}
& \omega_4 \, ( \, \pi_4 \, {\bm f} \, {\bf 1} \,,
\widehat{\lambda}^{\, 1} (x_1) \otimes \widehat{\lambda}^{\, 2} (x_2)
\otimes \widehat{\lambda}^{\, 3} (x_3) \otimes \widehat{\lambda}^{\, 4} (x_4) \, ) \\
& = {}-\hbar^2 \, [ \,
\widehat{S}^{\, 12} (x_1-x_2) \, \widehat{S}^{\, 34} (x_3-x_4)
+\widehat{S}^{\, 13} (x_1-x_3) \, \widehat{S}^{\, 24} (x_2-x_4) \\
& \qquad \qquad
+\widehat{S}^{\, 14} (x_1-x_4) \, \widehat{S}^{\, 23} (x_2-x_3) \, ] \,.
\end{split}
\end{equation}
This takes the form of
\begin{equation}
\begin{split}
& \omega_4 \, ( \, \pi_4 \, {\bm f} \, {\bf 1} \,,
\widehat{\lambda}^{\, 1} (x_1) \otimes \widehat{\lambda}^{\, 2} (x_2)
\otimes \widehat{\lambda}^{\, 3} (x_3) \otimes \widehat{\lambda}^{\, 4} (x_4) \, ) \\
& = \langle \, \widehat{\Psi}^1 (x_1) \, \widehat{\Psi}^2 (x_2) \, \rangle \,
\langle \, \widehat{\Psi}^3 (x_3) \, \widehat{\Psi}^4 (x_4) \, \rangle
+\langle \, \widehat{\Psi}^1 (x_1) \, \widehat{\Psi}^3 (x_3) \, \rangle \,
\langle \, \widehat{\Psi}^2 (x_2) \, \widehat{\Psi}^4 (x_4) \, \rangle \\
& \quad~ +\langle \, \widehat{\Psi}^1 (x_1) \, \widehat{\Psi}^4 (x_4) \, \rangle \,
\langle \, \widehat{\Psi}^2 (x_2) \, \widehat{\Psi}^3 (x_3) \, \rangle \,,
\end{split}
\end{equation}
and Wick's theorem for the four-point functions,
\begin{equation}
\begin{split}
\langle \, \widehat{\Psi}^1 (x_1) \, \widehat{\Psi}^2 (x_2) \,
\widehat{\Psi}^3 (x_3) \, \widehat{\Psi}^4 (x_4) \, \rangle
& = \langle \, \widehat{\Psi}^1 (x_1) \, \widehat{\Psi}^2 (x_2) \, \rangle \,
\langle \, \widehat{\Psi}^3 (x_3) \, \widehat{\Psi}^4 (x_4) \, \rangle \\
& \quad~ +\langle \, \widehat{\Psi}^1 (x_1) \, \widehat{\Psi}^3 (x_3) \, \rangle \,
\langle \, \widehat{\Psi}^2 (x_2) \, \widehat{\Psi}^4 (x_4) \, \rangle \\
& \quad~ +\langle \, \widehat{\Psi}^1 (x_1) \, \widehat{\Psi}^4 (x_4) \, \rangle \,
\langle \, \widehat{\Psi}^2 (x_2) \, \widehat{\Psi}^3 (x_3) \, \rangle \,,
\end{split}
\end{equation}
is correctly reproduced.

\subsection{The Schwinger-Dyson equations}

In the path integral formalism, correlation functions are defined by
\begin{equation}
\begin{split}
& \langle \, \Psi_{\alpha_1} (x_1) \, \ldots \Psi_{\alpha_n} (x_n) \,
\overline{\Psi}_{\beta_1} (y_1) \, \ldots \overline{\Psi}_{\beta_m} (y_m) \, \rangle \, \\
& = \frac{1}{Z} \int \mathcal{D} \Psi \, \mathcal{D} \overline{\Psi} \,
\Psi_{\alpha_1} (x_1) \, \ldots \Psi_{\alpha_n} (x_n) \,
\overline{\Psi}_{\beta_1} (y_1) \, \ldots \overline{\Psi}_{\beta_m} (y_m) \, e^{\frac{i}{\hbar}S},
\end{split}
\end{equation}
where
\begin{equation}
  Z = \int \mathcal{D}\Psi\,\mathcal{D}\overline{\Psi}\,e^{\frac{i}{\hbar}S}.
\end{equation}
Since
\begin{equation}
\frac{1}{Z} \int \mathcal{D} \Psi \, \mathcal{D} \overline{\Psi} \,
\frac{\delta}{\delta\overline{\Psi}_{\beta_m}(y_m)}
\Bigl[ \, \Psi_{\alpha_1} (x_1) \ldots \Psi_{\alpha_n} (x_n) \,
\overline{\Psi}_{\beta_1} (y_1) \ldots \overline{\Psi}_{\beta_{m-1}} (y_{m-1}) \,
e^{\frac{i}{\hbar}S} \, \Bigr] = 0
\end{equation}
and
\begin{equation}
\frac{1}{Z} \int \mathcal{D} \Psi \, \mathcal{D} \overline{\Psi} \,
\frac{\delta}{\delta\Psi_{\alpha_n}(x_n)}
\Bigl[ \, \Psi_{\alpha_1} (x_1) \ldots \Psi_{\alpha_{n-1}} (x_{n-1}) \,
\overline{\Psi}_{\beta_1} (y_1) \ldots \overline{\Psi}_{\beta_m} (y_m) \,
e^{\frac{i}{\hbar}S} \, \Bigr] = 0 \,,
\end{equation}
the Schwinger-Dyson equations are given by
\begin{equation}
\begin{split}
& \sum_{i=1}^{m-1} \, (-1)^{i-1} \, \delta_{\beta_i \beta_m} \, \delta^d (y_i-y_m) \,
\langle \, \Psi_{\alpha_1} (x_1) \ldots \,
\overline{\Psi}_{\beta_1} (y_1) \ldots \overline{\Psi}_{\beta_{i-1}} (y_{i-1}) \,
\overline{\Psi}_{\beta_{i+1}} (y_{i+1}) \ldots \overline{\Psi}_{\beta_{m-1}} (y_{m-1}) \, \rangle \\
& +(-1)^{m-1} \, \frac{i}{\hbar} \,
\langle \, \Psi_{\alpha_1} (x_1) \ldots \Psi_{\alpha_n} (x_n) \,
\overline{\Psi}_{\beta_1} (y_1) \ldots \overline{\Psi}_{\beta_{m-1}} (y_{m-1}) \,
\frac{\delta S}{\delta \overline{\Psi}_{\beta_m}(y_m)} \, \rangle = 0
\end{split}
\label{sd_dirac1}
\end{equation}
and
\begin{equation}
\begin{split}
& \sum_{i=1}^{n-1} \, (-1)^{i-1} \, \delta_{\alpha_i \alpha_n} \, \delta^d (x_i-x_n) \,
\langle \, \Psi_{\alpha_1} (x_1) \ldots \Psi_{\alpha_{i-1}} (x_{i-1}) \,
\Psi_{\alpha_{i+1}} (x_{i+1}) \ldots \Psi_{\alpha_{n-1}} (x_{n-1}) \,
\overline{\Psi}_{\beta_1} (y_1) \ldots \, \rangle \\
& +(-1)^{n+m-1} \, \frac{i}{\hbar} \,
\langle \, \Psi_{\alpha_1} (x_1) \ldots \Psi_{\alpha_{n-1}} (x_{n-1}) \,
\overline{\Psi}_{\beta_1} (y_1) \ldots \overline{\Psi}_{\beta_m} (y_m) \,
\frac{\delta S}{\delta \Psi_{\alpha_n}(x_n)} \, \rangle = 0 \,.
\label{sd_dirac2}
\end{split}
\end{equation}
Here and in what follows, degree-odd derivatives are left derivatives.
For example,
the functional derivatives of the action~\eqref{Dirac-action} for the free theory are given by
\begin{equation}
\begin{split}
\frac{\delta S}{\delta \Psi_\alpha (x)}
= \overline{\Psi}_\beta (x) \, ( \, i \overleftarrow{\partial\!\!\!/} +m \, )_{\beta \alpha} \,, \qquad
\frac{\delta S}{\delta \overline{\Psi}_\alpha (x)}
= ( \, i \, \partial\!\!\!/ -m \, )_{\alpha \beta} \, \Psi_\beta (x) \,.
\end{split}
\end{equation}
It is convenient to introduce functional derivatives
associated with $\widehat{\Psi}^i (x)$ as defined in Eq.~\eqref{Psi-hat}.
We define
\begin{equation}
\frac{\delta}{\delta \widehat{\Psi}^i (x)}
= \epsilon^i_\alpha \, \frac{\delta}{\delta \Psi_\alpha (x)}
+\overline{\epsilon}^{\, i}_\alpha \, \frac{\delta}{\delta \overline{\Psi}_\alpha (x)} \,.
\end{equation}
We then find
\begin{equation}
\frac{\delta}{\delta \widehat{\Psi}^i (x)} \, \widehat{\Psi}^j (y)
= \widehat{\delta}^{\, ij} (x-y) \,,
\end{equation}
where
\begin{equation}
\widehat{\delta}^{\, ij} (x-y)
= \delta^d (x-y) \,
( \, \overline{\epsilon}^{\, i}_\alpha \, \epsilon^j_\alpha
+\overline{\epsilon}^{\, j}_\alpha \, \epsilon^i_\alpha \, ) \,.
\end{equation}
Since
\begin{equation}
\frac{1}{Z} \int \mathcal{D} \Psi \, \mathcal{D} \overline{\Psi} \,
\frac{\delta}{\delta \widehat{\Psi}^n (x_n)} \,
\Bigl[ \, \widehat{\Psi}^1 (x_1) \, \widehat{\Psi}^2 (x_2) \, \ldots
\widehat{\Psi}^{n-1} (x_{n-1}) \, e^{\frac{i}{\hbar}S} \,
\Bigr] = 0 \,,
\end{equation}
the Schwinger-Dyson equations can be organized as
\begin{equation}
\begin{split}
& \sum_{i=1}^{n-1} \, \widehat{\delta}^{\, i n} (x_i-x_n) \,
\langle \, \widehat{\Psi}^1 (x_1) \, \ldots \, \widehat{\Psi}^{i-1} (x_{i-1}) \,
\widehat{\Psi}^{i+1} (x_{i+1}) \, \ldots \, \widehat{\Psi}^{n-1} (x_{n-1}) \, \rangle \\
& \qquad +\frac{i}{\hbar} \,
\langle \, \widehat{\Psi}^1 (x_1) \, \ldots \, \widehat{\Psi}^{n-1} (x_{n-1}) \,
\frac{\delta S}{\delta \widehat{\Psi}^n (x_n)} \, \rangle = 0 \,.
\end{split}
\label{Dirac-Schwinger-Dyson}
\end{equation}
Let us show that the Schwinger-Dyson equations are satisfied
for correlation functions from the formula~\eqref{extended-formula}
in the case of the free theory.
The proof for interacting theories is presented in Ref.~\cite{Konosu:2023rkm}.
Since
\begin{equation}
( \, {\bf{I}} +i \hbar \, {\bm{h}} \, {\bf U} \, ) \,
\frac{1}{{\bf{I}} +i \hbar \, \bm{h} \, \bf{U}} \, \bf{1} = \bf{1}
\end{equation}
and
\begin{equation}
\pi_n \, {\bf 1} = 0
\end{equation}
for $n \geq 1$, we have
\begin{equation}
\pi_n \,{\bm f} \, {\bf 1}
+i \hbar \,\pi_n \, {\bm h } \, {\bf U } \, {\bm f} \, {\bf 1 } = 0
\end{equation}
for $n \geq 1$.
Let us consider the relation
\begin{equation}
\begin{split}
& \omega_n \, ( \, \pi_n \, {\bm f} \, {\bf 1} \,,\,
\widehat{\lambda}^{\, 1} (x_1)  \otimes \widehat{\lambda}^{\, 2} (x_2)  \otimes \ldots
\otimes \widehat{\lambda}^{\, n} (x_n) \, ) \\
& +i \hbar \, \omega_n \, ( \, \pi_n \, {\bm h} \, {\bf U} \, {\bm f} \, {\bf 1} \,,\,
\widehat{\lambda}^{\, 1} (x_1)  \otimes \widehat{\lambda}^{\, 2} (x_2)  \otimes \ldots
\otimes \widehat{\lambda}^{\, n} (x_n) \, ) = 0 \,.
\end{split}
\label{Dirac-omega_n}
\end{equation}
The first term on the left-hand side corresponds to the $n$-point function given by
\begin{equation}
\omega_n \, ( \, \pi_n \, {\bm f} \, {\bf 1} \,,\,
\widehat{\lambda}^{\, 1} (x_1)  \otimes \widehat{\lambda}^{\, 2} (x_2)  \otimes \ldots
\otimes \widehat{\lambda}^{\, n} (x_n) \, )
= \langle \, \widehat{\Psi}^1 (x_1) \, \widehat{\Psi}^2 (x_2) \, \ldots \,
\widehat{\Psi}^n (x_n) \, \rangle \,.
\end{equation}
For the second term on the left-hand side,
\begin{equation}
i \hbar \, \omega_n \, ( \, \pi_n \, {\bm h} \, {\bf U} \, {\bm f} \, {\bf 1} \,,\,
\widehat{\lambda}^{\, 1} (x_1)  \otimes \widehat{\lambda}^{\, 2} (x_2)  \otimes \ldots
\otimes \widehat{\lambda}^{\, n} (x_n) \, ) \,,
\end{equation}
$\pi_n \, {\bm h} \, {\bf U}$ is given by
\begin{equation}
\begin{split}
\pi_1 \, {\bm h} \, {\bf U} & = 0 \,, \\
\pi_n \, {\bm h} \, {\bf U}
& = \int d^d x \sum_{i=1}^{n-1} \, ( \,
\mathbb{I}^{\otimes (i-1)} \otimes \theta_\alpha (x) \otimes \mathbb{I}^{\otimes (n-i-1)}
\otimes h \, \overline{\lambda}_\alpha (x) \, ) \,
\pi_{n-2} \\
& \quad~ +\int d^d x \sum_{i=1}^{n-1} \, ( \,
\mathbb{I}^{\otimes (i-1)} \otimes \overline{\theta}_\alpha (x) \otimes \mathbb{I}^{\otimes (n-i-1)}
\otimes h \, \lambda_\alpha (x) \, ) \,
\pi_{n-2}
\end{split}
\end{equation}
for $n > 1$.
Let us consider whether the symplectic form
\begin{equation}
\omega_n \, ( \, 
( \, \mathbb{I}^{\otimes (i-1)} \otimes \theta_\alpha (x) \otimes \mathbb{I}^{\otimes (n-i-1)}
\otimes h \, \overline{\lambda}_\alpha (x) \, ) \,
\pi_{n-2} \, {\bm f} \, {\bf 1} \,,\,
\widehat{\lambda}^{\, 1} (x_1) \otimes \ldots
\otimes \widehat{\lambda}^{\, n} (x_n) \, )
\label{term-to-factor}
\end{equation}
can be factorized as a product of
\begin{equation}
\omega_2 \, ( \, \theta_\alpha (x) \otimes h \, \overline{\lambda}_\alpha (x) \,,\,
\widehat{\lambda}^{\, i} (x_i) \otimes \widehat{\lambda}^{\, n} (x_n) \, )
\end{equation}
and
\begin{equation}
\omega_{n-2} \, ( \, \pi_{n-2} \, {\bm f} \, {\bf 1} \,,\,
\widehat{\lambda}^{\, 1} (x_1) \otimes \ldots
\widehat{\lambda}^{\, i-1} (x_{i-1})
\otimes \widehat{\lambda}^{\, i+1} (x_{i+1}) \otimes \ldots
\otimes \widehat{\lambda}^{\, n-1} (x_{n-1}) \, )
\end{equation}
as in the case of the scalar field.
When we expand Eq.~\eqref{term-to-factor},
\begin{equation}
\omega_n \, ( \, 
( \, \mathbb{I}^{\otimes (i-1)} \otimes \theta_\alpha (x) \otimes \mathbb{I}^{\otimes (n-i-1)}
\otimes h \, \overline{\lambda}_\alpha (x) \, ) \,
\pi_{n-2} \, {\bm f} \, {\bf 1} \,,\,
\overline{\lambda}_{\alpha_1} (x_1) \otimes \ldots
\otimes \overline{\lambda}_{\alpha_{n-1}} (x_{n-1}) \otimes \lambda_{\alpha_n} (x_n) \, )
\end{equation}
appears in front of $\epsilon^1_{\alpha_1} \ldots \epsilon^{n-1}_{\alpha_{n-1}}
\overline{\epsilon}^{\, n}_{\alpha_n}$.
Since $\pi_{n-2} \, {\bm f} \, {\bf 1}$ consists of tensor products
of $n-2$ degree-odd elements,
the degree-odd element $\theta_\alpha (x)$ passes $i-1$ degree-odd elements
to give $(-1)^{i-1}$
and the degree-odd element $h \, \overline{\lambda}_\alpha (x)$ passes
$n-2$ degree-odd elements to give $(-1)^{n-2}$.
This $\omega_n$ corresponds to the case
where
$\mathrm{deg} (\Phi_i) = 1$ mod~$2$
and $\mathrm{deg} (\widetilde{\Phi}_i) = 0$ mod~$2$ for any $i$ in Eq.~\eqref{omega_n}
so that $(-1)^\sigma = 1$,
and therefore $\omega_n$ is transformed to the product of $\omega_2$ and $\omega_{n-2}$
without any additional sign:
\begin{align}
& \omega_n \, ( \, 
( \, \mathbb{I}^{\otimes (i-1)} \otimes \theta_\alpha (x) \otimes \mathbb{I}^{\otimes (n-i-1)}
\otimes h \, \overline{\lambda}_\alpha (x) \, ) \,
\pi_{n-2} \, {\bm f} \, {\bf 1} \,,\,
\overline{\lambda}_{\alpha_1} (x_1) \otimes \ldots
\otimes \overline{\lambda}_{\alpha_{n-1}} (x_{n-1}) \otimes \lambda_{\alpha_n} (x_n) \, )
\nonumber \\
& = (-1)^{i-1} \, (-1)^{n-2} \,
\omega_2 \, ( \, \theta_\alpha (x) \otimes h \, \overline{\lambda}_\alpha (x) \,,\,
\overline{\lambda}_{\alpha_i} (x_i) \otimes \lambda_{\alpha_n} (x_n) \, )
\nonumber \\
& \quad~ \times \omega_{n-2} \, ( \, \pi_{n-2} \, {\bm f} \, {\bf 1} \,,\,
\overline{\lambda}_{\alpha_1} (x_1) \otimes \ldots \otimes
\overline{\lambda}_{\alpha_{i-1}} (x_{i-1}) \otimes
\overline{\lambda}_{\alpha_{i+1}} (x_{i+1}) \otimes \ldots
\otimes \overline{\lambda}_{\alpha_{n-1}} (x_{n-1}) \, ) \,.
\end{align}
On the other hand, we rearrange
$\epsilon^1_{\alpha_1} \ldots \epsilon^{n-1}_{\alpha_{n-1}}
\overline{\epsilon}^{\, n}_{\alpha_n}$ as
\begin{equation}
\epsilon^1_{\alpha_1} \ldots \epsilon^{n-1}_{\alpha_{n-1}}
\overline{\epsilon}^{\, n}_{\alpha_n}
= (-1)^{i-1} \, (-1)^{n-2} \,
\epsilon^i_{\alpha_i} \overline{\epsilon}^{\, n}_{\alpha_n}
\epsilon^1_{\alpha_1} \ldots \epsilon^{i-1}_{\alpha_{i-1}}
\epsilon^{i+1}_{\alpha_{i+1}} \ldots \epsilon^{n-1}_{\alpha_{n-1}} \,.
\end{equation}
The total sign factor is thus trivial,
\begin{equation}
(-1)^{i-1} \, (-1)^{n-2} \, (-1)^{i-1} \, (-1)^{n-2} = 1 \,,
\end{equation}
and this is the case for all the terms when we expand the term~\eqref{term-to-factor}.
We therefore find
\begin{equation}
\begin{split}
& \omega_n \, ( \, 
( \, \mathbb{I}^{\otimes (i-1)} \otimes \theta_\alpha (x) \otimes \mathbb{I}^{\otimes (n-i-1)}
\otimes h \, \overline{\lambda}_\alpha (x) \, ) \,
\pi_{n-2} \, {\bm f} \, {\bf 1} \,,\,
\widehat{\lambda}^{\, 1} (x_1) \otimes \ldots
\otimes \widehat{\lambda}^{\, n} (x_n) \, ) \\
& = \omega_2 \, ( \, \theta_\alpha (x) \otimes h \, \overline{\lambda}_\alpha (x) \,,\,
\widehat{\lambda}^{\, i} (x_i) \otimes \widehat{\lambda}^{\, n} (x_n) \, ) \\
& \quad~ \times
\omega_{n-2} \, ( \, \pi_{n-2} \, {\bm f} \, {\bf 1} \,,\,
\widehat{\lambda}^{\, 1} (x_1) \otimes \ldots \otimes
\widehat{\lambda}^{\, i-1} (x_{i-1})
\otimes \widehat{\lambda}^{\, i+1} (x_{i+1}) \otimes \ldots
\otimes \widehat{\lambda}^{\, n-1} (x_{n-1}) \, ) \,.
\end{split}
\end{equation}
Similarly, we have
\begin{equation}
\begin{split}
& \omega_n \, ( \, 
( \, \mathbb{I}^{\otimes (i-1)} \otimes \overline{\theta}_\alpha (x) \otimes \mathbb{I}^{\otimes (n-i-1)}
\otimes h \, \lambda_\alpha (x) \, ) \,
\pi_{n-2} \, {\bm f} \, {\bf 1} \,,\,
\widehat{\lambda}^{\, 1} (x_1) \otimes \ldots
\otimes \widehat{\lambda}^{\, n} (x_n) \, ) \\
& = \omega_2 \, ( \, \overline{\theta}_\alpha (x) \otimes h \, \lambda_\alpha (x) \,,\,
\widehat{\lambda}^{\, i} (x_i) \otimes \widehat{\lambda}^{\, n} (x_n) \, ) \\
& \quad~ \times \omega_{n-2} \, ( \, \pi_{n-2} \, {\bm f} \, {\bf 1} \,,\,
\widehat{\lambda}^{\, 1} (x_1) \otimes \ldots \otimes
\widehat{\lambda}^{\, i-1} (x_{i-1})
\otimes \widehat{\lambda}^{\, i+1} (x_{i+1}) \otimes \ldots
\otimes \widehat{\lambda}^{\, n-1} (x_{n-1}) \, ) \,,
\end{split}
\end{equation}
and the second term on the left-hand side of Eq.~\eqref{Dirac-omega_n} is given by
\begin{equation}
\begin{split}
& i \hbar \, \omega_n \, ( \, \pi_n \, {\bm h} \, {\bf U} \, {\bm f} \, {\bf 1} \,,\,
\widehat{\lambda}^{\, 1} (x_1)  \otimes \widehat{\lambda}^{\, 2} (x_2)  \otimes \ldots
\otimes \widehat{\lambda}^{\, n} (x_n) \, ) \\
& = i \hbar \, \sum_{i=1}^{n-1} \, \int d^d x \, \omega_2 \, ( \,
\theta_\alpha (x) \otimes h \, \overline{\lambda}_\alpha (x)
+\overline{\theta}_\alpha (x) \otimes h \, \lambda_\alpha (x) \,,\,
\widehat{\lambda}^{\, i} (x_i) \otimes \widehat{\lambda}^{\, n} (x_n) \, ) \\
& \qquad \times
\omega_{n-2} \, ( \, \pi_{n-2} \, {\bm f} \, {\bf 1} \,,\,
\widehat{\lambda}^{\, 1} (x_1) \otimes \ldots \otimes
\widehat{\lambda}^{\, i-1} (x_{i-1})
\otimes \widehat{\lambda}^{\, i+1} (x_{i+1}) \otimes \ldots
\otimes \widehat{\lambda}^{\, n-1} (x_{n-1}) \, ) \,.
\end{split}
\end{equation}
The factor $\omega_{n-2} \, ( \, \pi_{n-2} \, {\bm f} \, {\bf 1} \,,\,
\widehat{\lambda}^{\, 1} (x_1) \otimes \ldots \otimes
\widehat{\lambda}^{\, i-1} (x_{i-1})
\otimes \widehat{\lambda}^{\, i+1} (x_{i+1}) \otimes \ldots
\otimes \widehat{\lambda}^{\, n-1} (x_{n-1}) \, )$
on the right-hand side is the following $(n-2)$-point function:
\begin{equation}
\begin{split}
& \omega_{n-2} \, ( \, \pi_{n-2} \, {\bm f} \, {\bf 1} \,,\,
\widehat{\lambda}^{\, 1} (x_1) \otimes \ldots \otimes
\widehat{\lambda}^{\, i-1} (x_{i-1})
\otimes \widehat{\lambda}^{\, i+1} (x_{i+1}) \otimes \ldots
\otimes \widehat{\lambda}^{\, n-1} (x_{n-1}) \, ) \\
& = \langle\, \widehat{\Psi}^1 (x_1) \, \ldots \,
\widehat{\Psi}^{i-1} (x_{i-1}) \,
\widehat{\Psi}^{i+1} (x_{i+1}) \, \ldots \,
\widehat{\Psi}^{n-1} (x_{n-1}) \, \rangle \,.
\end{split}
\end{equation}
Since
\begin{equation}
\int d^d x \, \omega_2 \, ( \,
\theta_\alpha (x) \otimes h \, \overline{\lambda}_\alpha (x)
+\overline{\theta}_\alpha (x) \otimes h \, \lambda_\alpha (x) \,,\,
\widehat{\lambda}^{\, i} (x_i) \otimes \widehat{\lambda}^{\, n} (x_n) \, )
= \widehat{S}^{\, in} (x_i-x_n) \,,
\end{equation}
the second term on the left-hand side of Eq.~\eqref{Dirac-omega_n} is
\begin{equation}
\begin{split}
& i \hbar \, \omega_n \, ( \, \pi_n \, {\bm h} \, {\bf U} \, {\bm f} \, {\bf 1} \,,\,
\widehat{\lambda}^{\, 1} (x_1)  \otimes \widehat{\lambda}^{\, 2} (x_2)  \otimes \ldots
\otimes \widehat{\lambda}^{\, n} (x_n) \, ) \\
& = {}-\frac{\hbar}{i} \, \sum_{i=1}^{n-1} \widehat{S}^{\, in} (x_i-x_n) \,
\langle\, \widehat{\Psi}^1 (x_1) \, \ldots \,
\widehat{\Psi}^{i-1} (x_{i-1}) \,
\widehat{\Psi}^{i+1} (x_{i+1}) \, \ldots \,
\widehat{\Psi}^{n-1} (x_{n-1}) \, \rangle \,.
\end{split}
\end{equation}
The relation~\eqref{Dirac-omega_n} is translated into
\begin{equation}
\langle\, \widehat{\Psi}^1 (x_1) \, \rangle = 0
\end{equation}
and
\begin{equation}
\begin{split}
& \langle\, \widehat{\Psi}^1 (x_1) \, \widehat{\Psi}^2 (x_2) \, \ldots \,
\widehat{\Psi}^n (x_n) \, \rangle \\
& {}-\frac{\hbar}{i} \, \sum_{i=1}^{n-1} \widehat{S}^{\, in} (x_i-x_n) \,
\langle\, \widehat{\Psi}^1 (x_1) \, \ldots \,
\widehat{\Psi}^{i-1} (x_{i-1}) \,
\widehat{\Psi}^{i+1} (x_{i+1}) \, \ldots \,
\widehat{\Psi}^{n-1} (x_{n-1}) \, \rangle = 0
\end{split}
\label{Dirac-recursion-relation}
\end{equation}
for $n > 1$.

To derive the Schwinger-Dyson equations, we act
\begin{equation}
\epsilon^n_\alpha \, ( \, i \, \partial\!\!\!/_{x_n} +m \, )_{\beta \alpha} \,
\frac{\partial}{\partial \epsilon^n_\beta}
+\overline{\epsilon}^{\, n}_\alpha \, ( \, -i \, \partial\!\!\!/_{x_n} +m \, )_{\alpha \beta} \,
\frac{\partial}{\partial \overline{\epsilon}^{\, n}_\beta}
\label{Q-hat}
\end{equation}
on Eq.~\eqref{Dirac-recursion-relation}.
The action of this operator on $\widehat{S}^{\, in} (x_i-x_n)$ is given by
\begin{equation}
\biggl( \, \epsilon^n_\alpha \, ( \, i \, \partial\!\!\!/_{x_n} +m \, )_{\beta \alpha} \,
\frac{\partial}{\partial \epsilon^n_\beta}
+\overline{\epsilon}^{\, n}_\alpha \, ( \, -i \, \partial\!\!\!/_{x_n} +m \, )_{\alpha \beta} \,
\frac{\partial}{\partial \overline{\epsilon}^{\, n}_\beta} \, \biggr) \,
\widehat{S}^{\, in} (x_i-x_n)
= \widehat{\delta}^{\, in} (x_i-x_n) \,,
\end{equation}
and the action of this operator on $\widehat{\Psi}^n (x_n)$ is
\begin{equation}
\begin{split}
& \biggl( \, \epsilon^n_\alpha \, ( \, i \, \partial\!\!\!/_{x_n} +m \, )_{\beta \alpha} \,
\frac{\partial}{\partial \epsilon^n_\beta}
+\overline{\epsilon}^{\, n}_\alpha \, ( \, -i \, \partial\!\!\!/_{x_n} +m \, )_{\alpha \beta} \,
\frac{\partial}{\partial \overline{\epsilon}^{\, n}_\beta} \, \biggr) \,
\widehat{\Psi}^n (x_n) \\
& = \overline{\Psi}_\beta (x_n) \,
( \, i \overleftarrow{\partial\!\!\!/}\!_{x_n} +m \, )_{\beta \alpha} \, \epsilon^n_\alpha
+\overline{\epsilon}^{\, n}_\alpha
( \, -i \, \partial\!\!\!/_{x_n} +m \, )_{\alpha \beta} \, \Psi_\beta (x_n) \,.
\end{split}
\end{equation}
Since
\begin{equation}
\frac{\delta S}{\delta \widehat{\Psi}^n (x_n)}
= {}-\overline{\Psi}_\beta (x_n) \,
( \, i \overleftarrow{\partial\!\!\!/}\!_{x_n} +m \, )_{\beta \alpha} \, \epsilon^n_\alpha
+\overline{\epsilon}^{\, n}_\alpha
( \, i \, \partial\!\!\!/_{x_n} -m \, )_{\alpha \beta} \, \Psi_\beta (x_n)
\end{equation}
for the action~\eqref{Dirac-action} of the free theory, we find
\begin{equation}
\biggl( \, \epsilon^n_\alpha \, ( \, i \, \partial\!\!\!/_{x_n} +m \, )_{\beta \alpha} \,
\frac{\partial}{\partial \epsilon^n_\beta}
+\overline{\epsilon}^{\, n}_\alpha \, ( \, -i \, \partial\!\!\!/_{x_n} +m \, )_{\alpha \beta} \,
\frac{\partial}{\partial \overline{\epsilon}^{\, n}_\beta} \, \biggr) \,
\widehat{\Psi}^n (x_n)
= {}-\frac{\delta S}{\delta \widehat{\Psi}^n (x_n)} \,.
\end{equation}
Therefore, the operator~\eqref{Q-hat} acting on Eq.~\eqref{Dirac-recursion-relation} gives
\begin{equation}
\begin{split}
& {}-\langle\, \widehat{\Psi}^1 (x_1) \, \widehat{\Psi}^2 (x_2) \, \ldots \,
\widehat{\Psi}^{n-1} (x_{n-1}) \, \frac{\delta S}{\delta \widehat{\Psi}^n (x_n)} \, \rangle \\
& {}-\frac{\hbar}{i} \, \sum_{i=1}^{n-1} \widehat{\delta}^{\, in} (x_i-x_n) \,
\langle\, \widehat{\Psi}^1 (x_1) \, \ldots \,
\widehat{\Psi}^{i-1} (x_{i-1}) \,
\widehat{\Psi}^{i+1} (x_{i+1}) \, \ldots \,
\widehat{\Psi}^{n-1} (x_{n-1}) \, \rangle = 0 \,.
\end{split}
\end{equation}
We have thus reproduced the Schwinger-Dyson equations
organized in the form of Eq.~\eqref{Dirac-Schwinger-Dyson}.

\section{Conclusions and discussion}
\label{conclusion-section}

We extended the formula for correlation functions of scalar field theories
in terms of quantum $A_\infty$ algebras, presented in Ref.~\cite{Okawa:2022sjf},
to incorporate Dirac fields.
We used a description that is analogous to string field theory,
and a primary advantage of our approach is that
the formula for correlation functions takes the same form
for both scalar fields and Dirac fields. 
It is given by
\begin{equation}
\langle \, \Phi^{\otimes n} \, \rangle = \pi_n \, {\bm f} \, {\bf 1} \,,
\label{formula-conclusions}
\end{equation}
where
\begin{equation}
{\bm f} = \frac{1}{{\bf I} +{\bm h} \, {\bm m} +i \hbar \, {\bm h} \, {\bf U}} \,.
\end{equation}
For a scalar field $\varphi (x)$, $\Phi$ is given by
\begin{equation}
\Phi = \int d^d x \, \varphi (x) \, c(x) \,.
\end{equation}
For a Dirac field $\Psi (x)$, $\Phi$ is given by
\begin{equation}
\Phi = \int d^d x \, ( \, \overline{\theta}_\alpha (x) \, \Psi_\alpha (x)
+\overline{\Psi}_\alpha (x) \, \theta_\alpha (x) \, ) \,.
\end{equation}
We speculate that this formula for correlation functions
can be extended to gauge theories including string field theories
and it takes the same form in terms of $\Phi$
with different choices for the vector space $\mathcal{H}$.
We proved that correlation functions from our formula satisfy
the Schwinger-Dyson equations in the free theory.
The proof for interacting theories is presented in Ref.~\cite{Konosu:2023rkm}.

Let us discuss the relation of our formula to previous approaches
based on the Batalin-Vilkovisky formalism or its dual description
in terms of homotopy algebras.
As we mentioned in the introduction,
the formula for correlation functions in Ref.~\cite{Okawa:2022sjf} was found from the fact
that Feynman diagrams can be generated algebraically
in the framework of homotopy algebras~\cite{Kajiura:2003ax, Doubek:2017naz, Masuda:2020tfa}.
When we generate Feynman diagrams,
we have to choose a projection operator $P$,
which preserves the cohomology of $Q$.
On the other hand, it was recognized in Ref.~\cite{Okawa:2022sjf}
that the calculation of correlation functions
corresponds to the projection where $P=0$.\footnote{
This observation was also added to Ref.~\cite{Masuda:2020tfa}
when it was revised to the final version.
}
Apparently, the results in Refs.~\cite{Kajiura:2003ax, Doubek:2017naz, Masuda:2020tfa}
cannot be used to calculate
correlation functions of Lorentzian theories
where the cohomology of $Q$ is nontrivial.
The resolution in Ref.~\cite{Okawa:2022sjf} is related
to how we define the vacuum state in the path integral.
In the case of, e.g.~scalar field theories,
we choose the asymptotic behavior of the scalar field $\varphi$
such that the imaginary part of $\varphi$ is infinitesimally positive
in the limit ${\rm Re} \, \varphi \to -\infty$
and infinitesimally negative in the limit ${\rm Re} \, \varphi \to \infty$.
This corresponds to modifying $Q$ to $Q -i \epsilon$,
where the constant $\epsilon$ is infinitesimally positive. 
The cohomology of $Q -i \epsilon$ is then trivial,
and the operator $h$ associated with the projection where $P=0$
can be constructed using the Feynman propagator.
This reproduces the perturbation theory in the path integral
where the Feynman propagator is used in loop diagrams.

The discussion in Refs.~\cite{Doubek:2017naz, Masuda:2020tfa} was based
on Wick's theorem in the free theory,
and then they used the homological perturbation lemma
for generalization to interacting theories.
While we expect that the homological perturbation works formally,
we know in quantum field theory that
we need to renormalize the theory
by introducing counterterms
once we incorporate interactions.
Counterterms had not been considered in previous approaches based on homotopy algebras,
and the analysis including counterterms was first carried out
for scalar field theories in Ref.~\cite{Okawa:2022sjf}.
The analysis in Ref.~\cite{Okawa:2022sjf} revealed
that the structure corresponding to the one-particle irreducible effective action
plays an important role in renormalizing the theory.
Since the same set of Feynman diagrams are generated,
this is not surprising from the perspective of the path integral,
but this requires further studies to describe renormalization in the perspective of homotopy algebras.

In view of this situation for the analysis of interacting theories,
we consider that we need a basis that replaces the homological perturbation lemma.
In Ref.~\cite{Okawa:2022sjf}, the Schwinger-Dyson equations were taken as the basis,
and it was shown that correlation functions from the formula
based on quantum $A_\infty$ algebras satisfy the Schwinger-Dyson equations.
In this respect, the approach by Costello and Gwilliam~\cite{Costello:2016vjw, Costello:2021jvx}
based on factorization algebras is closely related to ours.
In Appendix~\ref{factorization-algebra-appendix}, we briefly explain the definition of correlation functions
in the approach by Costello and Gwilliam
and we show that their definition of correlation functions
coincides with the one in the formula based on quantum $A_\infty$ algebras
for Euclidean scalar field theories.
We demonstrate the equivalence for free theories,
but we can generalize the discussion to interacting theories.
While it would be important but challenging
to extend the whole framework of factorization algebras to Lorentzian theories,
the definition of correlation functions can be extended to Lorentzian theories,
and it is easy to see that the definition coincides with the one in the formula based on quantum $A_\infty$ algebras.

One advantage of the formula based on quantum $A_\infty$ algebras
is that it gives an {\it explicit} solution to the cohomology problem
in the approach by Costello and Gwilliam.
We emphasize that this is the case even for interacting theories.
While it would be difficult to consider the cohomology problem nonperturbatively,
the formula for correlation functions based on quantum $A_\infty$ algebras
can be valid nonperturbatively.
In fact, numerical evidence that nonperturbative correlation functions
can be obtained from the formula based on quantum $A_\infty$ algebras
for scalar field theories in zero dimensions
is presented in Ref.~\cite{Konosu:2024zrq}.
It is expected that the nonperturbative definition of correlation functions
involves the concept of Lefschetz thimbles,
and the analysis in Ref.~\cite{Konosu:2024zrq} indeed demonstrated
that the formula based on quantum $A_\infty$ algebras in general
gives correlation functions on a Lefschetz thimble.
It is difficult to define the path integral measure
in a mathematically rigorous way,
and the approach by Costello and Gwilliam based on factorization algebras
successfully circumvents the construction of the path integral measure
in defining perturbative quantum field theory.
Towards the nonperturbative definition of quantum field theory
without using the path integral measure,
we hope that the formula for correlation functions
based on quantum $A_\infty$ algebras will play
a role that complements the approach by Costello and Gwilliam based on factorization algebras.

While we chose $A_\infty$ algebras to describe correlation functions,
we can also use $L_\infty$ algebras
since Feynman diagrams can be algebraically generated
both by $A_\infty$ algebras and by $L_\infty$ algebras.
In fact,
$L_\infty$ algebras which involve symmetrization procedures
are more naturally related to the Batalin-Vilkovisky formalism,
and it is rather surprising that correlation functions
which are graded symmetric under the exchange of fields
are obtained in terms of $A_\infty$ algebras
without any symmetrization procedures.
This aspect was discussed in section~7 of Ref.~\cite{Okawa:2022sjf},
and we also added a discussion in the context of the approach by Costello and Gwilliam
in Appendix~\ref{factorization-algebra-appendix}.
This feature is important for us
because our primary motivation
is to apply our approach to open superstring field theory.
In the program of providing a framework to prove the AdS/CFT correspondence,
we are interested in the $1/N$ expansion
of correlation functions of gauge-invariant operators
in open superstring field theory~\cite{Okawa:2020llq},\footnote{
See recent papers~\cite{Maccaferri:2023gcg, Maccaferri:2023gof}
for interesting development which will be relevant to this program.
It would also be interesting to consider
the twisted holography~\cite{Costello:2016mgj, Costello:2017fbo, Costello:2018zrm} from our perspective,
and see Ref.~\cite{Zeng:2023qqp} for recent research.
}
and thus we prefer $A_\infty$ algebras to $L_\infty$ algebras.
It is encouraging that
our formula~\eqref{formula-conclusions} for correlation functions
is almost string-field-theory-like,
and we hope that the results of this paper will accelerate
research in this direction.

\bigskip
\noindent
{\normalfont \bfseries \large Funding}

\medskip
This work of Y.O.~was supported in part by Japan Society for the Promotion of Science (JSPS)
KAKENHI Grant Number JP17K05408.

\appendix

\renewcommand{\theequation}{\thesection.\arabic{equation}}

\section{Consistency of the definition of the symplectic form}
\label{omega-appendix}
\setcounter{equation}{0}

Suppose that the symplectic form $\omega$ satisfies
\begin{equation}
\omega \, ( \, e_1 \,,\, e_2 \, )
= {}-(-1)^{\mathrm{deg} (e_1) \, \mathrm{deg} (e_2)} \,
\omega \, ( \, e_2 \,,\, e_1 \, )
\end{equation}
for any pair $e_1$ and $e_2$ in the basis of $\mathcal{H}$.
The symplectic form $\omega \, ( \, e_1 \,,\, e_2 \, )$ can be a nonvanishing complex number
only when the degree of $e_1$ and the degree of $e_2$ are different.
Therefore, the symplectic form $\omega$ is considered to be degree odd,
and we should consider that a degree-odd object is inserted
somewhere in the symplectic form.
When $\Phi_1$ and $\Phi_2$ are given by
\begin{equation}
\Phi_1 = f_1 \, e_1 \,, \qquad \Phi_2 = f_2 \, e_2 \,,
\end{equation}
the degree-odd object may give signs
in the process of taking out $f_1$ and $f_2$ from
$\omega \, ( \, f_1 \, e_1 \,,\, f_2 \, e_2 \, )$
to obtain, e.g.~$f_1 \, f_2 \, \omega \, ( \, e_1 \,,\, e_2 \, )$.
Let us therefore consider whether the relation
\begin{equation}
\omega \, ( \, \Phi_1 \,,\, \Phi_2 \, )
= {}-(-1)^{\mathrm{deg} (\Phi_1) \, \mathrm{deg} (\Phi_2)} \,
\omega \, ( \, \Phi_2 \,,\, \Phi_1 \, )
\label{cyclicity}
\end{equation}
holds for the following three cases.\\

\noindent
(i) A degree-odd object is inserted to the left of $\Phi_1$.\\

In this case $\omega \, ( \, \Phi_1 \,,\, \Phi_2 \, )$ is given by
\begin{equation}
\omega \, ( \, \Phi_1 \,,\, \Phi_2 \, )
= \omega \, ( \, f_1 \, e_1 \,,\, f_2 \, e_2 \, )
= (-1)^{\mathrm{deg} (f_1)+\mathrm{deg} (f_2)+\mathrm{deg} (e_1) \, \mathrm{deg} (f_2)} \,
f_1 \, f_2 \, \omega \, ( \, e_1 \,,\, e_2 \, ) \,.
\end{equation}
On the other hand, the right-hand side of Eq.~\eqref{cyclicity} is given by
\begin{equation}
\begin{split}
& {}-(-1)^{\mathrm{deg} (\Phi_1) \, \mathrm{deg} (\Phi_2)} \,
\omega \, ( \, \Phi_2 \,,\, \Phi_1 \, )
= (-1)^{1+( \mathrm{deg} (f_1)+\mathrm{deg} (e_1) ) \, ( \mathrm{deg} (f_2)+\mathrm{deg} (e_2) )} \,
\omega \, ( \, f_2 \, e_2 \,,\, f_1 \, e_1 \, ) \\
& = (-1)^{1+( \mathrm{deg} (f_1)+\mathrm{deg} (e_1) ) \, ( \mathrm{deg} (f_2)+\mathrm{deg} (e_2) )
+\mathrm{deg} (f_2)+\mathrm{deg} (f_1)+\mathrm{deg} (e_2) \, \mathrm{deg} (f_1)} \,
f_2 \, f_1 \, \omega \, ( \, e_2 \,,\, e_1 \, ) \\
& = (-1)^{\mathrm{deg} (f_1)+\mathrm{deg} (f_2)+\mathrm{deg} (e_1) \, \mathrm{deg} (f_2)} \,
f_1 \, f_2 \, \omega \, ( \, e_1 \,,\, e_2 \, ) \,.
\end{split}
\end{equation}
We have therefore confirmed that the relation
\begin{equation}
\omega \, ( \, \Phi_1 \,,\, \Phi_2 \, )
= {}-(-1)^{\mathrm{deg} (\Phi_1) \, \mathrm{deg} (\Phi_2)} \,
\omega \, ( \, \Phi_2 \,,\, \Phi_1 \, )
\end{equation}
holds.\\

\noindent
(ii) A degree-odd object is inserted between $\Phi_1$ and $\Phi_2$.\\

In this case $\omega \, ( \, \Phi_1 \,,\, \Phi_2 \, )$ is given by
\begin{equation}
\omega \, ( \, \Phi_1 \,,\, \Phi_2 \, )
= \omega \, ( \, f_1 \, e_1 \,,\, f_2 \, e_2 \, )
= (-1)^{\mathrm{deg} (f_2)+\mathrm{deg} (e_1) \, \mathrm{deg} (f_2)} \,
f_1 \, f_2 \, \omega \, ( \, e_1 \,,\, e_2 \, ) \,.
\label{middle-1}
\end{equation}
On the other hand, the right-hand side of Eq.~\eqref{cyclicity} is given by
\begin{equation}
\begin{split}
& {}-(-1)^{\mathrm{deg} (\Phi_1) \, \mathrm{deg} (\Phi_2)} \,
\omega \, ( \, \Phi_2 \,,\, \Phi_1 \, )
= (-1)^{1+( \mathrm{deg} (f_1)+\mathrm{deg} (e_1) ) \, ( \mathrm{deg} (f_2)+\mathrm{deg} (e_2) )} \,
\omega \, ( \, f_2 \, e_2 \,,\, f_1 \, e_1 \, ) \\
& = (-1)^{1+( \mathrm{deg} (f_1)+\mathrm{deg} (e_1) ) \, ( \mathrm{deg} (f_2)+\mathrm{deg} (e_2) )
+\mathrm{deg} (f_1)+\mathrm{deg} (e_2) \, \mathrm{deg} (f_1)} \,
f_2 \, f_1 \, \omega \, ( \, e_2 \,,\, e_1 \, ) \\
& = (-1)^{\mathrm{deg} (f_1)+\mathrm{deg} (e_1) \, \mathrm{deg} (f_2)} \,
f_1 \, f_2 \, \omega \, ( \, e_1 \,,\, e_2 \, ) \,.
\end{split}
\label{middle-2}
\end{equation}
We compare Eqs.~\eqref{middle-1} and~\eqref{middle-2} to find that
\begin{equation}
\omega \, ( \, \Phi_1 \,,\, \Phi_2 \, )
\ne {}-(-1)^{\mathrm{deg} (\Phi_1) \, \mathrm{deg} (\Phi_2)} \,
\omega \, ( \, \Phi_2 \,,\, \Phi_1 \, )
\end{equation}
when $\mathrm{deg} (f_1) \ne \mathrm{deg} (f_2)$.\\

\noindent
(iii) A degree-odd object is inserted to the right of $\Phi_2$.\\

In this case $\omega \, ( \, \Phi_1 \,,\, \Phi_2 \, )$ is given by
\begin{equation}
\omega \, ( \, \Phi_1 \,,\, \Phi_2 \, )
= \omega \, ( \, f_1 \, e_1 \,,\, f_2 \, e_2 \, )
= (-1)^{\mathrm{deg} (e_1) \, \mathrm{deg} (f_2)} \,
f_1 \, f_2 \, \omega \, ( \, e_1 \,,\, e_2 \, ) \,.
\end{equation}
On the other hand, the right-hand side of Eq.~\eqref{cyclicity} is given by
\begin{equation}
\begin{split}
& {}-(-1)^{\mathrm{deg} (\Phi_1) \, \mathrm{deg} (\Phi_2)} \,
\omega \, ( \, \Phi_2 \,,\, \Phi_1 \, )
= (-1)^{1+( \mathrm{deg} (f_1)+\mathrm{deg} (e_1) ) \, ( \mathrm{deg} (f_2)+\mathrm{deg} (e_2) )} \,
\omega \, ( \, f_2 \, e_2 \,,\, f_1 \, e_1 \, ) \\
& = (-1)^{1+( \mathrm{deg} (f_1)+\mathrm{deg} (e_1) ) \, ( \mathrm{deg} (f_2)+\mathrm{deg} (e_2) )
+\mathrm{deg} (e_2) \, \mathrm{deg} (f_1)} \,
f_2 \, f_1 \, \omega \, ( \, e_2 \,,\, e_1 \, ) \\
& = (-1)^{\mathrm{deg} (e_1) \, \mathrm{deg} (f_2)} \,
f_1 \, f_2 \, \omega \, ( \, e_1 \,,\, e_2 \, ) \,.
\end{split}
\end{equation}
We have therefore confirmed that the relation
\begin{equation}
\omega \, ( \, \Phi_1 \,,\, \Phi_2 \, )
= {}-(-1)^{\mathrm{deg} (\Phi_1) \, \mathrm{deg} (\Phi_2)} \,
\omega \, ( \, \Phi_2 \,,\, \Phi_1 \, )
\end{equation}
holds.\\

We have found that case (i) and case (iii) are consistent.
In this paper, we choose the convention
\begin{equation}
\omega \, ( \, f_1 \, e_1 \,,\, f_2 \, e_2 \, )
= (-1)^{\mathrm{deg} (f_1)+\mathrm{deg} (f_2)+\mathrm{deg} (e_1) \, \mathrm{deg} (f_2)} \,
f_1 \, f_2 \, \omega \, ( \, e_1 \,,\, e_2 \, ) \,.
\end{equation}
We may also want to transform $\omega \, ( \, f_1 \, e_1 \,,\, f_2 \, e_2 \, )$
to $f_1 \, \omega \, ( \, e_1 \,,\, e_2 \, ) \,f_2$
or to $\omega \, ( \, e_1 \,,\, e_2 \, ) \, f_1 \, f_2$.
In those cases we find
\begin{equation}
\begin{split}
\omega \, ( \, f_1 \, e_1 \,,\, f_2 \, e_2 \, )
& = (-1)^{\mathrm{deg} (f_1)+\mathrm{deg} (e_2) \, \mathrm{deg} (f_2)} \,
f_1 \, \omega \, ( \, e_1 \,,\, e_2 \, ) \, f_2 \,, \\
\omega \, ( \, f_1 \, e_1 \,,\, f_2 \, e_2 \, )
& = (-1)^{\mathrm{deg} (e_1) \, \mathrm{deg} (f_1)+\mathrm{deg} (e_2) \, \mathrm{deg} (f_1)
+\mathrm{deg} (e_2) \, \mathrm{deg} (f_2)} \,
\omega \, ( \, e_1 \,,\, e_2 \, ) \, f_1 \, f_2 \,.
\end{split}
\end{equation}
They are related to $f_1 \, f_2 \, \omega \, ( \, e_1 \,,\, e_2 \, )$ as
\begin{equation}
\begin{split}
f_1 \, \omega \, ( \, e_1 \,,\, e_2 \, ) \, f_2
& = (-1)^{( \mathrm{deg} (e_1)+\mathrm{deg} (e_2)+1 ) \, \mathrm{deg} (f_2)}
f_1 \, f_2 \, \omega \, ( \, e_1 \,,\, e_2 \, ) \,, \\
\omega \, ( \, e_1 \,,\, e_2 \, ) \, f_1 \, f_2
& = (-1)^{( \mathrm{deg} (e_1)+\mathrm{deg} (e_2)+1 ) \, ( \mathrm{deg} (f_1)+\mathrm{deg} (f_2) )}
f_1 \, f_2 \, \omega \, ( \, e_1 \,,\, e_2 \, ) \,,
\end{split}
\end{equation}
but the symplectic form $\omega \, ( \, e_1 \,,\, e_2 \, )$ can be nonvanishing
only when $\mathrm{deg} (e_1)+\mathrm{deg} (e_2)+1 = 0 \mod 2$,
so $f_1 \, f_2 \, \omega \, ( \, e_1 \,,\, e_2 \, )$,
$f_1 \, \omega \, ( \, e_1 \,,\, e_2 \, ) \, f_2$,
and $f_1 \, f_2 \, \omega \, ( \, e_1 \,,\, e_2 \, )$ all coincide.

\section{Correlation functions in factorization algebras}
\label{factorization-algebra-appendix}
\setcounter{equation}{0}

It is difficult to define the path integral measure in a mathematically rigorous way.
Costello and Gwilliam have developed a formalism that circumvents the construction of the path integral measure
in defining perturbative quantum field theory
based on factorization algebras~\cite{Costello:2016vjw, Costello:2021jvx}.
We briefly explain the definition of correlation functions
in the approach by Costello and Gwilliam in Appendix~\ref{the-divergence-theorem-appendix},
and we demonstrate the equivalence of their definition
with the one in the formula for correlation functions based on quantum $A_\infty$ algebras
in Appendix~\ref{equivalence-appendix}.

\subsection{The Schwinger-Dyson equations as the divergence theorem}
\label{the-divergence-theorem-appendix}

Consider a scalar field $f (\sigma)$ in a 3D space
of Euclidean signature with coordinates $(\sigma^1, \sigma^2, \sigma^3 \,)$
and metric $g_{ij} (\sigma)$.
We denote the determinant of the metric by $g (\sigma)$.
We require the scalar field to decay sufficiently fast
as $(\sigma^1)^2+(\sigma^2)^2+(\sigma^3)^2 \to \infty$.
Then the divergence theorem states
that the integral of the scalar field given by
\begin{equation}
\int d^3 \sigma \, f (\sigma) \, \sqrt{g (\sigma)}
\end{equation}
vanishes when $f (\sigma)$ is the divergence of a vector field $v (\sigma)$,
\begin{equation}
f (\sigma) = \nabla_i \, v^i (\sigma)
\end{equation}
with $v^i (\sigma) \to 0$ as $(\sigma^1)^2+(\sigma^2)^2+(\sigma^3)^2 \to \infty$,
where the vector field $v (\sigma)$ is given by
\begin{equation}
v (\sigma) = v^i (\sigma) \, \partial_i
\end{equation}
and the divergence of $v (\sigma)$ is
\begin{equation}
\nabla_i \, v^i (\sigma)
= \partial_i \, v^i (\sigma) +\Gamma^{\, i}_{\,\,\, ij} (\sigma ) \, v^{\, j} (\sigma)
= \partial_i \, v^i (\sigma) +v^i (\sigma) \, \partial_i \ln \sqrt{g (\sigma)} \,.
\end{equation}
This can be seen as follows:
\begin{equation}
\int d^3 \sigma \, f (\sigma) \, \sqrt{g (\sigma)}
= \int d^3 \sigma \, \nabla_i \, v^i (\sigma) \, \sqrt{g (\sigma)}
= \int d^3 \sigma \, \partial_i \, \bigl( \, v^i (\sigma) \, \sqrt{g (\sigma)} \,\, \bigr) = 0 \,.
\end{equation}

Let us generalize the integral in three dimensions to the path integral of Euclidean scalar field theories.
We restrict our attention to polynomial scalar fields and polynomial vector fields.
A polynomial scalar field homogeneous of degree $n$ takes the form
\begin{align}
f (\sigma)
& = ( \, f_{11} \, \sigma^1 +f_{12} \, \sigma^2 +f_{13} \, \sigma^3 \, ) \,
( \, f_{21} \, \sigma^1 +f_{22} \, \sigma^2 +f_{23} \, \sigma^3 \, ) \, \ldots
( \, f_{n1} \, \sigma^1 +f_{n2} \, \sigma^2 +f_{n3} \, \sigma^3 \, )
\nonumber \\
& = \prod_{k=1}^n \, \Bigl( \, \sum_{i_k=1}^3 \, f_{k \, i_k} \, \sigma^{i_k} \, \Bigr) \,,
\label{polynomial-scalar-field}
\end{align}
where $f_{k \, i_k}$ are constants.
Here we explicitly wrote the summation over $i_k$
to prepare for the generalization to the path integral.
A polynomial vector field homogeneous of degree $n-1$ takes the form
\begin{align}
v (\sigma)
& = ( \, h_{11} \, \sigma^1 +h_{12} \, \sigma^2 +h_{13} \, \sigma^3 \, ) \, \ldots
( \, h_{n-1 \, 1} \, \sigma^1 +h_{n-1 \, 2} \, \sigma^2 +h_{n-1 \, 3} \, \sigma^3 \, ) \,
( \, \phi^1 \partial_1 +\phi^2 \partial_2 +\phi^3 \partial_3 \, )
\nonumber \\
& = \Biggl[ \, \prod_{k=1}^{n-1} \, \Bigl( \, \sum_{i_k=1}^3 \, h_{k \, i_k} \, \sigma^{i_k} \, \Bigr) \, \Biggr] \,
\Bigl( \, \sum_{i=1}^3 \, \phi^i \partial_i \, \Bigr) \,,
\label{polynomial-vector-field}
\end{align}
where $h_{k \, i_k}$ and $\phi^i$ are constants.
The divergence of this polynomial vector field is given by
\begin{equation}
\begin{split}
\sum_{i=1}^{n-1} \, \Biggl[ \, \prod_{k \ne i} \, \Bigl( \, \sum_{i_k=1}^3 \, h_{k \, i_k} \, \sigma^{i_k} \, \Bigr) \, \Biggr] \,
\Bigl( \, \sum_{j=1}^3 h_{ij} \, \phi^{\, j} \, \Bigr)
+\Biggl[ \, \prod_{k=1}^{n-1} \, \Bigl( \, \sum_{i_k=1}^3 \, h_{k \, i_k} \, \sigma^{i_k} \, \Bigr) \, \Biggr]
\Bigl( \, \sum_{i=1}^3 \, \phi^i \partial_i \ln \sqrt{g (\sigma)} \, \, \Bigr) \,.
\end{split}
\end{equation}

For the path integral of a free scalar field $\varphi (x)$ in $d$ dimensions,
we generalize these by 
\begin{equation}
\sigma^i \to \varphi (x) \,, \qquad
\sum_{i=1}^3 \to \int d^d x \,.
\end{equation}
The polynomial scalar field in Eq.~\eqref{polynomial-scalar-field}
and the polynomial vector field in Eq.~\eqref{polynomial-vector-field}
then generalize to
\begin{equation}
\prod_{k=1}^n \, \biggl( \, \int d^d x_k \, f_k (x_k) \, \varphi (x_k) \, \biggr)
\end{equation}
and
\begin{equation}
\Biggl[ \, \prod_{k=1}^{n-1} \, \biggl( \, \int d^d x_k \, h_k (x_k) \, \varphi (x_k) \, \biggr) \, \Biggr] \,
\biggl( \, \int d^d x_n \, \phi (x_n) \, \frac{\delta}{\delta \varphi (x_n)} \, \biggr) \,,
\end{equation}
respectively.\footnote{
The scalar field $\varphi (x)$ plays the role of coordinates
in the polynomial scalar field and the polynomial vector field.
The two different usages of {\it scalar field} should be distinguished.
}
Consider the path integral of the polynomial scalar field given by
\begin{equation}
\int \mathcal{D} \varphi \, \prod_{k=1}^n \, \biggl( \, \int d^d x_k \, f_k (x_k) \, \varphi (x_k) \, \biggr) \,
e^{-\frac{S}{\hbar}} \,,
\end{equation}
where the action $S$ in the Euclidean case is given by
\begin{equation}
S = \frac{1}{2} \int d^d x \, \bigl[ \, \partial_\mu \varphi (x) \, \partial_\mu \varphi (x)
+m^2 \, \varphi (x)^2 \, \bigr] \,.
\end{equation}
Since $\ln \sqrt{g (\sigma)}$ corresponds to $-S/\hbar \,$,
the divergence of the polynomial vector field should be
\begin{equation}
\begin{split}
& \sum_{i=1}^{n-1} \, \Biggl[ \, \prod_{k \ne i} \,
\biggl( \, \int d^d x_k \, h_k (x_k) \, \varphi (x_k) \, \biggr) \, \Biggr] \,
\biggl( \, \int d^d x_n \, h_i (x_n) \, \phi (x_n) \, \biggr) \\
& {}-\frac{1}{\hbar} \, \Biggl[ \, \prod_{k=1}^{n-1} \,
\biggl( \, \int d^d x_k \, h_k (x_k) \, \varphi (x_k) \, \biggr) \, \Biggr] \,
\biggl( \, \int d^d x_n \, \phi (x_n) \, \frac{\delta S}{\delta \varphi (x_n)} \, \biggr) \,.
\end{split}
\label{vanishing-divergence}
\end{equation}
The path integral of the divergence of a vector field vanishes.
This corresponds to the Schwinger-Dyson equations:
\begin{equation}
\begin{split}
\sum_{i=1}^{n-1} \,
\langle \, \varphi (x_1) \ldots \varphi (x_{i-1}) \, \varphi (x_{i+1}) \ldots \varphi (x_{n-1}) \, \rangle \,
\delta^d (x_i-x_n) \\
{}-\frac{1}{\hbar} \, \langle \, \varphi (x_1) \, \varphi (x_2) \ldots \varphi (x_{n-1}) \,
\frac{\delta S}{\delta \varphi (x_n)} \rangle = 0 \,.
\end{split}
\label{Schwinger-Dyson-appendix}
\end{equation}
We multiply the left-hand side of Eq.~\eqref{Schwinger-Dyson-appendix}
without brackets
by $h_1 (x_1) \, \ldots \, h_{n-1} (x_{n-1}) \, \phi (x_n)$
and integrate the resulting expression over $x_1$, $x_2$, \ldots \,, $x_{n-1}$, and $x_n$
to obtain Eq.~\eqref{vanishing-divergence}.

Motivated by this, Costello and Gwilliam defined the space of observables
associated with an open set $U$ of the spacetime manifold
in Ref.~\cite{Costello:2016vjw} as the quotient space
derived from the set of polynomial scalar fields
\begin{equation}
\prod_{k=1}^n \, \biggl( \, \int d^d x_k \, f_k (x_k) \, \varphi (x_k) \, \biggr)
\end{equation}
for functions $f_1$, $f_2$, \ldots \,, $f_{n-1}$, and $f_n$ which have compact support in $U$
with the equivalence relation
\begin{align}
& \prod_{k=1}^n \, \biggl( \, \int d^d x_k \, f_k (x_k) \, \varphi (x_k) \, \biggr) \nonumber \\
& \sim \prod_{k=1}^n \, \biggl( \, \int d^d x_k \, f_k (x_k) \, \varphi (x_k) \, \biggr)
+\hbar \sum_{i=1}^{n-1} \, \Biggl[ \, \prod_{k \ne i} \,
\biggl( \, \int d^d x_k \, h_k (x_k) \, \varphi (x_k) \, \biggr) \, \Biggr] \,
\biggl( \, \int d^d x_n \, h_i (x_n) \, \phi (x_n) \, \biggr) \nonumber \\
& \quad~ {}-\Biggl[ \, \prod_{k=1}^{n-1} \,
\biggl( \, \int d^d x_k \, h_k (x_k) \, \varphi (x_k) \, \biggr) \, \Biggr] \,
\biggl( \, \int d^d x_n \, \phi (x_n) \, \frac{\delta S}{\delta \varphi (x_n)} \, \biggr)
\label{equivalence-relation}
\end{align}
for any functions $h_1$, $h_2$, \ldots \,, $h_{n-1}$, and $\phi$
which have compact support in $U$.
When $U$ is the whole space $\mathbb{R}^d$,
we require $\varphi (x)$ to decay sufficiently fast as $| x | \to \infty$
and then the quotient space is isomorphic to $\mathbb{R}$,
which is identified with the correlation function.
For example, the quadratic observable
\begin{equation}
\int d^d x_1 \, f_1 (x_1) \, \varphi (x_1)
\int d^d x_2 \, f_2 (x_2) \, \varphi (x_2)
\end{equation}
is equivalent to $1$ multiplied by
\begin{equation}
\hbar \int d^d x_1 \, d^d x_2 \, f_1 (x_1) \, \Delta (x_1-x_2) \, f_2 (x_2) \,,
\end{equation}
where $\Delta (x-y)$ in the Euclidean case given by
\begin{equation}
\Delta (x-y)
= \int \frac{d^d k}{(2 \pi)^d} \,
\frac{e^{ik \, (x-y)}}{k^2+m^2}
\end{equation}
satisfies
\begin{equation}
\int d^d x \, \frac{\delta S}{\delta \varphi (x)} \, \Delta (x-y) = \varphi (y) \,.
\end{equation}
This can be shown as follows:
\begin{equation}
\begin{split}
& \int d^d x_1 \, f_1 (x_1) \, \varphi (x_1)
\int d^d x_2 \, f_2 (x_2) \, \varphi (x_2) \\
& \sim \int d^d x_1 \, f_1 (x_1) \, \varphi (x_1)
\int d^d x_2 \, f_2 (x_2) \, \varphi (x_2) \\
& \quad~ +\hbar \int d^d x \, h (x) \, \phi (x)
-\int d^d x \, h (x) \, \varphi (x)
\int d^d y \, \phi (y) \, \frac{\delta S}{\delta \varphi (y)} \\
& = \hbar \int d^d x_1 \, d^d x_2 \, f_1 (x_1) \, \Delta (x_1-x_2) \, f_2 (x_2)
\end{split}
\end{equation}
with
\begin{equation}
h (x) = f_1 (x) \,, \qquad
\phi (x) = \int d^d y \, \Delta (x-y) \, f_2 (y) \,.
\end{equation}
We therefore have
\begin{equation}
\int d^d x_1 \, d^d x_2 \, f_1 (x_1)  \, f_2 (x_2) \,
\langle \, \varphi (x_1) \, \varphi (x_2) \, \rangle
= \hbar \int d^d x_1 \, d^d x_2 \, f_1 (x_1) \, \Delta (x_1-x_2) \, f_2 (x_2) \,.
\end{equation}

\subsection{The relation to the formula based on quantum $A_\infty$ algebras}
\label{equivalence-appendix}

The observable is specified by the set of functions
$f_1$, $f_2$, \ldots \,, $f_{n-1}$, and $f_n$.
We associate this observable
\begin{equation}
\prod_{k=1}^n \, \biggl( \, \int d^d x_k \, f_k (x_k) \, \varphi (x_k) \, \biggr)
\label{observables}
\end{equation}
with the element
\begin{equation}
\int d^d x_1 \, f_1 (x_1) \, d (x_1) 
\otimes \int d^d x_2 \, f_2 (x_2) \, d (x_2)
\otimes \ldots
\otimes \int d^d x_n \, f_n (x_n) \, d (x_n)
\end{equation}
in $\mathcal{H}_2^{\otimes n}$, where
\begin{equation}
\mathcal{H}_2^{\otimes n}
= \underbrace{\, \mathcal{H}_2 \otimes\mathcal{H}_2 \otimes \ldots \otimes \mathcal{H}_2 \,}_{n} \,.
\end{equation}
Then an element $\mathbb{E}$ in $T\mathcal{H}_1$ gives a linear map
from the set of observables to $\mathbb{C}$ via
\begin{equation}
\omega_n \, \biggl( \, \pi_n \, \mathbb{E} \,,
\int d^d x_1 \, f_1 (x_1) \, d (x_1) 
\otimes \int d^d x_2 \, f_2 (x_2) \, d (x_2)
\otimes \ldots
\otimes \int d^d x_n \, f_n (x_n) \, d (x_n) \, \biggr) \,.
\end{equation}
The factor
\begin{equation}
\int d^d x \, \phi (x) \, \frac{\delta S}{\delta \varphi (x)}
\end{equation}
in~\eqref{vanishing-divergence} is linear in $\varphi (x)$ for the free theory.
Its explicit form is given by
\begin{equation}
\int d^d x \, \phi (x) \, \frac{\delta S}{\delta \varphi (x)}
= \int d^d x \, \phi (x) \, ( {}-\partial^2 +m^2 \, ) \, \varphi (x)
= \int d^d x \, \Bigl[ \, ( {}-\partial^2 +m^2 \, ) \, \phi (x) \, \Bigr] \, \varphi (x) \,,
\end{equation}
and we associate this factor with
\begin{equation}
\int d^d x \, \Bigl[ \, ( {}-\partial^2 +m^2 \, ) \, \phi (x) \, \Bigr] \, d (x)
= \int d^d x \, \phi (x) \, ( {}-\partial^2 +m^2 \, ) \, d (x) \,.
\end{equation}
Since the action of $Q$ on $c(x)$ in the Euclidean case is given by
\begin{equation}
Q \, c(x) = ( {}-\partial^2 +m^2 \, ) \,  d(x) \,,
\end{equation}
we have
\begin{equation}
\int d^d x \, \phi (x) \, ( {}-\partial^2 +m^2 \, ) \, d (x)
= Q \int d^d x \, \phi (x) \, c(x) \,.
\end{equation}
We therefore associate the observable given by
\begin{equation}
\Biggl[ \, \prod_{k=1}^{n-1} \,
\biggl( \, \int d^d x_k \, h_k (x_k) \, \varphi (x_k) \, \biggr) \, \Biggr] \,
\biggl( \, \int d^d x_n \, \phi (x_n) \, \frac{\delta S}{\delta \varphi (x_n)} \, \biggr)
\end{equation}
with the element
\begin{equation}
\int d^d x_1 \, h_1 (x_1) \, d (x_1) 
\otimes \ldots
\otimes \int d^d x_{n-1} \, h_{n-1} (x_{n-1}) \, d (x_{n-1})
\otimes Q \int d^d x_n \, \phi (x_n) \, c (x_n)
\label{variation-observable}
\end{equation}
in $\mathcal{H}_2^{\otimes n}$.

A polynomial vector field is specified
by a set of functions $h_1$, $h_2$, \ldots \,, $h_{n-1}$, and $\phi$.
We associate this polynomial vector field with the element
\begin{equation}
\int d^d x_1 \, h_1 (x_1) \, d (x_1) 
\otimes \ldots
\otimes \int d^d x_{n-1} \, h_{n-1} (x_{n-1}) \, d (x_{n-1})
\otimes \int d^d x_n \, \phi (x_n) \, c (x_n)
\end{equation}
in $\mathcal{H}_2^{\otimes (n-1)} \otimes \mathcal{H}_1$.
Let us consider the quantity
\begin{equation}
\omega_n \, \biggl( \, \pi_n \, {\bf Q} \, \mathbb{E} \,,
\int d^d x_1 \, h_1 (x_1) \, d (x_1) 
\otimes \ldots
\otimes \int d^d x_{n-1} \, h_{n-1} (x_{n-1}) \, d (x_{n-1})
\otimes \int d^d x_n \, \phi (x_n) \, c (x_n) \, \biggr) \,.
\label{QE}
\end{equation}
When we expand $\pi_n \, \mathbb{E}$ using the basis vectors
$c (y_1) \otimes c (y_2) \otimes \ldots \otimes c (y_n)$,
only the terms of the form
$c (y_1) \otimes c (y_2) \otimes \ldots \otimes Q \, c (y_n)$
in the expansion of $\pi_n \, {\bf Q} \, \mathbb{E}$
contribute in Eq.~\eqref{QE}.
Using Eqs.~\eqref{omega_n} and~\eqref{Q-cyclicity}, we find
\begin{equation}
\begin{split}
& \omega_n \, ( \, c (y_1) \otimes \ldots \otimes c (y_{n-1}) \otimes Q \, c (y_n) \,,\,
d (x_1) \otimes \ldots \otimes d (x_{n-1}) \otimes c (x_n) \, ) \\
& = (-1)^{n-1} \,
\omega \, ( \, c (y_1) \,,\, d (x_1) \, ) \, \ldots \, \omega \, ( \, c (y_{n-1}) \,,\, d (x_{n-1}) \, ) \,\,
\omega \, ( \, Q \, c (y_n) \,,\, c (x_n) \, ) \\
& = (-1)^n \,
\omega \, ( \, c (y_1) \,,\, d (x_1) \, ) \, \ldots \, \omega \, ( \, c (y_{n-1}) \,,\, d (x_{n-1}) \, ) \,\,
\omega \, ( \, c (y_n) \,,\, Q \, c (x_n) \, ) \\
& = (-1)^n \, \omega_n \, ( \, c (y_1) \otimes \ldots \otimes c (y_{n-1}) \otimes c (y_n) \,,\,
d (x_1) \otimes \ldots \otimes d (x_{n-1}) \otimes Q \, c (x_n) \, ) \,.
\end{split}
\end{equation}
We therefore have
\begin{equation}
\begin{split}
& \omega_n \, \biggl( \, \pi_n \, {\bf Q} \, \mathbb{E} \,,
\int d^d x_1 \, h_1 (x_1) \, d (x_1) 
\otimes \ldots
\otimes \int d^d x_n \, \phi (x_n) \, c (x_n) \, \biggr) \\
& = (-1)^n \, \omega_n \, \biggl( \, \pi_n \, \mathbb{E} \,,
\int d^d x_1 \, h_1 (x_1) \, d (x_1) 
\otimes \ldots
\otimes \, Q \int d^d x_n \, \phi (x_n) \, c (x_n) \, \biggr) \,.
\end{split}
\end{equation}
It follows from Eq.~\eqref{variation-observable}
that this is the complex number we obtain
by the linear map associated with $\mathbb{E}$ from the observable
\begin{equation}
(-1)^n \, \Biggl[ \, \prod_{k=1}^{n-1} \,
\biggl( \, \int d^d x_k \, h_k (x_k) \, \varphi (x_k) \, \biggr) \, \Biggr] \,
\biggl( \, \int d^d x_n \, \phi (x_n) \, \frac{\delta S}{\delta \varphi (x_n)} \, \biggr) \,.
\end{equation}
Let us also consider the quantity
\begin{equation}
\omega_n \, \biggl( \, \pi_n \, {\bf U} \, \mathbb{E} \,,
\int d^d x_1 \, h_1 (x_1) \, d (x_1) 
\otimes \ldots
\otimes \int d^d x_{n-1} \, h_{n-1} (x_{n-1}) \, d (x_{n-1})
\otimes \int d^d x_n \, \phi (x_n) \, c (x_n) \, \biggr) \,.
\label{UE}
\end{equation}
When we expand $\pi_{n-2} \, \mathbb{E}$ using the basis vectors
$c (y_1) \otimes c (y_2) \otimes  \ldots \otimes c (y_{n-2})$,
only the terms of the form
$c (y_1) \otimes \ldots \otimes c (y_{i-1})
\otimes c (y) \otimes c (y_i) \otimes \ldots \otimes c (y_{n-2}) \otimes d (y)$
in the expansion of $\pi_n \, {\bf U} \, \mathbb{E}$
contribute in Eq.~\eqref{UE}.
Using Eq.~\eqref{omega_n}, we find
\begin{align}
& \omega_n \, ( \, c (y_1) \otimes \ldots \otimes c (y_{i-1})
\otimes c (y) \otimes c (y_i) \otimes \ldots \otimes c (y_{n-2}) \otimes d (y) \,,\,
d (x_1) \otimes \ldots \otimes d (x_{n-1}) \otimes c (x_n) \, ) \nonumber \\
& = (-1)^{n-1} \, \omega \, ( \, c (y) \,,\, d (x_i) \, ) \,\,
\omega \, ( \, d (y) \,,\, c (x_n) \, ) \nonumber \\
& \quad~ \times
\omega \, ( \, c (y_1) \,,\, d (x_1) \, ) \, \ldots \,
\omega \, ( \, c (y_{i-1}) \,,\, d (x_{i-1}) \, ) \,\,
\omega \, ( \, c (y_i) \,,\, d (x_{i+1}) \, ) \, \ldots \,
\omega \, ( \, c (y_{n-2}) \,,\, d (x_{n-1}) \, ) \nonumber \\
& = (-1)^{n-1} \, \omega \, ( \, c (y) \,,\, d (x_i) \, ) \,\,
\omega \, ( \, d (y) \,,\, c (x_n) \, ) \nonumber \\
& \quad~ \times
\omega_{n-2} \, ( \, c (y_1) \otimes \ldots \otimes c (y_{n-2}) \,,\,
d (x_1) \otimes \ldots \otimes d (x_{i-1})
\otimes d (x_{i+1}) \otimes \ldots \otimes d (x_{n-1}) \, ) \,.
\end{align}
Since
\begin{equation}
\int d^d y \, d^d x_i \, d^d x_n \, h_i (x_i) \, \phi (x_n) \,\,
\omega \, ( \, c (y) \,,\, d (x_i) \, ) \,\, \omega \, ( \, d (y) \,,\, c (x_n) \, )
= {}-\int d^d x_n \, h_i (x_n) \, \phi (x_n) \,,
\end{equation}
we have
\begin{equation}
\begin{split}
& \omega_n \, \biggl( \, \pi_n \, {\bf U} \, \mathbb{E} \,,
\int d^d x_1 \, h_1 (x_1) \, d (x_1) 
\otimes \ldots
\otimes \int d^d x_n \, \phi (x_n) \, c (x_n) \, \biggr) \\
& = (-1)^n \sum_{i=1}^{n-1} \int d^d x_n \, h_i (x_n) \, \phi (x_n) \\
& \quad~ \times \omega_{n-2} \, \biggl( \, \pi_{n-2} \, \mathbb{E} \,,
\int d^d x_1 \, h_1 (x_1) \, d (x_1) \otimes \ldots
\otimes \int d^d x_{i-1} \, h_{i-1} (x_{i-1}) \, d (x_{i-1}) \\
& \qquad \qquad \qquad~ \otimes \int d^d x_{i+1} \, h_{i+1} (x_{i+1}) \, d (x_{i+1}) \otimes \ldots
\otimes \int d^d x_{n-1} \, h_{n-1} (x_{n-1}) \, \biggr) \,.
\end{split}
\end{equation}
This is the complex number we obtain by the linear map associated with $\mathbb{E}$ from the observable
\begin{equation}
(-1)^n \sum_{i=1}^{n-1} \, \Biggl[ \, \prod_{k \ne i} \,
\biggl( \, \int d^d x_k \, h_k (x_k) \, \varphi (x_k) \, \biggr) \, \Biggr] \,
\biggl( \, \int d^d x_n \, h_i (x_n) \, \phi (x_n) \, \biggr) \,.
\end{equation}
If the relation
\begin{equation}
{\bf Q} \, \mathbb{E} = \hbar \, {\bf U} \, \mathbb{E}
\end{equation}
is satisfied, the observable
\begin{equation}
(-1)^n \, \Biggl[ \, \prod_{k=1}^{n-1} \,
\biggl( \, \int d^d x_k \, h_k (x_k) \, \varphi (x_k) \, \biggr) \, \Biggr] \,
\biggl( \, \int d^d x_n \, \phi (x_n) \, \frac{\delta S}{\delta \varphi (x_n)} \, \biggr)
\end{equation}
and the observable
\begin{equation}
\hbar \, (-1)^n \sum_{i=1}^{n-1} \, \Biggl[ \, \prod_{k \ne i} \,
\biggl( \, \int d^d x_k \, h_k (x_k) \, \varphi (x_k) \, \biggr) \, \Biggr] \,
\biggl( \, \int d^d x_n \, h_i (x_n) \, \phi (x_n) \, \biggr)
\end{equation}
are mapped to the same complex number.
In other words, the observable
\begin{equation}
\begin{split}
\hbar \sum_{i=1}^{n-1} \, \Biggl[ \, \prod_{k \ne i} \,
\biggl( \, \int d^d x_k \, h_k (x_k) \, \varphi (x_k) \, \biggr) \, \Biggr] \,
\biggl( \, \int d^d x_n \, h_i (x_n) \, \phi (x_n) \, \biggr) \\
{}-\Biggl[ \, \prod_{k=1}^{n-1} \,
\biggl( \, \int d^d x_k \, h_k (x_k) \, \varphi (x_k) \, \biggr) \, \Biggr] \,
\biggl( \, \int d^d x_n \, \phi (x_n) \, \frac{\delta S}{\delta \varphi (x_n)} \, \biggr)
\end{split}
\end{equation}
is mapped to zero. This means that the linear map
\begin{equation}
\omega_n \, \biggl( \, \pi_n \, \mathbb{E} \,,
\int d^d x_1 \, f_1 (x_1) \, d (x_1) 
\otimes \int d^d x_2 \, f_2 (x_2) \, d (x_2)
\otimes \ldots
\otimes \int d^d x_n \, f_n (x_n) \, d (x_n) \, \biggr)
\end{equation}
is compatible with the equivalence relation~\eqref{equivalence-relation},
and it gives the correlation function
under the definition by Costello and Gwilliam.
On the other hand, it has been shown in Ref.~\cite{Konosu:2024dpo} that
\begin{equation}
( \, {\bf Q} -\hbar \, {\bf U} \, ) \, \frac{1}{{\bf I} -\hbar \, {\bm h} \, {\bf U}} \, {\bf 1} = 0 \,. 
\end{equation}
Therefore, the correlation function under the definition by Costello and Gwilliam
is given by
\begin{equation}
\omega_n \, \biggl( \, \pi_n \, \frac{1}{{\bf I} -\hbar \, {\bm h} \, {\bf U}} \, {\bf 1} \,,
\int d^d x_1 \, f_1 (x_1) \, d (x_1) 
\otimes \ldots
\otimes \int d^d x_n \, f_n (x_n) \, d (x_n) \, \biggr) \,.
\end{equation}

We can summarize this in the following way.
When $\mathbb{E}$ in $T \mathcal{H}_1$ satisfies the relation
\begin{equation}
( \, {\bf Q} -\hbar \, {\bf U} \, ) \, \mathbb{E} = 0 \,,
\end{equation}
a {\it state} which is a linear map from an observable $\mathcal{O}$ to a complex number
is given by
\begin{equation}
\langle \, \mathcal{O} \, \rangle = \omega \, ( \, \mathbb{E} \,, \mathcal{O} \, ) \,,
\end{equation}
where
\begin{equation}
\omega \, ( \, \mathcal{A} \,, \mathcal{B} \, )
= \sum_{n=0}^\infty \omega_n \, ( \, \pi_n \, \mathcal{A} \,, \pi_n \, \mathcal{B} \, )
\end{equation}
with $\omega_0 \, ( \, {\bf 1}, {\bf 1} \, ) = 1$.
We can generalize this to interacting theories.
When $\mathbb{E}$ in $T \mathcal{H}_1$ satisfies the relation
\begin{equation}
( \, {\bf Q} +{\bm m} -\hbar \, {\bf U} \, ) \, \mathbb{E} = 0 \,,
\label{state-condition-Euclidean}
\end{equation}
a state is given by
\begin{equation}
\langle \, \mathcal{O} \, \rangle = \omega \, ( \, \mathbb{E} \,, \mathcal{O} \, ) \,.
\end{equation}
This can also be generalized to Lorentzian theories.
When $\mathbb{E}$ in $T \mathcal{H}_1$ satisfies the relation
\begin{equation}
( \, {\bf Q} +{\bm m} +i \hbar \, {\bf U} \, ) \, \mathbb{E} = 0 \,,
\label{state-condition-Lorentzian}
\end{equation}
a state is given by
\begin{equation}
\langle \, \mathcal{O} \, \rangle = \omega \, ( \, \mathbb{E} \,, \mathcal{O} \, ) \,.
\end{equation}
The cohomology problem by Costello and Gwilliam is thus translated
into the construction of $\mathbb{E}$ in $T \mathcal{H}_1$
which satisfies Eq.~\eqref{state-condition-Euclidean} for Euclidean theories
and Eq.~\eqref{state-condition-Lorentzian} for Lorentzian theories.
As shown in Ref.~\cite{Konosu:2024dpo},
\begin{equation}
\mathbb{E} = \frac{1}{{\bf I} +{\bm h} \, {\bm m} -\hbar \, {\bm h} \, {\bf U}} \, {\bf 1}
\label{Euclidean-E}
\end{equation}
satisfies Eq.~\eqref{state-condition-Euclidean} and
\begin{equation}
\mathbb{E} = \frac{1}{{\bf I} +{\bm h} \, {\bm m} +i \hbar \, {\bm h} \, {\bf U}} \, {\bf 1}
\label{Lorentzian-E}
\end{equation}
satisfies Eq.~\eqref{state-condition-Lorentzian}
so that the formula for correlation functions
based on quantum $A_\infty$ algebras gives
an {\it explicit} solution to the cohomology problem by Costello and Gwilliam.\footnote{
Since $( \, {\bf Q} +{\bm m} -\hbar \, {\bf U} \, )^2 = 0$
and $( \, {\bf Q} +{\bm m} +i \hbar \, {\bf U} \, )^2 = 0$,
the construction of $\mathbb{E}$ satisfying Eq.~\eqref{state-condition-Euclidean}
or~\eqref{state-condition-Lorentzian} is also a cohomology problem.
However, there are no exact states in $T \mathcal{H}_1$
so any solution to Eq.~\eqref{state-condition-Euclidean} or~\eqref{state-condition-Lorentzian}
is in the cohomology.
}

The observables homogeneous of degree $n$ in Eq.~\eqref{observables}
can be generalized to a class of observables given by
\begin{equation}
\int d^d x_1 d^d x_2 \ldots d^d x_n \, f (x_1, x_2, \ldots , x_n) \,
\varphi (x_1) \, \varphi (x_2) \ldots \varphi (x_n) \,.
\label{generalized-observables}
\end{equation}
On the other hand, a general element of $\mathcal{H}_2^{\otimes n}$ can be expanded as
\begin{equation}
\int d^d x_1 d^d x_2 \ldots d^d x_n \, f (x_1, x_2, \ldots , x_n) \,
d (x_1) \otimes d (x_2) \otimes \ldots \otimes d (x_n) \,.
\label{H_2^n-expansion}
\end{equation}
While $f (x_1, x_2, \ldots , x_n)$ in Eq.~\eqref{generalized-observables}
is a totally symmetric function of $x_1$, $x_2$, \ldots \,, $x_{n-1}$, and $x_n$ by construction,
$f (x_1, x_2, \ldots , x_n)$ in Eq.~\eqref{H_2^n-expansion} is not symmetric.
It turned out, however, that
\begin{equation}
\omega_n \, \biggl( \, \pi_n \, \mathbb{E} \,, \int d^d x_1 d^d x_2 \ldots d^d x_n \, f (x_1, x_2, \ldots , x_n) \,
d (x_1) \otimes d (x_2) \otimes \ldots \otimes d (x_n) \, \biggr)
\end{equation}
with $\mathbb{E}$ in Eq.~\eqref{Euclidean-E} for Euclidean theories
or $\mathbb{E}$ in Eq.~\eqref{Lorentzian-E} for Lorentzian theories
can be nonvanishing only when $f (x_1, x_2, \ldots , x_n)$
is a totally symmetric function of $x_1$, $x_2$, \ldots \,, $x_{n-1}$, and $x_n$.
For $A_\infty$ algebras, we use the tensor algebra instead of the symmetric algebra
so this is a special property of $\mathbb{E}$.
This is not manifest in the conditions of Eqs.~\eqref{state-condition-Euclidean}
and~\eqref{state-condition-Lorentzian},
but the solution~$\mathbb{E}$ has to have this property
in order for the associated state to give symmetric correlation functions.
It is surprising that the solution~$\mathbb{E}$
in Eq.~\eqref{Euclidean-E} or in Eq.~\eqref{Lorentzian-E}
has this property without any symmetrization procedures.
The formula would be simpler compared to the approach based on $L_\infty$ algebras because of this property,
but this can be a more important advantage of using $A_\infty$ algebras
when we consider the $1/N$ expansion.

\end{document}